\documentclass[prb,showpacs,preprintnumbers,amsmath,amssymb]{revtex4}

\usepackage{epsfig}% Include figure files
\usepackage{graphicx}% Include figure files
\usepackage{dcolumn}% Align table columns on decimal point
\usepackage{bm}% bold math

\newcommand{\EQ}{\begin{equation}}
\newcommand{\EN}{\end{equation}}
\newcommand{\ea}{\end{eqnarray}}
\newcommand{\ba}{\begin{eqnarray}}

\newcommand{\bear}{\begin{eqnarray}}
\newcommand{\ear}{\end{eqnarray}}

\begin{document}

\title{$d$-wave superconductivity in Hubbard model on the square lattice perturbed by
weak 3D uniaxial anisotropy}
\author{J. M. P. Carmelo} 
\affiliation{GCEP-Centre of Physics, University of Minho, Campus Gualtar, P-4710-057 Braga, Portugal}

\date{January 2010}
%\date{\today}

%%%%%%%%%%%%%%%%%%%%%%%%%%%%%%%%%%%%%%%%%%%%%%%%%%%%%%%%%%%%%%%%%%%%%%%%%%
%                              abstract                                  %
%%%%%%%%%%%%%%%%%%%%%%%%%%%%%%%%%%%%%%%%%%%%%%%%%%%%%%%%%%%%%%%%%%%%%%%%%%

\begin{abstract}
The Hubbard model on a square lattice is one of the most studied condensed-matter quantum problems.
Here we find evidence that for intermediate $U/4t$ values and a hole-concentration
range $x\in (x_c,x_*)$ the ground state of the Hubbard model on the square lattice perturbed by
weak three-dimensional (3D) uniaxial anisotropy has long-range $d$-wave superconducting order. 
Here $t$ is the effective nearest-neighbor 
transfer integral and $U$ the effective on-site repulsion. The lower critical concentration $x_c$
involves the Ginzburg number Gi and is approximately given by $x_c\approx {\rm Gi}+x_0\approx 0.05$.
Here $x_0<{\rm Gi}$ is a small critical hole concentration that
marks a sharp quantum phase transition from a Mott-Hubbard insulator with long-range antiferromagnetic
order for $x<x_0$ to an Anderson insulator with short-range incommensurate spiral spin
order for $x\in (x_0,x_c)$. The value of the critical hole concentration $x_*$ depends on $U/4t$
and is given by $x_*\approx 0.27$ for $U/4t\approx 1.525$.
The long-range $d$-wave superconducting order emerges below a critical 
temperature $T_c$ for a hole concentration range centered at $x_{op}= (x_c+x_*)/2\approx 0.16$. 
It results from the effects of the residual 
interactions of the charge $c$ and spin-neutral two-spinon $s1$ fermions of Ref. \cite{companion2}, as a 
by-product of the short-range spin correlations. The spin subsystem provides through such 
interactions the energy needed for the effective pairing coupling between the $c$ fermions of the 
virtual-electron pair configurations. 
\end{abstract}

\maketitle
%%%%%%%%%%%%%%%%%%%%%%%%%%%%%%%%%%%%%%%%%%%%%%%%%%%%%%%%%%%%%%%%%%%%%%%%%%
%                              body of paper                             
%%%%%%%%%%%%%%%%%%%%%%%%%%%%%%%%%%%%%%%%%%%%%%%%%%%%%%%%%%%%%%%%%%%%%%%%%%
%\tableofcontents

\section{Introduction}

The Hubbard model on a square lattice \cite{companion2} is one of the most studied 
condensed-matter quantum problems. However, it has no exact solution and many open 
questions about its properties remain unsolved. Besides being a many-electron
problem with physical interest in its own right, there is some consensus that  
it is the simplest toy model for describing the effects of electronics 
correlations in the cuprate superconductors 
\cite{two-gaps,k-r-spaces,duality,2D-MIT,Basov,ARPES-review,Tsuei,pseudogap-review,vortices-RMP}
and their Mott-Hubbard insulators parent compounds \cite{companion2,Cuprates-insulating-phase,LCO-neutr-scatt}.
In addition, the model can be experimentally realized with unprecedented 
precision in systems of ultra-cold fermionic atoms on an optical lattice. 
One may expect very detailed experimental results over a wide range of 
parameters to be available \cite{Zoller}. The pairing of fermions lies at
the heart of both superconductivity studied in this paper and 
superfluidity. Recent studies of a system of $^6$Li ultra-cold atoms
observed a zero-temperature quantum phase transition from a
fully paired state to a partially polarized normal state \cite{pairing-cold-at}.
Hence, our studies are of interest for both cuprate
superconductors and systems of ultra-cold fermionic atoms on an optical lattice.

The virtual-electron pairing mechanism studied in this paper is consistent with
the evidence provided in Refs. \cite{Yu-09,spin-super} that unconventional superconductivity 
is in different classes of systems such as cuprate superconductors, heavy-fermion 
superconductors, and iron arsenides mediated by magnetic fluctuations.
Our investigations have as starting point the square-lattice quantum liquid
of Ref. \cite{companion2}, which refers to the Hubbard model in the {\it one- and two-electron 
subspace} \cite{companion2}. It is spanned by the ground state and the excited states 
generated by application onto it of one- and two-electron operators. 
For such a square-lattice quantum liquid only the charge $c$ fermions
and spin-neutral two-spinon $s1$ fermions play an active role.
The $c$ and $s1$ fermion description and the related 
general rotated-electron description are consistent with the global 
$SO(3)\times SO(3)\times U(1)=[SO(4)\times U(1)]/Z_2$ symmetry
found recently in Ref. \cite{bipartite} for the Hubbard model
on any bipartite lattice and thus on the square lattice. Such a global symmetry is an extension 
of the $SO(4)$ symmetry known to occur for the model on such lattices \cite{Zhang}. 

An extended presentation of the results of this manuscript, including 
their application to the study of the unusual properties of the
hole-doped cuprate superconductors, can be found in Ref. \cite{cuprates0}. 

The Hubbard model on a square lattice with 
torus periodic boundary conditions, spacing $a$, 
$N_a^2\equiv [N_a]^2$ sites, lattice edge length $L=N_a\,a$,
and $N_a\gg1$ even reads,
\begin{equation}
\hat{H} = -t\sum_{\langle\vec{r}_j\vec{r}_{j'}\rangle}\sum_{\sigma =\uparrow
,\downarrow}[c_{\vec{r}_j,\sigma}^{\dag}\,c_{\vec{r}_{j'},\sigma}+h.c.] + 
U\,[N_a^2-\hat{Q}]/2 \, ; \hspace{0.35cm}
{\hat{Q}} = \sum_{j=1}^{N_a^2}\sum_{\sigma =\uparrow
,\downarrow}\,n_{\vec{r}_j,\sigma}\,(1- n_{\vec{r}_j,-\sigma}) \, .
\label{H}
\end{equation}
Here $n_{{\vec{r}}_j,\sigma} = c_{\vec{r}_j,\sigma}^{\dag} c_{\vec{r}_j,\sigma}$
and $-\sigma=\uparrow$ (and $-\sigma=\downarrow$)
for $\sigma =\downarrow$ (and $\sigma =\uparrow$)
counts the number of electron singly occupied sites. Consistently, the operator
${\hat{D}}=[{\hat{N}}-{\hat{Q}}]/2$ where 
${\hat{N}} = \sum_{\sigma}{\hat{N}}_{\sigma}$ and ${\hat{N}}_{\sigma}=\sum_{j=1}^{N_a^2}
n_{{\vec{r}}_j,\sigma}$ counts that of electron doubly
occupied sites. We denote the $\eta$-spin (and spin) value of the energy 
eigenstates by $S_{\eta}$ (and $S_s$) and the corresponding 
projection by $S^z_{\eta}= -[N_a^2-N]/2$ (and $S^z_s= -[N_{\uparrow}-
N_{\downarrow}]/2$). We focus our attention on initial ground states with 
hole concentration $x=[N_a^2-N]/N_a^2\geq 0$ and spin
density $m=[N_{\uparrow}-N_{\downarrow}]/N_a^2=0$ and
their excited states belonging to the one- and two-electron
subspace.

As found in Ref. \cite{companion2}, one can perform an
extended Jordan-Wigner transformation \cite{J-W,Wang} that maps the 
$s1$ bond particles onto $s1$ fermions with 
operators $f^{\dag}_{{\vec{r}}_{j},s1}$ and $f_{{\vec{r}}_{j},s1}$
whose real-space coordinates ${\vec{r}}_j$ where
$j=1,...,N^2_{a_{s1}}$ are those of
the square $s1$ effective lattice with
$N^2_{a_{s1}}\approx (1-x)N^2_a/2$ sites.  
Such operators are related to the corresponding 
$s1$ bond-particle operators $g^{\dag}_{{\vec{r}}_{j},s1}$
and $g_{{\vec{r}}_{j},s1}$, respectively, defined in Ref. 
\cite{companion1b} as follows,
\begin{eqnarray}
f^{\dag}_{{\vec{r}}_{j},s1} & = & e^{i\phi_{j,s1}}\,
g^{\dag}_{{\vec{r}}_{j},s1} 
\, ; \hspace{0.35cm} 
f_{{\vec{r}}_{j},s1} = e^{-i\phi_{j,s1}}\,
g_{{\vec{r}}_{j},s1} \, , 
\nonumber \\
\phi_{j,s1} & = & \sum_{j'\neq j}f^{\dag}_{{\vec{r}}_{j'},s1}
f_{{\vec{r}}_{j'},s1}\,\phi_{j',j,s1} \, ; \hspace{0.35cm} 
\phi_{j',j,s1} = \arctan \left({{x_{j'}}_2-{x_{j}}_2\over {x_{j'}}_1-{x_{j}}_1}\right) \, .
\label{JW-f+}
\end{eqnarray}
Here the phase $\phi_{j',j,s1}$ is defined in the range $\phi_{j',j,s1}\in (0,2\pi)$ and ${x_{j}}_1$ and ${x_{j}}_2$
(and ${x_{j'}}_1$ and ${x_{j'}}_2$) are the Cartesian components of
the vector ${\vec{r}}_{j}$ (and ${\vec{r}}_{j'}$). The phase $\phi_{j,s1}$
is that created by a gauge field whose fictitious magnetic field and corresponding
effective vector potential read,
\begin{eqnarray}
{\vec{B}}_{s1} ({\vec{r}}_j) & = & {\vec{\nabla}}_{\vec{r}_j}\times {\vec{A}}_{s1} ({\vec{r}}_j)
= \Phi_0\sum_{j'\neq j} n_{\vec{r}_{j'},s1}\,\delta ({\vec{r}}_{j'}-{\vec{r}}_{j})\,{\vec{e}}_{x_3} \, ,
\nonumber \\
{\vec{A}}_{s1} ({\vec{r}}_j) & = & \Phi_0\sum_{j'\neq j}
{n_{\vec{r}_{j'},s1}
\over \vert{\vec{r}}_{j}-{\vec{r}}_{j'}\vert}\,
{\vec{e}}_{\phi_{j',j}+\pi/2} \, ,
\label{A-j-s1}
\end{eqnarray}   
respectively. In these expressions ${\vec{e}}_{x_3}$ is a unit vector perpendicular to the plane, $\Phi_0=1$ in our units, 
and $n_{\vec{r}_j,s1} = f_{\vec{r}_j,s1}^{\dag}\,f_{\vec{r}_j,s1}$ is the $s1$ fermion local density operator.
In equation (\ref{A-j-s1}) and the remaining of this paper we denote by,
\begin{equation}
{\vec{e}}_{\phi}  =
\left[\begin{array}{c}
\cos \phi \\
\sin \phi
\end{array} \right] \, ,
\label{unit-vector}
\end{equation}
an in-plane unit vector whose direction is defined by the angle $\phi$. The 
effective potential (\ref{A-j-s1}) generates long-range interactions between
the $s1$ fermions. In addition, each spin-neutral
two-spinon $s1$ fermion has in average a flux tube of one flux quantum 
attached to it. 

The components ${q_j}_{x_1}$ and ${q_j}_{x_2}$ of the discrete momenta
${\vec{q}}_j$ of the $s1$ fermions are the eigenvalues of the two corresponding 
$s1$ translation generators in the presence of the fictitious magnetic field ${\vec{B}}_{s1}$ of Eq. (\ref{A-j-s1}). 
The $s1$ fermion operators are defined in and act onto subspaces
spanned by mutually neutral states \cite{companion2}. Therefore,
such operators commute and the two components ${q_j}_{x_1}$ and 
${q_j}_{x_2}$ can be simultaneously specified. A property that plays a central role in the studies of
this paper and Ref. \cite{cuprates} is that for vanishing spin density $m=0$ ground states at
finite hole concentrations $x>0$
the $s1$ fermion momentum band is full and for
one-electron and two-electron excited states 
displays a single hole and none or two holes,
respectively. The $s1$ - $s1$ fermion  
interactions associated with the effective vector potential of Eq.  
(\ref{A-j-s1}) are stronger than those
that arise between the emerging $s1$ fermions and
pre-existing $c$ fermions. In spite of that, the former do not lead to
$s1$ - $s1$ fermion inelastic scattering. The
obvious reason is that due to phase-space restrictions
associated with the exclusion principle and
energy and momentum conservation requirements
there are no available momentum values
in the $s1$ band for excited-state occupancy
configurations. 

According to Mermin-Wagner-Berezinskii Theorem 
\cite{MW,Mer,Bere-70,Bere}, in 2D quantum systems destruction of the fluctuations of 
long-range orders occurs at finite temperatures.
To search for long-range superconducting orders at finite temperature we add 
a small three-dimensional (3D) uniaxial anisotropy perturbation to the square-lattice Hamiltonian (\ref{H}),
\begin{equation}
\hat{H}_{3D} = \hat{H} + \hat{H}_{\perp} \, ; \hspace{0.35cm}
\hat{H}_{\perp} = -t_{\perp}\sum_{\langle\vec{r}_j\vec{r}_{j'};\perp\rangle}\sum_{\sigma =\uparrow
,\downarrow}[c_{\vec{r}_j,\sigma}^{\dag}\,c_{\vec{r}_{j'},\sigma}+h.c.] \, ; \hspace{0.35cm}
M = {1\over 2t_{\perp}} \, .
\label{M-t}
\end{equation}
Here the sum $\sum_{\langle\vec{r}_j\vec{r}_{j'};\perp\rangle}$ runs over first-neighboring
sites on nearest-neighboring planes, $t_{\perp}\ll t$ is a small transfer integral associated with 
electron hopping between such planes, and the mass $M$ is given in units of lattice spacing 
and Planck constant one. For the Hamiltonian $\hat{H}_{3D}$ of Eq. (\ref{M-t}) $N$ denotes the expectation 
value of the number of electrons per plane. In addition to weak 3D uniaxial anisotropy, in this
paper we find evidence that for the hole concentration range $x\in (x_c,x_*)$ the effects of 
the cuprates intrinsic disorder are very weak and can be accounted for by an oversimplified scheme 
involving a single effective suppression coefficient $\gamma_d$. Here $x_c\approx 0.05$
is a critical hole concentration introduced below and $x_*$ is the critical hole concentration 
above which there is no short-range spin order at zero temperature \cite{companion2}. 

In this paper we consider three completely different types of anisotropy: (i) The 3D uniaxial anisotropy
associated with the small Hamiltonian term $\hat{H}_{\perp}$ of Eq. (\ref{M-t}); (ii) The in-plane anisotropy as
measured by the values of the Fermi-velocity anisotropy coefficient $\eta_{\Delta}= {\rm max}\,V^{\Delta}_{Bs1}/V_{Fc}$
and Fermi-energy anisotropy coefficient $\eta_0= \vert\Delta\vert/W_c^h$ introduced in Ref. \cite{companion2}
and given in Section II -- It refers to the anisotropy of the Fermi line associated with the dependence of its 
finite energy width on the Fermi angle $\phi$ given in Eq. (\ref{phi-F}) of Appendix A;
(iii) Finally, the in-plane superfluid-density anisotropy, which refers to the in-plane anisotropy of the superfluid density
of some cuprate superconductors. These three types of anisotropy are in the following in general called
3D uniaxial anisotropy, anisotropy, and superfluid-density anisotropy, respectively.
  
The effects of the 3D uniaxial anisotropy perturbation $\hat{H}_{\perp}$ on the 
square-lattice Hamiltonian (\ref{H}) can for very small anisotropy parameter $\varepsilon^2 =m_c^*/M\ll 1$ 
be ignored for most properties. Here $m_c^*$ is the $c$ fermion mass of expression (\ref{bands}) of 
Appendix A and $M=1/2t_{\perp}\gg 1/2t$ the effective mass given in Eq. (\ref{M-t}) associated with electron hopping between 
nearest-neighboring planes. The $c$ fermion mass becomes infinite in the limit $U/4t\rightarrow 0$. Hence in
spite of $t_{\perp}/t\ll 1$ for the original Hamiltonian $\hat{H}_{3D}$ of Eq. (\ref{M-t}), which is written in terms of
electron operators, for small $U/4t\ll u_0\approx 1.3$ the 3D uniaxial anisotropy parameter 
reaches values $\varepsilon^2 =m_c^*/M\geq 1$ larger than one upon decreasing $U/4t$. Indeed, rather than the bare electronic 
transfer integral $t$ and corresponding mass $m_c^{\infty}=1/2t$, the relevant parameter for that 
quantum liquid is the $c$ fermion mass $m_c^* =1/(2r_c t)$. That follows from the occupancy configurations 
of the $c$ fermions generating the charge degrees of freedom of states close to the energy eigenstates 
of $\hat{H}_{3D}$. The parameter $r_c=m_c^{\infty}/m_c^*$ in the $c$ fermion mass expression is the mass ratio 
given in Eq. (\ref{m*c/mc-UL}) of Appendix A. According to that equation, its limiting values are $r_c\rightarrow 0$ 
and $r_c\rightarrow 1$ for $U/4t\rightarrow 0$ and $U/4t\rightarrow\infty$, respectively.
Our study refers to the range $U/4t>u_0$ for which approximately $r_c\in (0.74, 1.00)$.
For it that $t_{\perp}/t\ll 1$ for the original Hamiltonian $\hat{H}_{3D}$ of Eq. (\ref{M-t}) implies 
as well that $\varepsilon^2 =m_c^*/M\ll 1$. Here $u_0\approx 0.13$ is the $U/4t$ value at
which the important energy scale $\Delta_0$ of Eq. (\ref{Delta-0}) of Appendix A reaches
its maximum magnitude.

The studies of this paper focus on the Hamiltonian $\hat{H}_{3D}$ of Eq. (\ref{M-t}) in 
the one- and two-electron subspace \cite{companion2}. For hole concentrations $x\in (x_c,x_*)$
we call {\it virtual-electron pair quantum liquid} (VEP quantum liquid) that quantum problem under the weak 
suppression effects considered below. The main effect of the small perturbation $\hat{H}_{\perp}$ 
of the VEP quantum liquid is for approximately $U/4t>u_0$ the emergence of the 3D uniaxial 
anisotropy parameter $\varepsilon^2 =m_c^*/M\ll 1$ in the expressions 
of some physical quantities sensitive to the thermal and quantum fluctuations \cite{vortices-RMP}.
Specific examples are the critical hole concentration $x_c\approx {\rm Gi}+x_0$, the phases 
$\theta$ of the virtual-electron pairs considered below, and 
related quantities. Here Gi is the Ginzburg number \cite{vortices-RMP} and $x_0<{\rm Gi}$ is a small critical hole concentration that
marks a sharp quantum phase transition from a Mott-Hubbard insulator with long-range antiferromagnetic
order for $x<x_0$ to a Anderson insulator with short-range incommensurate spiral spin
order for $x\in (x_0,x_c)$. As discussed in Appendix B, the strong intrinsic-disorder hole-trapping effects  
present in the hole-doped cuprate superconductors for $x\in (0,x_c)$ render $x_0$
a critical hole concentration. Moreover, such effects,
which are active mainly for hole concentrations below $x_c$, shift the critical hole concentration 
from $x_c\approx {\rm Gi}$ to $x_c\approx {\rm Gi}+x_0$. This is their only effect 
on the physics for $x\in (x_c,x_*)$. 

Concerning the short-range spin order occurring for $x\in (x_0,x_*)$, for both $x\in (x_0,x_c)$ and
$0<(x-x_c)\ll 1$ it is a incommensurate spiral spin order as that of the square-lattice
quantum liquid of Ref. \cite{companion2} for small $x$. Further details on that order for instance
for $x$ near the optimal hole concentration $x_{op}=(x_c+x_*)/2$ and $U/4t\in (u_0,u_1)$ remains an interesting open problem.
We find in this paper that for both $0<(x-x_c)\ll 1$ and $0<(x_*-x)\ll 1$ the quantum fluctuations are large,
so that the VEP quantum-liquid ground state refers to a liquid. In turn, it is found that at and near 
$x\approx x_{op}$ such fluctuations are smaller. Hence the ground state is expected to be intermediate between a
liquid, where such fluctuations are large, and a crystal, where they are small. The results
of Ref. \cite{cuprates} seem to indicate that the VEP quantum-liquid physics is for the
latter range of $x$ values controlled by a quantum critical point. 

The possibility of using $U/4t$ as a tuning parameter plays a central
role in our scheme. Indeed, the change in the $U/4t$ ratio is strongly linked to a
change to the electron-lattice coupling, since increasing $U/4t$
also involves increasing the strength of the periodic potential.
In contrast to large $U/4t$ values for which the electron-lattice coupling
strength further increases and the important energy scale
$\Delta_0$ of Eq. (\ref{Delta-0}) of Appendix A becomes small, for
intermediate $U/4t\in (u_0,u_1)$ values that strength remains much
smaller, hence freeing the correlated VEP quantum liquid from its rigid link to the underlying
lattice. Intermediate $U/4t$ values then refer to relatively high 
effective correlation strength without an increase in the
effective strength of the periodic potential. The magnitude of the
energy scale $\Delta_0$ provides a measure of such an
correlation strength effectiveness, vanishing both for $U/4t\rightarrow 0$
and $U/4t\rightarrow\infty$ and being largest for  
$U/4t\in (u_0,u_1)$. Such a correlation strength effectiveness is probably behind the exotic physics 
emerging for both intermediate $U/4t\in (u_0,u_1)$ values and hole concentrations at and near 
$x_{op}=(x_c+x_*)/2$, which as mentioned above is according to the studies of
Ref. \cite{cuprates} controlled by a quantum critical point. 

Hence often our results focus on the intermediate-$U/4t$ range $U/4t\in (u_0,u_1)$ 
and hole concentrations $x\in (x_c,x_*)$. As discussed in Ref. \cite{cuprates0}, the value of the parameter 
$\varepsilon^2 =m_c^*/M\ll 1$ is set so that $x_c\approx {\rm Gi}+x_0\approx 0.05$.
In turn, that of the zero-temperature upper hole-concentration $x_*$ is insensitive
to $\varepsilon^2$ and rather depends on $U/4t$. Alike
for the square-lattice quantum liquid of Ref. \cite{companion2}, it changes smoothly
from $x_*\approx 0.23$ for $u_0 \approx 1.3$ to $x_*\approx 0.28$ for $u_0 \approx 1.6$. 
Specifically, the value $x_*=0.27$ appropriate to the families of hole-doped cuprates considered
in this paper corresponds to $U/4t\approx u_*=1.525$.
Provided that $\varepsilon^2 =m_c^*/M\ll 1$, it is expected that the momentum occupancy configurations of 
the $c$ and $s1$ fermions that generate the energy eigenstates of the $\varepsilon^2 =0$ square-lattice
quantum liquid of Ref. \cite{companion2} are for $x\in (x_c,x_*)$ close to energy eigenstates of the
VEP quantum liquid. It then follows that the interactions of such objects remain being residual. 

The phases of the phase-coherent virtual-electron pair configurations studied in later sections 
have the form $\theta=\theta_0 +\theta_1$. Their fluctuations play an important role in the 
VEP quantum liquid physics. At zero temperature the fluctuations of the phases $\theta_0$ and $\theta_1$ become
large for $x\rightarrow x_c$ and $x\rightarrow x_*$, respectively.  
Expression in terms of the rotated-electron operators 
\cite{companion2} of the effective microscopic Hamiltonian describing the
quantum fluctuations of the VEP quantum liquid leads to a problem with basic similarities to that
considered in Ref. \cite{duality}. Fortunately, the very involved problem
of the derivation of an effective action for the phases $\theta$ can be
mapped onto a corresponding problem already investigated in that
reference. That is achieved on replacing in the Hamiltonian (1)
of Ref. \cite{duality} electron operators by rotated electron operators
\cite{companion2}. The transformation that
relates rotated electrons to electrons is unitary, so that the effective 
action for the phases $\theta$ considered in this paper
is valid for approximately $U/4t>u_0$ and specifically at $U/4t\approx u_*=1.525$. 
Expression of the quantum problem that describes the fluctuations of the phases 
$\theta=\theta_0 +\theta_1$ of the virtual-electron pairs in terms of
$c$ and $s1$ fermion operators reveals the central role plaid by
the $c$ - $s1$ fermion interactions in the emergence of a long-range superconducting
order for $x\in (x_c,x_*)$.

Indeed, strong evidence is found below that for intermediate $U/4t$ values,
vanishing spin density $m=0$, and hole concentrations in the range $x\in (x_c,x_*)$ the ground 
state of the VEP quantum liquid has a $d$-wave long-range superconducting order, coexisting with its short-range spin order. Superconductivity
emerges below a $x$ dependent critical temperature $T_c$ 
and results from the effects of the residual $c$ - $s1$ fermion interactions, as a by-product of the short-range spin correlations.
The spin subsystem provides through such interactions 
the energy needed for the effective pairing coupling
between the $c$ fermions of the virtual-electron pair configurations
introduced in the following. The suppression effects due to intrinsic disorder
or superfluid density anisotropy slightly lessen the magnitude of
$T_c$ and related physical quantities. Otherwise the $\gamma_d=1$ physics remains
unchanged under these small effects.

Our scheme is used in Refs. \cite{cuprates0,cuprates} in the further understanding of the mechanisms
behind the unusual properties of hole-doped cuprates. The studies of these references focus on five 
representative systems of the hole-doped cuprates for which $x_c\approx 0.05$ and $x_*\approx 0.27$: 
YBa$_2$Cu$_3$O$_{6+\delta}$ (YBCO 123), Bi$_{2}$Sr$_2$CaCu$_2$O$_{8+\delta}$ (Bi 2212), 
HgBa$_2$CuO$_{4+\delta}$ (Hg 1201), Tl$_2$Ba$_2$CuO$_{6+\delta}$ (Tl 2201), and 
La$_{2-x}$Sr$_x$CuO$_4$ (LSCO). In Sertion III-D it is found that for $\gamma_d\approx 1$
some of the VEP quantum liquid expressions are valid provided that $x_c/x_*<1/4$. 
The critical hole concentration values $x_c\approx 0.05$ and $x_*\approx 0.27$ 
are compatible with such a requirement. One finds $(x_c-x_0)\approx {\rm Gi}\approx 0.026$ for the four representative hole-doped
cuprates other than LSCO and $(x_c-x_0)\approx {\rm Gi}\approx 0.037$ for LSCO. The relation 
${\rm Gi}= G/\varepsilon^2=G\,[M/m_c^*]$ then implies that our scheme is not valid for the 
$\varepsilon^2 =m_c^*/M=0$ 2D square-lattice quantum liquid studied in Ref. \cite{companion2} 
Indeed, such a problem corresponds to vanishing values of $\varepsilon^2 =m_c^*/M$, such that 
$1/\varepsilon^2=M/m_c^*\rightarrow\infty$ and thus $x_c/x_*>1/4$. Fortunately, one finds 
that $G$ is for the five representative systems very small, $G\approx 10^{-5} - 10^{-4}$.
This allows that the ratio $\varepsilon^2 =m_c^*/M\approx 10^{-4} - 10^{-2}$ is smaller than ${\rm Gi}$. 
Interestingly, our results provide evidence that the zero-temperature hole-concentration width
$(x_*-x_c)$ of the superconducting dome decreases upon decreasing the
3D uniaxial anisotropy. This is consistent with the weak 3D uniaxial anisotropy effects associated with
the small Hamiltonian term $\hat{H}_{\perp}$ of Eq. (\ref{M-t}) being needed for
the occurrence of superconductivity in the quantum liquid under investigation in this paper. 

Our preliminary studies on the five representative systems focus mainly onto the
hole concentration range $x\in (x_c,x_*)$ where $x_c\approx 0.05$ and 
$x_*\approx 0.27$ at $U/4t\approx u_*=1.525$ for which the strong hole trapped 
effects caused by intrinsic disorder discussed in Appendix B are not active.
For that $x$ range the suppression effects originated by intrinsic disorder or in-plane superfluid density anisotropy 
are very weak for the four representative hole-doped cuprates other than the 
random alloy LSCO. Indeed, the minimum magnitude $\gamma_d^{min}$ reached at 
$x_{op} = (x_c+x_*)/2\approx 0.16$ of the corresponding {\it suppression coefficient} 
$\gamma_d =T_c/T_c\vert_ {\alpha_d=0}$ is found in Ref. \cite{cuprates0} to belong to the range $\gamma_d^{min} \in (0.94,0.98)$,
whereas $\gamma_d =1$ both for $0<(x-x_c)\ll 1$ and $0<(x_*-x)\ll 1$. 
Our investigations and those of Ref. \cite{cuprates} provide evidence that for $x\in (x_c,x_*)$  the interplay
of the electronic correlations described in this paper by the $c$ - $s1$ fermion residual interactions
with the weak effects of the 3D uniaxial anisotropy perturbation $\hat{H}_{\perp}$ of Eq. (\ref{M-t})
is behind the unusual universal properties of the hole-doped cuprate superconductors
\cite{two-gaps,k-r-spaces,duality,2D-MIT,Basov,ARPES-review,Tsuei,pseudogap-review}.

For the VEP quantum liquid referring to the range $x\in (x_c,x_*)$
the $x$ dependence of the fluctuations of the phases $\theta=\theta_0 +\theta_1$ is for intermediate $U/4t$
values found to imply naturally a dome-like dependence on $x$ for the critical temperature $T_c$. 
It reaches its maximum magnitude $T^{max}_c$ at the above-mentioned optimal hole concentration 
$x_{op}=(x_c+x_*)/2\approx 0.16$ for $x_c= 0.05$ and $x_*= 0.27$.
The energy parameter $\Delta_0$ appearing in the critical temperature and other
energy scales expressions plays
an important role in the square-lattice quantum liquid \cite{companion2}.
It vanishes both in the limits $U/4t\rightarrow 0$ and $U/4t\rightarrow\infty$
and goes through a maximum magnitude ${\rm max}\,\{\Delta_0\} \approx t/\pi$ at 
$U/4t=u_0\approx 1.302$. For instance, it is found in Ref. \cite{cuprates0} to control
the dependence of $T^{max}_c$ on pressure $P$. Under application of $P$ it first increases until it reaches a maximum 
magnitude at a pressure value $P=P_0$. A generalized formula valid for arbitrary magnitudes 
of the critical hole concentrations $x_c$ and $x_*$ and pressure $P\in (0,P_0)$ is introduced. 
At $P=0$ it becomes the empirical formula $(1 - T_c/T^{max}_c) = 82.64\,(x-0.16)^2$ found by many authors to 
apply to several families of hole-doped cuprates for the range $x\in (x_c,x_*)$ \cite{parabolic0,Tc-1/8}.
Importantly, an universal ratio $T^{max}_c (P_0)/T^{max}_c (0)\approx 1.26$ is predicted
for hole-doped cuprates with $P=0$ and $T=0$ critical hole concentrations $x_c\approx 0.05$ 
and $x_*\approx 0.27$. The validity of such a theoretical prediction is confirmed
for two systems with completely different values of $T^{max}_c (0)$ \cite{LBCO-pressure,YBCO-pressure}. 

Moreover, the low-temperature incommensurate peaks in the inelastic neutron scattering of LSCO 
observed at momenta $\delta \vec{P}=[\pi\pm 2\pi x,\pi]$ and $\delta \vec{P}=[\pi ,\pi\pm 2\pi x]$
for the hole concentration range $x\in (0.05,0.12)$
\cite{Cuprates-insulating-phase,neutron-Yamada,LCO-x-depen,LSCO-Nd} follow within
the VEP quantum liquid physics from a suitable generalization of the 
$s1$ fermion microscopic processes found in Ref. \cite{companion2} to lead at $x=0$
to the sharpest features in the neutron scattering of LCO. The incommensurate character 
of the LSCO peaks results from the contraction of the $s1$ momentum band boundary line
upon increasing the hole concentration $x$.

The VEP quantum liquid is consistent with the coexisting two-gap scenario \cite{two-gaps,Delta-no-super,two-gap-Tanaka,two-gap-Tacon}: 
A dome-like superconducting energy scale $2\vert\Omega\vert=4k_B T_c/(1-x_c\,T_c/x_*\,T^{max}_c)$ and 
pseudogap $2\vert\Delta\vert = (1-x/x_*)\,2\Delta_0$, over the whole dome $x\in (x_c,x_*)$. 
The energy parameters $2\vert\Delta\vert$ and $\vert\Delta\vert$ are the maximum magnitudes of the spinon pairing energy 
$2\vert\Delta_{s1} ({\vec{q}}^{\,d}_{Bs1})\vert =2\vert\Delta\vert\vert\cos 2\phi\vert$ and
corresponding one-electron gap $\vert\Delta\vert\vert\cos 2\phi\vert$, respectively. Here
${\vec{q}}^{\,d}_{Bs1}$ is a $s1$ band boundary line momentum and $d=\pm 1$ the doublicity \cite{companion2}. 
It is reached at the values $\phi=0,\pi/2$ of the Fermi angle 
$\phi$ of Eq. (\ref{phi-F}) of Appendix A. The energy scale $2\vert\Omega\vert$ is the maximum magnitude reached 
at $\phi=\pi/4$ of the superconducting virtual-electron pairing energy
 $2\vert\Omega_{s1} ({\vec{q}}^{\,d}_{Bs1})\vert =2\vert\Omega\vert
\vert\sin 2\phi\vert$ introduced in this paper. Both the magnitudes and $x$ dependences
of the energy scales $2\vert\Delta\vert$ and $2\vert\Omega\vert$ and of the superfluid density predicted
by the VEP quantum liquid scheme agree with the corresponding experimental results of the five 
representative cuprate superconductors. That for $x\approx 0.04<x_c$ only the $d$-wave gap
$\approx\vert\Delta\vert\vert\cos 2\phi\vert$ is observed in the experiments of Ref. \cite{nodall}
is not inconsistent with our two-gap picture. Indeed, $2\vert\Omega\vert$ emerges for $x>x_c$.  

The paper is organized as follows. The energy scales of the quantum liquid
and other basic physical quantities are discussed in Section II. In Section III evidence 
is found that the ground state of the VEP quantum liquid has for intermediate
$U/4t$ values and hole concentrations in the range $x\in (x_c,x_*)$ a $d$-wave
long-range superconducting order coexisting with its short-range spin order. 
A corresponding effective superconductivity theory is then introduced. 
The results presented in Section IV include the study 
of the VEP quantum liquid general energy functional, dependence
on the hole concentration $x$ of the superfluid density, and
role of the $c$ - $s1$ fermion residual interactions in the
$c$ fermion strong effective coupling. Finally, 
Section V contains the concluding remarks.

\section{Energy scales, crossover hole concentrations, and Fermi momentum}

In this section and in Appendix A some of the results of Refs. \cite{companion2,companion1b} 
needed for our studies are shortly reviewed. That as considered in Ref. \cite{companion2} the angle
between the $c$ Fermi hole momentum and $s1$ boundary line momentum 
of the $c$ fermion and $s1$ fermion hole, respectively, created upon an one-electron addition excitation
is exactly $\pi/2$ is a good approximation for the study of some
properties. However, having in view the study of the one-electron scattering properties
fulfilled in Ref. \cite{cuprates}, in this paper we account for the small deviations of that angle from $\pi/2$.

Concerning our short review of the results of Refs. \cite{companion2,companion1b},
there occurs for the square-lattice quantum liquid a sharp quantum phase transition such that 
the $m=0$ ground state has a long-range antiferromagnetic order at $x=0$ and
a short-range spiral-incommensurate spin order with strong antiferromagnetic
correlations for $0<x\ll 1$.  As a result of it, the maximum magnitude $2\vert\Delta\vert$ 
of the $s1$ fermion spinon-pairing energy has a singular behavior at $x=0$. 
Indeed, one has that $2\Delta\vert_{x=0}=\mu^0$ for $x=0$ and $m=0$ whereas
$\lim_{x\rightarrow 0}2\vert\Delta\vert =2\Delta_0$. 
Here $\mu^0$ is the chemical potential $\mu$ in the limit
$\mu^0\equiv \lim_{x\rightarrow 0}\mu$
whose behaviors are given in Eq. (\ref{DMH}) of Appendix A.
It is one half the Mott-Hubbard gap, consistently with
for $x=0$ and $m=0$ the chemical potential 
belonging to the range $\mu \in (-\mu^0,\mu^0)$.
In spite of $2\mu^0$ being the charge Mott-Hubbard gap,
it also refers to the spin degrees of freedom \cite{companion2}:
The energy scale $2\Delta\vert_{x=0}=\mu^0$ is 
the excitation energy below which the long-range 
antiferromagnetic order survives for $x=0$, 
$m=0$, and zero temperature $T=0$. 
In turn, the energy scale $\lim_{x\rightarrow 0}2\vert\Delta\vert =2\Delta_0$
plays a key role in the $x>0$ and $x>x_0$ physics of the square-lattice quantum liquid
of Ref. \cite{companion2} and the VEP quantum liquid, respectively.
The small hole concentration $x_0<x_c$ is given below.  
According to the analysis of Appendix B, the hole-traping effects caused for
hole concentrations below $x_c$ by the cuprates intrinsic disorder are strong 
and shift that sharp quantum phase transition from $x=0$ to $x=x_0$. 

The approximate limiting behaviors of the energy scale $2\Delta_0$ are provided 
in Eq. (\ref{Delta-0}) of Appendix A. As a function of $u=U/4t$, it is such that
$\partial \Delta_0 (u)/\partial u =0$ at $u=U/4t=u_0$. This is consistent with
it reaching a maximum magnitude $t/\pi$ at that $U/4t$ value. It can be expressed
in terms of the $s1$ fermion energy dispersion bandwidth $W_{s1}^0$ as,
\begin{eqnarray}
\Delta_0 & = & r_s\,4W_{s1}^0 = 4W_{s1}^0\,e^{-\lambda_s} 
\, ; \hspace{0.35cm} 
W_{s1}^0=\lim_{x\rightarrow 0}W_{s1} =W_{s1}\vert_{x=0} \, ,
\nonumber \\
W_{s1} & = & [\epsilon_{s1} ({\vec{q}}^{\,N\,d}_{Bs1})-\epsilon_{s1} (0)]
\, ; \hspace{0.35cm} 
\lambda_s = \vert\ln (\Delta_0/4W_{s1}^0)\vert \, .
\label{Delta-0-gen}
\end{eqnarray}
Here $\epsilon_{s1} ({\vec{q}})$ is the $s1$ fermion energy dispersion derived for
the square-lattice quantum liquid in Ref. \cite{companion2} whose generalized 
expression is  for the VEP quantum liquid introduced below and
${\vec{q}}^{\,N\,d}_{Bs1}$ is the $s1$ boundary line nodal momentum defined in
that reference.

For the Hubbard model on the square lattice at zero temperature $T=0$ a small 
concentration $x$ of holes prevents the occurrence of long-range antiferromagnetic order \cite{companion2}.   
At vanishing spin density $m=0$ and both for small finite temperature $T>0$ and 
zero hole concentration $x=0$ and for vanishing or small finite temperature $T\geq 0$ 
and small finite hole concentration $0<x\ll 1$ the system is driven into
a renormalized classical regime where the 
$T=0$ and $x=0$ long-range antiferromagnetic 
order is replaced by a quasi-long-range spiral-incommensurate 
spin order as that studied in Ref. \cite{Principles} for simpler spin
systems. For hole concentrations obeying the inequality
$0<x<x_*$ the system is in a short-range
spin ordered phase whose order parameter is the 
maximum magnitude of the $s1$ fermion spinon-pairing
energy $2\vert\Delta\vert =2\Delta_0\,(1-x/x_*)$. Following the
analysis of Appendix B, for the quantum problem considered
here such an order occurs for a smaller range $x\in (x_0,x_*)$.
In either case the above $2\vert\Delta\vert$ expression
is valid approximately for the range $U/4t\in (u_0, u_{\pi})$
where $u_{\pi}>u_1$ is the $U/4t$ value at which $x_*=1/\pi\approx 0.32$.
For approximately $U/4t>u_1$ and thus including at $U/4t\approx u_{\pi}$ one has that $r_s<r_c<2r_s$ and the 
approximate expression $r_s\approx e^{-4t\,u_0/U}$ of
Eq. (\ref{m*c/mc-UL}) of Appendix A is not valid. Often our results refer to
the smaller range $U/4t\in (u_0,u_1)$ for which the relation 
$r_c\approx 2r_s\approx 2e^{-4t\,u_0/U}$ approximately holds.

Alike the quasiparticle mass ratios in Fermi-liquid theory \cite{Pines}, 
the $U/4t$-dependent spin ratio $r_s=\Delta_0/4W_{s1}^0$ and charge 
mass ratio $r_c=m^{\infty}_c/m^{*}_{c}$,
both given in Eq. (\ref{m*c/mc-UL}) of Appendix A, control the effects
of the electronic correlations in many physical quantities. These ratios
play an important role in the physics of both the square-lattice quantum liquid of
Ref. \cite{companion2} and VEP quantum liquid. In contrast to a Fermi liquid, here the mass 
$m^{\infty}_c=\lim_{U/4t\rightarrow\infty}m^{*}_{c}=1/2t$, which 
refers to the limit of infinite on-site interaction, plays the role 
of bare mass. In turn, the VEP quantum liquid mass ratio $1/\varepsilon^2 =M/m^{*}_{c}$ involving
the mass $M$ of Eq. (\ref{M-t}) and the $c$ fermion mass $m_c^*$ of expression (\ref{bands}) of 
Appendix A controls the effects of the weak 3D uniaxial anisotropy. 

Unlike the Fermi-liquid quasiparticles, the
$c$ and $s1$ fermions do not evolve into electrons upon adiabatically turning
off the interaction $U$. Instead, upon turning
off adiabatically the parameter $4t^2/U$ the $c$ fermions evolve into
the spinless fermions that describe the charge degrees
of freedom of the electrons that singly occupy sites within the energy-eigenstate
configurations of the state basis used in the studies of
Ref. \cite{companion2}. Furthermore, for $4t/U\rightarrow 0$
their spin degrees of freedom are described by the 
spin-neutral two-spinon $s1$ fermions whose energy-dispersion bandwidth 
vanishes in that limit. Within a mean-field approximation for the 
fictitious magnetic field ${\vec{B}}_{s1} ({\vec{r}}_j)$ of Eq. (\ref{A-j-s1}) 
the $s1$ fermion occupancy configurations that generate the
$m=0$ ground states are in that limit those of a full lowest Landau level with
$N_{s1}=N_{a_{s1}}^2=N/2$ one-$s1$-fermion degenerate states of the 2D quantum  
Hall effect \cite{companion2}. Here $N_{a_{s1}}^2$ is the number of 
both sites of the $s1$ effective square lattice and $s1$ band discrete
momentum values. 

In turn, for $4t^2/U>0$ the $s1$ fermion dispersion acquires a finite energy bandwidth. 
Both the square-lattice quantum liquid of Ref. \cite{companion2} and the
VEP quantum liquid are defined in the one- and two-electron subspace.
For finite $U/4t$ values the $c$ and $s1$ fermions 
describe the charge and spin degrees of freedom, respectively, of the occupancy 
configurations of the rotated electrons that generate the states
that span that subspace. In it only
the $c$ and $s1$ fermions play an active role. The $c$ momentum band has the same shape 
and area $(2\pi/L)^2\,N_a^2$ as the first Brillouin zone. The $s1$ momentum band is exotic and
such that its momentum area and shape are subspace dependent.
For ground states corresponding to hole concentrations $x\in (x_c,x_*)$ and  
spin density $m=0$ the $c$ fermions have for the square-lattice quantum liquid
of Ref. \cite{companion2} an isotropic hole like $c$ Fermi line. The hole momenta
${\vec{q}}_{Fc}^{\,h\,d}$ belonging to that line, defined in Eq. (\ref{kF-qFc-qBs1}), 
are centered at the momentum $-\vec{\pi}$. 
Out of the $c$ band $N_a^2$ discrete hole
momentum values ${\vec{q}}^{\,h}_{j}$ where $j=1,...,N_a^2$, $N_c=2S_c$ are 
filled and $N_c^h=[N_a^2-2S_c]$ unfilled and
correspond to $c$ fermions and $c$ fermion holes,
respectively. For the ground state one has that $N_c=2S_c=N=(1-x)\,N_a^2$
and the hole $c$ Fermi line encloses a momentum area 
$[N_a^2-N_c]\,[2\pi/L]^2=[N_a^2-2S_c]\,[2\pi/L]^2=4\pi^2 x$.

For ground states corresponding to vanishing spin density $m=0$ the $s1$ 
band is full. Hence the number of holes in the $s1$ band vanishes, $N^h_{s1}=0$,
and the number of $s1$ fermions reads $N_{s1}=N/2=(1-x)\,N_a^2/2$.
The $s1$ momentum band is particle like so that its boundary line is 
centered at ${\vec{q}}=0$ and encloses a momentum area $2\pi^2 (1-x)$. 
The momenta ${\vec{q}}^{\,d}_{Bs1}$ of the $s1$ band boundary line are given 
in the following. For the states that span
the one- and two-electron subspace the number of holes in the
$s1$ momentum band is given by $N^h_{s1}=1$ and
$N^h_{s1}=0$ or $N^h_{s1}=2$, respectively \cite{companion2}.

For each $[N_{\uparrow},N_{\downarrow}]=[N/2,N/2-1]$-electron ground state there is a corresponding
zero-spin-density and $[N_{\uparrow},N_{\downarrow}]=[N/2,N/2]$-electron ground state with one more  
spin-down electron. We say that such a ground state is the {\it $m=0$ 
generating ground state} of both the $[N/2,N/2-1]$-electron ground state and its 
$[N/2,N/2-1]$-electron excited states of small momentum and
low energy. The reduced one-electron subspace is spanned
by a $[N/2,N/2-1]$-electron ground state and its $[N/2,N/2-1]$-electron excited states of small momentum and low energy. 
The doublicity $d=\pm 1$ labels the $s1$ fermions whose momentum occupancy configurations 
generate the states that span that subspace.
The $s1$ band momenta ${\vec{q}}$ of the reduced one-electron subspace 
$s1$ band are related to the $s1$ band momenta ${\vec{q}}_0$
of the corresponding $m=0$ generating ground state as \cite{companion2},
\begin{equation}
{\vec{q}} = A^{d}_{s1}\,{\vec{q}}_0 
\, ; \hspace{0.35cm}
\lim_{x\rightarrow 0}A^d_{s1}={\bf I} \, .
\label{qs1-qhc}
\end{equation}
Here we neglect terms of order $1/N_a^2$, which vanish in the thermodynamic limit.
${\bf I}$ denotes the $2\times 2$ unit matrix and for $x\in (x_c,x_*)$ and momenta
${\vec{q}} $ at or near the $s1$ boundary line one has that $A^{d}_{s1} = A^{d}_{F}$.
The $F$ rotation matrix $A^{d}_{F}$ is given below. 

The Fermi-velocity anisotropy coefficient $\eta_{\Delta}= {\rm max}\,V^{\Delta}_{Bs1}/V_{Fc}$,
Fermi-energy anisotropy coefficient $\eta_0= \vert\Delta\vert/W_c^h$, and hole
concentrations $x_{\Delta}$ and $x_0$ at which $\eta_{\Delta} =1$
and $\eta_0=1$, respectively, play an important role. For approximately $U/4t\in (u_0,u_1)$ 
such quantities read \cite{companion2}, 
\begin{eqnarray}
\eta_{\Delta} & = & {\rm max}\,r_{\Delta} \approx \sqrt{x\pi\over 2}\eta_0
\, ; \hspace{0.35cm} 
r_{\Delta} = {V^{\Delta}_{Bs1}\over V_{Fc}}
\, ; \hspace{0.35cm} \eta_0 =
{\vert\Delta\vert\over W_c^h} \approx \sqrt{2x_{\Delta}\over \pi}\left({1\over x}-{1\over x_*}\right) 
\, ; \hspace{0.35cm}
x_{\Delta} = {1\over 2\pi}\left({\Delta_0\over 4r_c t}\right)^2 \, , 
\nonumber \\ 
x_0 & = & {\Delta_0\over t} \left({x_*\over (2r_c)^2 + \Delta_0/t}\right) \approx {\Delta_0\over t}{1\over (2\pi)^2x_*}
\, ; \hspace{0.35cm} x_{c1} = {1\over 8} \, .
\label{x-h}
\end{eqnarray}
Here $\Delta_0 $ is the energy parameter of Eq. (\ref{Delta-0-gen}),
$r_c\approx 2r_s\approx 2e^{-4t\,u_0/U}$ where $u_0\approx 1.302$, and $W_c^h\approx x\,2\pi/m_c^*$ is
the unfilled $c$ fermion sea energy bandwidth. Following the discussions of Appendix B, the hole trapping
effects caused by strong intrinsic disorder render $x_0$ a critical hole concentration. It marks
a sharp transition between zero-temperature states with long-range antiferromagnetic order
and short-range spin order. In Ref. \cite{companion2} $x_{c1}$ was defined as the hole concentration at which 
the equality $\eta_{\Delta} =2x_0$ is satisfied. For $U/4t\in (u_0,u_1)$ that formula leads
to $x_{c1}\approx 1/8$. Since $x_{c1}$ is a mere crossover hole concentration in this paper we
use the magnitude $x_{c1}=1/8$, valid for the interaction range $U/4t\in (u_0,u_1)$.
The absolute value $V^{\Delta}_{Bs1}$ of the $s1$ fermion velocity ${\vec{V}}^{\,\Delta}_{s1} (\vec{q})$ 
of Eq. (\ref{g-velocities-def}) of Appendix A at the $s1$ boundary line and that
of the $c$ fermion velocity at hole momenta belonging to the $c$ Fermi line 
also appearing in the expressions of Eq. (\ref{x-h}) read,
\begin{equation}
V^{\Delta}_{Bs1} \equiv V^{\Delta}_{s1} ({\vec{q}}^{\,d}_{Bs1}) 
= {\vert\Delta\vert\over\sqrt{2}}\vert\sin 2\phi\vert \, ;
\hspace{0.35cm} V_{Fc} = {q^h_{Fc} \over m_c^*}\approx {\sqrt{x\pi}\,2\over m_c^*} \, ,
\label{pairing-en-v-Delta}
\end{equation}
respectively. The dependence on the Fermi angle
$\phi\in (0,2\pi)$ defined in Eq. (\ref{phi-F}) of Appendix A of the
absolute value $V^{\Delta}_{Bs1}$ of the $s1$ fermion velocity 
confirms the anisotropic character of the $s1$ boundary line.
In turn, that for hole concentrations $x\in (x_c,x_*)$ the $c$ Fermi 
line is isotropic follows from the independence of the angle $\phi$ of 
the $c$ fermion velocity.

For $x\in (x_c,x_*)$ and $U/4t$ values approximately in the range
$U/4t\in (u_0,u_1)$ the Fermi-line level of anisotropy is controlled by the interplay of the $s1$ boundary
line anisotropy and $c$ Fermi line isotropy. The Fermi energy has the form
$E_F=\mu +\delta E_F$ where $\mu\approx {\breve{\mu}}^0+W_c^h$,
${\breve{\mu}}^0$ is given in Eq. (\ref{mu-x-AL}) of Appendix B,
and below it is found that the square-lattice quantum liquid 
expression $\delta E_F\approx \vert\Delta\vert\vert\cos 2\phi\vert$ acquires an
extra term due to the superconducting fluctuations of the VEP
quantum liquid studied in this paper.

According to the criteria of Ref. \cite{companion2}, for the approximate range
$U/4t\in (u_0,u_1)$ of intermediate $U/4t$ values the Fermi line is strongly anisotropic for 
hole concentrations below $x_{c1}=1/8$, has some Fermi-energy anisotropy yet the Fermi velocity 
is nearly isotropic for the hole concentration range $x\in (x_{c1},x_{c2})$, and is nearly isotropic 
for the $x$ range $x\in (x_{c2},x_*)$. Here the hole concentration 
$x_{c2}\approx ([2\gamma_0 +1]/[3\gamma_0 +1])\,x_*$ where $\gamma_0=(1-x_0/x_*)$
is introduced below in Section III-E. The Fermi line is hole
and particle like for $x<x_h$ and $x>x_h$, respectively. The value of the hole concentration 
$x_h\geq x_{c2}$ is not accurately known \cite{companion2}. Likely it belongs to the range $x_h\in (x_{c2}, x_*)$.  
Provided that $x_h<x_*$, the angle $\phi_{AN}$ of Eq. (\ref{phi-F}) of Appendix A vanishes for hole concentrations 
$x\leq x_h$ and is small for $x\in (x_h,x_*)$. Hence for hole concentrations in the range $x\in (x_h,x_*)$ we consider only
corrections up to first order in $\phi_{AN}$, so that 
${\rm max}\,\{\delta E_F\}\approx \vert\Delta\vert$ and 
${\rm min}\,\{V^{\Delta}_{Bs1}\}\approx [\phi_{AN}\,\sqrt{2}]\vert\Delta\vert$.

The hole Fermi momentum $\vec{k}_F^h=\vec{k}_F+\vec{\pi}$ 
given in Eq. (\ref{phi-F}) of Appendix A where $\vec{\pi}=[\pi,\pi]$ can be 
expressed in terms of the corresponding $c$ hole
Fermi momentum ${\vec{q}}_{Fc}^{\,h\,d}$ and
$s1$ boundary-line momentum ${\vec{q}}^{\,d}_{Bs1}$
as follows \cite{companion2},
\begin{eqnarray}
\vec{k}_F^h & = & {\vec{q}}_{Fc}^{\,h\,d} - {\vec{q}}^{\,d}_{Bs1}
= k^h_{F} (\phi)\,{\vec{e}}_{\phi} \, ; \hspace{0.35cm}
{\vec{q}}^{\,h\,d}_{Fc} = {\vec{q}}_{Fc}^{\,d} + \vec{\pi} \, ,
\nonumber \\
{\vec{q}}_{Fc}^{\,h\,d} & = & q^h_{Fc} (\phi^{d}_c)\,{\vec{e}}_{\phi^{d}_c} 
\, ; \hspace{0.35cm}
{\vec{q}}^{\,d}_{Bs1} = q_{Bs1} (\phi)\,{\vec{e}}_{\phi^{d}_{s1}} 
\, , \hspace{0.15cm} d = \pm 1 \, .
\label{kF-qFc-qBs1}
\end{eqnarray}
These general expressions are for one-electron excited states valid for $x\in (x_c,x_*)$.
For $x\in (x_{c1},x_{c2})$ the specific general dependences on the Fermi angle $\phi$ of the angles 
$\phi^{d}_c\in (0,2\pi)$ and $\phi^{d}_{s1}\in (0,2\pi)$ are, 
\begin{equation}
\phi^{d}_c = [\phi -d\pi_{cs}/2 + \phi^{d}_{F}] \, ; \hspace{0.35cm}
\phi^{d}_{s1} = [\phi_{s1} + \phi^{d}_{F}] = [\phi + \pi + \phi^{d}_{F}]
\, , \hspace{0.15cm} d = \pm 1 \, .
\label{phiF-c-s1}
\end{equation}
In the studies of Ref. \cite{companion2} it is considered that for the 
Fermi velocity isotropic $x$ range $x\in (x_{c1},x_{c2})$ the 
$c$ Fermi hole momentum ${\vec{q}}_{Fc}^{\,h\,d}$ and
$s1$ boundary line momentum ${\vec{q}}^{\,d}_{Bs1}$
of the $c$ fermion and $s1$ fermion hole, respectively,
created upon an one-electron addition excitation are perpendicular,
so that ${\vec{q}}_{Fc}^{\,h\,d}\cdot {\vec{q}}^{\,d}_{Bs1}=0$.
Since $\pi_{cs}/2$ is the angle between the momentum-space directions 
of ${\vec{q}}_{Fc}^{\,h\,d}$ and ${\vec{q}}^{\,d}_{Bs1}$, this is equivalent to considering that 
$\pi_{cs}/2=\pi/2$ in the $\phi^{d}_c$ expression 
of Eq. (\ref{phiF-c-s1}). Hence, in spite of the $c$ momentum
band remaining unaltered upon such an excitation, for
states belonging to the reduced one-electron subspace
there are two alternative $c$ Fermi momenta. Those are associated
with the two doublicity values $d=\pm 1$, respectively. Their
expression is given in Eq. (\ref{kF-qFc-qBs1}). 
Specifically, the doublicity $d=\pm 1$ of a $c$ fermion created
or annihilated under an one-electron excitation equals
that of the corresponding $s1$ fermion hole created
under the same excitation. 

The concept of doublicity remains valid when the angle $\pi_{cs}/2$ 
between the momentum-space directions 
of ${\vec{q}}_{Fc}^{\,h\,d}$ and ${\vec{q}}^{\,d}_{Bs1}$
appearing in the $\phi^{d}_c$ expression of Eq. (\ref{phiF-c-s1}) is $\phi$
dependent, provided that the following integrals involving 
the corresponding function $\pi_{cs}=\pi_{cs}(\phi)$ and the
$c$ Fermi line hole momentum and $s1$ boundary line momentum
absolute values $q^h_{Fc}(\phi)$ and $q_{Bs1}(\phi)$, respectively, vanish,
\begin{equation}
\int_{0}^{\pi/4}d\phi \,q^h_{Fc}(\phi)q_{Bs1}(\phi)\cos\left({\pi_{cs}(\phi)\over 2}\right) = 
\int_{\pi/4}^{\pi/2}d\phi \,q^h_{Fc}(\phi)q_{Bs1}(\phi)\cos\left({\pi_{cs}(\phi)\over 2}\right) = 0 \, .
\label{pi-cs--sum-rule}
\end{equation} 
This ensures that the Fermi line centered at $-\vec{\pi}$
encloses the correct momentum area,
\begin{eqnarray}
\int_{0}^{2\pi}{d\phi\over 2\pi}\,
\pi\left[k^h_{F} (\phi)\right]^2 & = & \int_{0}^{2\pi}{d\phi\over 2\pi}\,
\pi\left(\left[q^h_{Fc}(\phi)\right]^2 + 
\left[q_{Bs1}(\phi)\right]^2+2q^h_{Fc}(\phi)q_{Bs1}(\phi)\cos\left({\pi_{cs}(\phi)\over 2}\right)\right) 
\nonumber \\
& = & x\,4\pi^2 + (1-x)\,2\pi^2 = (1+x)\,2\pi^2 \, .
\label{KF-sum-rule}
\end{eqnarray}
Here $x\,4\pi^2$ and $(1-x)\,2\pi^2$ are the momentum areas enclosed by
the $c$ Fermi line and $s1$ boundary line when centered at $-\vec{\pi}$
and $\vec{0}=[0,0]$, respectively \cite{companion2}. In this paper we consider the general case
for which the angle $\pi_{cs}/2=\pi_{cs}(\phi)/2$ is for $x\in (x_{c1},x_{c2})$ a periodic function of
$\phi$ of period $\pi/2$. In addition, for $\phi\in (0,\pi/4)$ it obeys the equality 
$\pi_{cs}(\phi)/2=\pi_{cs}(\pi/2-\phi)/2$ where $\pi/2-\phi\in (\pi/4,\pi/2)$. It
follows that for $x\in (x_{c1},x_{c2})$ the one-electron $F$ angle $\phi^{d}_{F}(\phi)$ 
appearing in Eq. (\ref{phiF-c-s1}) reads,
\begin{equation}
\phi^{d}_{F} (\phi) = d \arctan 
\left({q^h_{Fc} (\phi)\sin \left({\pi_{cs}(\phi)\over 2}\right)\over q_{Bs1} (\phi)+q^h_{Fc} (\phi)\cos\left({\pi_{cs}(\phi)\over 2}\right)}\right) 
\, , \hspace{0.15cm} d = \pm 1 \, .
\label{phi-c-s1}
\end{equation}
It can have two values, $\phi_{F}^{-1}$ and $\phi_{F}^{+1}$. The two corresponding rotations refer 
to the doublicity $d=-1$ and $d=+1$, respectively. 
For $\pi_{cs}/2=\pi/2$ this general expression recovers that found in Ref. \cite{companion2}.
The $F$ angle $\phi^{d}_{F}(\phi)$ is associated with the one-electron $F$ rotation matrix $A^{d}_{F}$
such that,
\begin{equation}
{\vec{q}}^{\,d}_{Bs1} = A^{d}_{F}\,{\vec{q}}_{Bs1}
\, ; \hspace{0.35cm}
A^{d}_{F} = \left[
\begin{array}{cc}
\cos \phi^{d}_{F} & -\sin \phi^{d}_{F}  \\
\sin \phi^{d}_{F} & \cos \phi^{d}_{F} 
\end{array}\right] \, , \hspace{0.15cm} d = \pm 1 \, .
\label{A-c-s1}
\end{equation}
Alike in Ref. \cite{companion2}, we denote the auxiliary momentum of the $s1$ boundary momentum ${\vec{q}}^{\,d}_{Bs1}$
by ${\vec{q}}_{Bs1}$. For $x\in (x_{c1},x_{c2})$ and $\vec{q}$ at or near the $s1$ boundary line the matrix $A^d_{s1}$ of Eq. (\ref{qs1-qhc}) is 
orthogonal and equals the $F$ rotation matrix $A^{d}_{F}$ given here. 

If follows from the above expressions that for the hole-concentration range $x\in (x_{c1},x_{c2})$ the absolute value 
$k^h_{F} (\phi)$ of the hole Fermi momentum given in Eq. (\ref{kF-qFc-qBs1}) and Eq. (\ref{phi-F}) of Appendix A reads,
\begin{equation}
k^h_{F} (\phi) = \sqrt{[q^h_{Fc}(\phi)]^2+[q_{Bs1}(\phi)]^2 +2q^h_{Fc}(\phi)q_{Bs1}(\phi)\cos\left({\pi_{cs}(\phi)\over 2}\right)} \, .
\label{KF-h-absolute}
\end{equation}
Again, for $\pi_{cs}/2=\pi/2$ this recovers the corresponding expression of Ref. \cite{companion2}. The hole Fermi 
momentum $\vec{k}_F^h$ of Eq. (\ref{kF-qFc-qBs1}) for any value of the Fermi angle
$\phi$ and the nodal Fermi momentum $\vec{k}_F^{\,N}=\vec{k}_F^{h\,N}-\vec{\pi}$ for $\phi=\pi/4$ 
can be expressed as follows, 
\begin{equation} 
\vec{k}^{\,h}_F (\phi) = - {k^h_{F} (\phi)\over q_{Bs1} (\phi)}\,{\vec{q}}_{Bs1}(\phi)
\, ; \hspace{0.35cm} 
\vec{k}_F^{\,N} = - {k_{F}^N \over q_{Bs1}^N}\,{\vec{q}}_{Bs1}^{\,N} 
\, ; \hspace{0.35cm} q_{Bs1} (\phi) = q_{Bs1} (\pi/2-\phi) \, .
\label{khF-kF}
\end{equation}
Here $\vec{k}^{\,h}_F (\phi)$ and $\vec{k}_F^{\,N}$ are centered at $-\vec{\pi}$
and $\vec{0}$, respectively, and $k_{F}^N$ is the absolute value of the vector $\vec{k}_F^{\,N}$. 
Note that the auxiliary $s1$ boundary momentum ${\vec{q}}_{Bs1}$ of Eq.
(\ref{A-c-s1}) has been constructed to inherently pointing in the same direction as 
$\vec{k}^{\,h}_F$. This justifies the validity of the equalities (\ref{khF-kF}), which
involve the auxiliary $s1$ boundary-line momentum rather than
the corresponding $s1$ boundary-line momentum.

The ratios $r_s$ and $r_c$ of Eq.(\ref{m*c/mc-UL}) of Appendix A, which for $U/4t\in (u_0,u_1)$
are related as $r_c = \pi x_* \approx 2r_s$, control the effects of electronic correlations
in many quantities. For instance, for such a $U/4t$ range they control the range 
width max$(\pi_{cs}/2)$-min$(\pi_{cs}/2)=r_s$ of the angle $\pi_{cs}/2$. Indeed, $\pi_{cs}/2\in ([1-x_A]\,\pi/2,[1+x_A]\,\pi/2)$.
Here $x_A\approx x_*/2= 0.135$ is a crossover hole concentration slightly larger 
than $x_{c1}=1/8=0.125$. According to the studies of Ref. \cite{cuprates} it
marks the emergence of a scale invariance in the VEP quantum liquid, which dominates the physics in a 
hole concentration range $x\in (x_A,x_{c2})$. The smallest and largest $\pi_{cs}/2$ magnitudes are 
reached for hole Fermi momenta pointing in the nodal and anti-nodal directions, 
\begin{equation}
\pi_{cs}(\pi/4)/2 = [1-x_A]\,\pi/2 
\, ; \hspace{0.3cm}
\pi_{cs}(0)/2 = \pi_{cs}(\pi/2) = [1+x_A]\,\pi/2 
\, ; \hspace{0.3cm} x_A \approx {x_*\over 2} \, ,
\label{pi-cs-N-AN}
\end{equation}
respectively. For $U/4t\approx 1.525$ this gives $\pi_{cs}(\pi/4)/2\approx 0.43\,\pi$ and
$\pi_{cs}(0)/2= \pi_{cs}(\pi/2)/2 \approx 0.57\,\pi$. Hence $\pi_{cs}/2\in (0.43\,\pi,0.57\,\pi)$ has 
indeed values near $\pi/2$. Both this and that its average value is $\pi/2$ justifies the 
approximation of Ref. \cite{companion2} that it reads $\pi/2$. 

Except for the angle $\pi_{cs}/2=\pi/2$ being replaced by a $\phi$ dependent angle 
$\pi_{cs}/2\in ([1-x_A]\,\pi/2,[1+x_A]\,\pi/2)$, the expressions and physics reported in 
Ref. \cite{companion2} remain valid. Moreover, the $c$ fermion energy dispersion 
$\epsilon_{c} ({\vec{q}}^{\,h})$ provided in Eq. (\ref{c-band}) of Appendix A is not changed 
by the VEP quantum liquid superconducting fluctuations studied in this paper. In contrast, 
the $s1$ fermion energy dispersion $\epsilon_{s1} ({\vec{q}})$ is, as found in the following. 

\section{Long-range $d$-wave superconducting order}

Here evidence is provided that the ground state of the VEP quantum liquid is superconducting
for zero spin density $m=0$, hole concentrations in the range $x\in (x_c,x_*)$, and intermediate $U/4t$ values. 
We recall that for approximately $U/4t>u_0$ the VEP quantum liquid refers to the Hamiltonian $\hat{H}_{3D}$ of 
Eq. (\ref{M-t}) in the one- and two-electron subspace \cite{companion2} for $t_{\perp}/t\ll 1$ plus
the very small suppression effects considered below. For such a $U/4t$ range of values, that the inequality $t_{\perp}/t\ll 1$ holds 
assures that the anisotropy parameter $\varepsilon^2 =m_c^*/M$ is very small as well.  

We start by considering the ground state of the $\gamma_d=1$ and $t_{\perp}/t=0$ 
square-lattice quantum liquid of Ref. \cite{companion2} at vanishing spin density $m=0$. We find that a necessary condition for it being
that of a $d$-wave superconductor is fulfilled for finite hole concentrations below $x_*$. 
Such a result is valid for the VEP quantum liquid as well provided that the only
significant effects of the weak 3D uniaxial anisotropy and suppression effects are the emergence
of the small anisotropic parameter $\varepsilon^2 =m_c^*/M$ and parameter $\gamma_d\approx 1$
in the expression of several physical quantities associated with quantum and
thermal fluctuations. For the five representative systems the parameter $\varepsilon^2 =m_c^*/M$ is 
found in Ref. \cite{cuprates0} to belong to the range $\varepsilon^2\in (3\times 10^{-4},1\times 10^{-2})$.
For such a small values of that parameter and $\gamma_d\approx 1$ the energy eigenstates 
of the $\gamma_d=1$ and $t_{\perp}/t=0$ square-lattice quantum 
liquid studied in Ref. \cite{companion2} are expected to be a good
approximation for those of the VEP quantum liquid.

Consistently with the Mermin-Wagner-Berezinskii Theorem \cite{MW,Mer,Bere-70,Bere}, 
the introduction of the weak 3D uniaxial anisotropy prevents the destruction of the fluctuations of 
long-range orders at finite temperature. For $t_{\perp}/t\ll 1$ very small and approximately
$U/4t>u_0$ we find indeed strong evidence of virtual-electron-pair phase coherence below some critical temperature $T_c$.  
(The concept of a virtual electron pair is introduced below.) 
At zero temperature such a virtual-electron-pair phase coherence
occurs for the hole-concentration range $x\in (x_c,x_*)$. Here the lower critical concentration 
$x_c\approx {\rm Gi}+x_0$ where ${\rm Gi} =G\,[M/m_c^*]=G/\varepsilon^2$ is fully determined by the  
anisotropic parameter $\varepsilon^2 =m_c^*/M\ll 1$ and $x_0$ emerges due to the hole-trapping effects 
of Appendix B. The magnitude of the proportionality constant $G$ is found in Ref. \cite{cuprates0}
to be in the range $G\in (1\times 10^{-5},3\times 10^{-4})$ for the five representative systems.
Hence it is so small that  ${\rm Gi}\approx 10^{-2}$ in spite of $m_c^*/M\ll 1$ being much smaller than ${\rm Gi}$. 
The weak 3D uniaxial anisotropy allows considering the pairing energy within a small coherent volume. Such an
elementary volume controls the magnitude of parameters associated with the thermal 
and quantum fluctuations of the square-lattice quantum liquid 
perturbed by weak 3D uniaxial anisotropy. The suppression effects slightly lessen
the $T_c$ magnitude, which is linear in $\gamma_d\approx 1$.

The physical picture that emerges from our study is that 
superconductivity arises in the VEP quantum liquid as a by-product of the short-range spin 
correlations. We then construct a consistent scheme 
concerning the $d$-wave pairing mechanism, phase-coherent-pair superconducting order, 
and corresponding order parameter, which follows from the
properties of that quantum liquid. Our results are inconclusive on whether the ground state
of the 2D Hubbard model on the square lattice is superconducting. They
seem to indicate that some small 3D uniaxial anisotropy is needed for the emergence 
of superconductivity. 

\subsection{A necessary condition for the ground state at zero spin density being
that of a $d$-wave superconductor}

Let us consider the $t_{\perp}/t=0$ and $\gamma_d= 1$ Hubbard model on the square lattice (\ref{H}). 
The one- and two-electron subspace is spanned by states whose deviation $\delta N_c^h$ in the 
number of $c$ band holes and number $N_{s1}^h$ of $s1$ band holes read \cite{companion2}, 
\begin{equation}
\delta N_c^h = - \delta N = 0, \mp 1, \mp 2 \, ; \hspace{0.35cm}
N_{s1}^h = \pm (\delta N_{\uparrow} - \delta N_{\downarrow})
+ 2L_{s,\,\mp 1/2} + 2N_{s2} = L_{s,\,-1/2} + L_{s,\,+1/2} + 2N_{s2} = 0, 1, 2 \, .
\label{deltaNcs1}
\end{equation}
Here $\delta N= [\delta N_{\uparrow} +\delta N_{\downarrow}]$ is the deviation in the number of electrons, 
$\delta N_{\uparrow}$ and $\delta N_{\downarrow}$ those in
the number of spin-projection $\uparrow$ and $\downarrow$ 
electrons, respectively, $N_{s2}$ the number of the excited-state $s2$
fermions, and $L_{s,\,\pm1/2}$ that of independent
spinons of spin projection $\pm1/2$. 

For finite hole concentrations below $x_*$ and intermediate $U/4t$ values the use of the
energy functional introduced in Ref. \cite{companion2} reveals that
excited states with $s1$ band hole numbers $N_{s1}^h=0$ and $N_{s1}^h=1,2$ 
and involving addition to or removal from the $c$ Fermi line of $c$ fermions refer to a gapless
excitation branch and have an energy gap, respectively. For the latter excited states
the energy gap vanishes only if the auxiliary momentum of the
$s1$ band hole ($N_{s1}^h=1$) or both $s1$ holes ($N_{s1}^h=2$) points in the 
nodal directions. Use of Eq. (\ref{deltaNcs1}) then reveals
that addition or removal of (i) one electron and (ii) two electrons 
with the same spin projection to or from the hole-like Fermi
line whose hole Fermi momenta are given in Eq.
(\ref{kF-qFc-qBs1}) involves creation of (i) one and (ii) two holes,
respectively, in the $s1$ momentum band. Except for the excitation-momentum nodal directions
these excited states have a finite energy gap. In contrast, note that addition or removal 
of two electrons of opposite spin projection to or from 
that Fermi line leads to a final excited state whose $s1$ band
is full alike that of the initial ground state. Indeed such excitations correspond
to $N_{s2}=L_{s,\,\mp 1/2} = 0$ in Eq. (\ref{deltaNcs1}), so that
$N_{s1}^h =0$ for $\delta N_{\uparrow} = \delta N_{\downarrow}=\pm 1$.
Hence the latter processes refer to a gapless branch of two-electron excitations. 
Specifically, for excitations whose $s1$ holes are created at the $s1$ boundary line
the energies of such processes read to first order in the $c$ and $s1$ hole momentum 
distribution-function deviations,
\begin{eqnarray}
\delta E & = & \delta E_F (\phi) = \vert\Delta\vert\vert\cos 2\phi\vert
\, ; \hspace{0.1cm} \delta N_{\sigma}=\pm 1 \, , \hspace{0.1cm}
\delta N_{-\sigma} = 0 \, ,
\nonumber \\
\delta E & = & \delta E_F (\phi) + \delta E_F (\phi') 
= \vert\Delta\vert\,[\vert\cos 2\phi\vert + \vert\cos 2\phi'\vert] \, ; \hspace{0.1cm}
\delta N_{\sigma} = \pm 2 \, , \hspace{0.1cm}
\delta N_{-\sigma} = 0 \, ,
\nonumber \\
\delta E & = & 0 \, ; \hspace{0.1cm} \delta N_{\uparrow}=\delta N_{\downarrow} =\pm 1 \, .
\label{DE-3proc}
\end{eqnarray}
Such general spectra refer to vanishing
spin densities $m=0$ and approximately $U/4t \in (u_0,u_{\pi})$. (For
the square-lattice quantum liquid of Ref. \cite{companion2} and the 
VEP quantum liquid they refer to $x\in (0,x_*)$ and $x\in (x_c,x_*)$, respectively.)
In the expressions provided in Eq. (\ref{DE-3proc}),
$\delta E_F = \vert\Delta_{s1} ({\vec{q}}^{\,d}_{Bs1})\vert$ is the anisotropic 
Fermi-energy term and $\vert\Delta_{s1} ({\vec{q}}^{\,d}_{Bs1})\vert$ is the $s1$ fermion pairing
energy per spinon given in Eq. (\ref{bands-bipt}) of Appendix A. The Fermi energy
reads $E_F=\mu +\delta E_F$ where $\mu$ is the zero-temperature
chemical potential. The anisotropic Fermi energy $\delta E_F$ 
vanishes for $x>x_*$ since then ${\rm max}\,\{\delta E_F\}=\vert\Delta\vert =0$. It follows that 
for the hole concentration range $x\in (x_*,1)$ the ground state is
a disordered state without short-range spin order. Consistently, for that range of $x$ values 
the $s1$ fermion spinon-pairing energy
$2\vert\Delta_{s1} ({\vec{q}}^{\,d}_{Bs1})\vert$ vanishes
for all momentum values.

The numbers of $s1$ fermions and $s1$ fermion holes in the $s1$ momentum band 
equal those of occupied and unoccupied sites, respectively, in the $s1$ effective lattice. 
Hence in the following we discuss the structure of the different energy spectra
$\delta E$ of Eq. (\ref{DE-3proc}) in terms of the numbers of sites, occupied sites,
and unoccupied sites of that lattice. Those equal the corresponding
numbers of discrete momentum values, filled discrete momentum values,
and unfilled discrete momentum values, respectively, of the $s1$ fermion band.
That the energy $\delta E$ of Eq. (\ref{DE-3proc}) vanishes for excitations involving creation 
of two electrons of opposite spin projection and except for the nodal directions is 
for $x\in (x_c,x_*)$ finite and given by $\delta E = [\delta E_F (\phi) + \delta E_F (\phi')]$
for those involving creation of two electrons with the same spin projection is not a trivial result. 
(The additional term to $\delta E_F = \vert\Delta_{s1} ({\vec{q}}^{\,d}_{Bs1})\vert$
found below in Section IV-A as a result of the superconducting fluctuations does 
not change the basic property that $\delta E=0$ and
$\delta E>0$ for excitations involving creation of two electrons of opposite spin 
projection and the same spin projection, respectively.)
That behavior follows from the number $N_{a_{s1}}^2$ of sites of 
the $s1$ effective lattice being for the Hubbard model on the square
lattice a subspace-dependent functional \cite{companion2}. 
For the one- and two-electron subspace referring both to the square-lattice quantum liquid
of Ref. \cite{companion2} and the VEP quantum liquid, the expressions of the number deviations $\delta N_{a_{s1}}^2$ and $\delta N_{s1}$ of 
sites and occupied sites, respectively, of the $s1$ effective lattice read,
\begin{equation}
\delta N_{a_{s1}}^2 = \delta N_{\uparrow} + L_{s,\,-1/2}  \, ;
\hspace{0.35cm}
\delta N_{s1} = \delta N_{\downarrow} - L_{s,\,-1/2} - 2N_{s2}  \, .
\label{Nas1-Nhs1}
\end{equation}
Such expressions are consistent with that of $N^h_{s1}= [\delta N_{a_{s1}}^2-\delta N_{s1}]$, given in Eq. (\ref{deltaNcs1}).
For the deviation numbers and numbers $\delta N_{\uparrow}=\delta N_{\downarrow}=\pm 1$
and $L_{s,\,\pm 1/2}=N_{s2}=0$ of the $\delta N=\pm 2$ excited states
for which the two added or
removed electrons are in a spin-singlet configuration one finds
from the use of the expressions provided in Eqs. (\ref{deltaNcs1}) and (\ref{Nas1-Nhs1})
that the corresponding deviations in the numbers of sites and occupied sites
of the $s1$ effective lattice read $\delta N_{a_{s1}}^2=\pm 1$ 
and $\delta N_{s1}= \pm 1$, respectively. Therefore, the
deviation in the number of unoccupied sites vanishes,
$\delta N^h_{s1}=[\delta N_{a_{s1}}^2-\delta N_{s1}]=0$.
Indeed, under such excitations the creation or annihilation of one $s1$ fermion
is exactly cancelled by an increase or decrease, respectively, in
the number of sites of the $s1$ effective lattice. As a result
the number of unoccupied sites of the corresponding excited states remains being zero,
as for the initial ground state. Use of the
energy functional introduced in Ref. \cite{companion2}
and given below in Section IV-A then leads to the gapless spectrum 
$\delta E=0$ of Eq. (\ref{DE-3proc}). 

In turn, for the deviation numbers and numbers $\delta N_{\uparrow}=\pm 2$, $\delta N_{\downarrow}=0$,
$L_{s,\,\pm 1/2}=2$, and $L_{s,\,\mp 1/2}=N_{s2}=0$ or $\delta N_{\downarrow}=\pm 2$, $\delta N_{\uparrow}=0$,
$L_{s,\,\mp 1/2}=2$, and $L_{s,\,\pm 1/2}=N_{s2}=0$ of the 
$\delta N=\pm 2$ excited states for which the two added or
removed electrons are in a spin-triplet configuration one finds
from the use of the expressions provided in Eqs. (\ref{deltaNcs1}) and (\ref{Nas1-Nhs1}) 
that the corresponding deviations in the numbers of sites and occupied sites 
of the $s1$ effective lattice read $\delta N_{a_{s1}}^2=[\pm 1+1]$ 
and $\delta N_{s1}= [\pm 1-1]$, respectively. It then follows that the
deviation in the number of unoccupied sites are given by
$\delta N^h_{s1}=[\delta N_{a_{s1}}^2-\delta N_{s1}]=2$.
In contrast to the above excitations involving a 
spin-singlet electron pair, the creation or annihilation 
of one $s1$ fermion is not cancelled under
the latter excitations by an increase or decrease, respectively, in
the number of sites of the $s1$ effective lattice. As a result, 
its number of unoccupied sites increases from zero to two for 
the excited states. Use of the general energy functional 
of Ref. \cite{companion2} then leads to the gapped spectrum 
given in Eq. (\ref{DE-3proc}) for these excitations. (Use of the
modified VEP quantum liquid functional of Section IV-A leads
to a gapped spectrum including an additional term, which as
mentioned above does not change the physics discussed here.)
 
A similar analysis can be carried out for the $\delta N=\pm 1$ 
electron and $\delta N=0$ spin excitations. Their
energy spectrum is in general gapped.
The $d$-wave-like structure of the one- and two-electron spectra
of the Hubbard model on the square lattice given in Eq. 
(\ref{DE-3proc}) follows from the momentum dependence of
the $s1$ fermion dispersion studied in Ref. \cite{companion2}.
Whether the general energy spectrum (\ref{DE-3proc}) is
gapless or displays a gap is fully controlled by the interplay between
the deviations in the numbers of sites and occupied sites, respectively, 
of the $s1$ effective lattice. That lattice is exotic in that its
number of sites is subspace dependent. The spectrum (\ref{DE-3proc})
refers to a gapless branch of excitations whenever the corresponding
creation or annihilation of $s1$ fermions
is exactly cancelled by an increase or decrease, respectively, in
the number of sites of the $s1$ effective lattice so that 
its number of unoccupied sites remains being zero
as for the initial ground state. Above we gave the
example of creation or annihilation of a spin-singlet
electron pair. Such a canceling occurs for creation or annihilation 
of any finite number of such spin-singlet
electron pairs. 

In turn, for one-electron excitations and creation or annihilation of spin-triplet
electron pairs the spectrum of Eq. (\ref{DE-3proc}) is in general gapped except
for some momentum directions. The $d$-wave-like structure of that spectrum is a necessary
condition for the ground state of the model being for vanishing spin density $m=0$, 
finite hole concentrations $x\in (x_c,x_*)$, and thus finite $2\vert\Delta\vert$ that of
a $d$-wave superconductor. However, it is not a sufficient condition for the occurrence of coherent pairing
needed for the macroscopic condensate. Below we find strong evidence 
that for intermediate $U/4t$ values, vanishing spin density $m=0$, and the hole concentration 
range $x\in (x_c,x_*)$ the sufficient condition of phase coherence is met by the VEP quantum liquid. 
At zero temperature and both for hole concentrations in the ranges $x\in (x_0,x_c)$ and 
$x\in (x_c,x_*)$ there is short-range spin order. For $x\in (x_0,x_c)$ 
pairing correlations occur that due to strong phase fluctuations do not lead to coherent 
pairing and superconductivity.  

Finally, we emphasize that the number deviation and number expressions
provided in Eqs. (\ref{deltaNcs1}) and (\ref{Nas1-Nhs1}) also hold
for $D=1$ spatial dimensions with $N_{a_{s1}}^2$ replaced by $N_{a_{s1}}$. 
However, for the 1D Hubbard model
there is no short-range spin order for any range of $x$
values so that $\vert\Delta\vert=0$ in Eq. (\ref{DE-3proc})
and $\delta E=0$ for all one- and two-electron excitations
under consideration. Consistently, the ground
state of the 1D model is not superconducting. 
Such an analysis reveals that a necessary
condition for the occurrence of a superconductivity
order in the VEP quantum liquid is the 
occurrence of short-range spin order associated with
the finite energy parameter $2\vert\Delta\vert>0$. 
That strongly suggests that the occurrence of 
superconductivity in such a system 
is a by-product of the short-range spin correlations
associated with the energy scale $2\vert\Delta\vert$.
However, our results are inconclusive on whether for
the Hubbard model on the square lattice the phase
coherence needed for the occurrence of long-range superconducting order
does occur. 

\subsection{Short-range spin order, the pseudogap
energy scale, and the pseudogap temperature $T^*$}

Here we provide some basic information about the energy scales associated with the short-range spin 
order, which for vanishing spin density $m=0$ and hole concentrations in the range $x\in (x_0,x_*)$  refers to 
finite temperatures below a pseudogap temperature $T^*$. Indeed, the VEP quantum liquid scheme accounts 
for the hole trapping effects reported in Appendix B, so that the short-range spin order occurs for the 
range $x\in (x_0,x_*)$ rather than $x\in (0,x_*)$ for the square-lattice quantum liquid
of Ref. \cite{companion2}. Evidence is provided in Ref. \cite{cuprates0} that for the hole concentration range $x\in (x_c,x_*)$ 
the suppression effects due to intrinsic disorder or superfluid density 
anisotropy are for the four representative systems other than LSCO very small. 
In contrast, the LSCO cation-randomness effects introduced in Ref. \cite{cuprates0} are not small. 
Fortunately, they are merely accounted for by multiplying the energy scale $\Delta_0$ and 
related magnitudes by a factor of $1/2$, leaving the pseudogap temperature $T^*$ considered
in the following unaltered. Its magnitude remains unaltered as well under the
suppression effects. The results of this paper and Ref. \cite{cuprates} confirm 
that our oversimplified description of the effects intrinsic disorder or superfluid density 
anisotropy and LSCO randomness in terms of suppression effects and 
cation-randomness effects, respectively, leads to agreement between 
theory and experiments on the five representative systems.

At zero temperature the energy order parameter $2\vert\Delta\vert$ of the short-range
spin correlations is the maximum magnitude of the $s1$ fermion spinon-pairing energy 
of Eq. (\ref{D-x}) of Appendix A \cite{companion2}.
Upon increasing $x$ the residual $c$ - $s1$ fermion interactions tend to 
suppress the short-range spin correlations. For approximately $U/4t \in (u_0, u_{\pi})$ 
and hole concentrations $x\in (x_0,x_*)$ such an effect
leads to the linear decreasing $2\vert\Delta\vert \approx (1-x/x_*)\,2\Delta_0$
of the zero-temperature order parameter of the corresponding short-range spin
order. That energy parameter plays the role
of pseudogap energy scale, as it controls the
magnitude of the pseudogap temperature $T^*$ above 
which there is no short-range spin order, associated with $s1$ fermion
spinon pairing. However, there are several definitions of $T^*$ 
corresponding to effects of the pseudogap associated with
$2\vert\Delta\vert$ on different
physical quantities. Indeed, such effects may appear
at different temperatures. 

For the square-lattice quantum liquid of Ref. \cite{companion2}, 
for which the short-range spin order occurs for $0<x<x_*$,
we identify in the limit $x\rightarrow 0$ the temperature $T^*$ with the 
temperature $T_x$ of Ref. \cite{Hubbard-T*-x=0}. In that limit it is related to the
energy parameter $\Delta_0=\lim_{x\rightarrow 0}\vert\Delta\vert$ as $T_x\approx \Delta_0/k_B$ \cite{companion2}.
For the quantum problem considered here and approximately $U/4t\in (u_0,u_{\pi})$ the 
corresponding expression valid for finite hole concentrations in the range $x\in (x_0,x_*)$ is then,
\begin{equation}
T^* \approx \left(1-{x\over x_*}\right) {\Delta_0\over k_B}  \, .
\label{Dp-T*}
\end{equation}
This is actually the maximum magnitude that $T^*$ can achieve. Its magnitudes as
obtained from different properties should obey the inequality $T^* \leq (1-x/x_*)[\Delta_0/k_B]$.
Hence the zero-temperature magnitude of the short-range spin order parameter
$2\vert\Delta\vert \approx (1-x/x_*)\,2\Delta_0$ controls the range of the 
pseudogap temperature $T^*\approx 2\vert\Delta\vert/2k_B$. For temperatures above $T^*$ the system is
at zero spin density $m=0$ and for finite hole concentrations in the range $x\in (x_0,x_*)$
driven into a spin disordered state without short-range spin order.

\subsection{Selected Hamiltonian terms: The VEP quantum-liquid microscopic Hamiltonian}

For the hole concentration ranges $x\in (0,x_c)$ and $x\in (x_c,x_*)$ the quantum problem
considered in this paper is qualitatively different. For $x\in (0,x_c)$ it refers to that reported 
in Appendix B, for which the effects of intrinsic disorder are strong. For $x\in (x_c,x_*)$ the VEP 
quantum-liquid suppression effects are found in Ref. \cite{cuprates0} to be very
small. That under the LSCO cation-randomness effects introduced in that section
the energy scale $\Delta_0$ is lessened by a factor two is behind the hole 
concentration $x_0$ of Eq. (\ref{x-h}) being different for the parameters appropriate to LSCO 
and the remaining four representative systems: It reads 
$x_0\approx 0.013$ and $x_0\approx 0.024$, respectively. 

Expression of the $t_{\perp}/t\ll 1$ Hamiltonian $\hat{H}_{3D}$ of Eq. (\ref{M-t}) 
in terms of rotated-electron creation and annihilation operators 
leads to an infinite number of terms \cite{companion2}. Although in terms
of electron operators the Hamiltonian has only on-site interactions, 
in terms of rotated electron operators there emerge effective interactions
involving rotated electrons on different sites. Only a small number of such
Hamiltonian terms are relevant to the physics of the present quantum problem.
However, the appropriate selection of the latter Hamiltonian terms is a problem of
huge complexity. 

Here we use as criterion for that selection the general strongly correlated 
microscopic Hamiltonian that, according to the analysis of 
Ref. \cite{duality}, almost certainly underlies the essential physics of the representative 
hole-doped cuprates. For  $U/4t>u_0$ this leads to a simplified
effective Hamiltonian with the same general form as that given in Eq. (1) of Ref. \cite{duality},
but with the electron creation and annihilation operators replaced by corresponding rotated electron
operators. Its kinetic-energy terms include those generated from the operator $\hat{H}_{\perp}$ given in 
Eq. (\ref{M-t}). In Section IV-A we introduce the general energy functional corresponding
to such a microscopic Hamiltonian. The small hole concentration denoted by $x_0$ in Ref. \cite{duality}, 
which reads $x_0\approx 0.01$ for LSCO,
is identified here with the hole concentration $x_0$ of Eq. (\ref{x-h}). 
For the VEP quantum liquid $x_0$ is the critical hole-concentration 
at which there occurs a sharp quantum phase transition from the Mott-Hubbard insulator 
with long-range antiferromagnetic order to a short-range incommensurate-spiral spin ordered state. 
That in Ref. \cite{cuprates0} it is found to read $x_0\approx 0.013$ for LSCO and as given in
Eq. (\ref{x-h}) is proportional to $\Delta_0/t$ is consistent with the results of Ref. \cite{duality}. 

The selected Hamiltonian terms of our microscopic Hamiltonian are basically the same 
as those of the microscopic Hamiltonian (1) of Ref. \cite{duality}, with the electron
operators replaced by rotated-electron operators and
without the term containing the parameter $U_p$. Indeed, after expression
of the rotated-electron operators in terms of $c$ and $s1$ fermion
operators the one- and two-electron subspace with no rotated-electron
double occupancy is well-defined: The generators of the states
that span such a subspace have simple expressions in terms of
$c$ and $s1$ fermion operators \cite{companion2}. As a result there is no need of
introducing artificial Hamiltonian terms to impose the lack of double occupancy. In our
case that is achieved by defining the microscopic Hamiltonian in the one- and 
two-electron subspace. Indeed, the corresponding VEP quantum liquid is
defined in that subspace. Below we express the terms of the microscopic Hamiltonian that control
the thermal and quantum fluctuations in terms of $c$ and $s1$ fermion operators.
The kinetic-energy terms of the VEP quantum-liquid microscopic Hamiltonian
involve both the $U/4t$-dependent $c$ fermion mass $m_c^*$ and the much larger
mass $M$ of Eq. (\ref{M-t}). 

The advantage of the rotated-electron operator description of Ref.
\cite{companion2} is that it has been constructed to inherently single and
double rotated-electron occupancies being good quantum numbers
for $U/4t>0$. Hence for such operators the lack of rotated-electron
double occupancy is exact for the range $U/4t>u_0$ of the
VEP quantum liquid. The transformation that
relates rotated electrons to electrons is unitary, so that the effective 
action for the phases $\theta$ considered in this paper
is valid for approximately $U/4t>u_0$ and specifically at 
$U/4t\approx u_*=1.525$. This is important in view
of the results of Refs. \cite{cuprates0,cuprates}. 
Such results provide evidence that within the description
of the properties of several classes of hole-doped cuprates with
$x_c\approx 0.05$ and $x_*\approx 0.27$ by the VEP quantum liquid the appropriate value
of $U/4t$ is not large and reads $U/4t\approx u_* =1.525\in (u_0,u_1)$.
This includes the five representative systems.
The intermediate value $U/4t\approx 1.525$ is also that found in Ref. \cite{companion2}
to be appropriated for the description of the spin-wave spectrum of the
$x=0$ parent compound La$_2$CuO$_4$ (LCO) \cite{LCO-neutr-scatt}.

Most of the VEP quantum-liquid microscopic Hamiltonian terms refer to in-plane processes.
The very small $t_{\perp}$-dependent terms generated from the operator $\hat{H}_{\perp}$ 
given in Eq. (\ref{M-t}) control the magnitude of the hole concentration $x_c$ and thus have 
effects on the virtual-electron pair phases $\theta_{j,0}$ and $\theta_{j,1}$
introduced below. In the superconducting
phase considered in the following such terms allow the Josephson tunneling through nearest-neighboring planes 
of a very small density of vanishing-energy spin-singlet electron pairs. For each square-lattice plane such
excitations correspond to the gapless excited states of the general spectrum (\ref{DE-3proc}), which
refer to creation or annihilation of spin-singlet electron pairs.
The small average numbers of such pairs that leave and arrive to a given square-lattice plane are
identical. The main role of that tunneling is to allow for thermal and quantum fluctuations 
associated with in-plane long-range superconducting order at finite temperatures.

The general energy functional introduced in Section IV-A corresponds to the
VEP quantum-liquid microscopic Hamiltonian,
yet it has an in-plane character. It includes implicitly the needed 3D uniaxial anisotropy effects. Indeed, 
part of such effects are effectively described by the use of a suitable mean-field approximation for 
the square-lattice physics. Specifically and alike in the microscopic Hamiltonian of Ref. \cite{duality}, in 
some of the selected Hamiltonian terms studied below a complex gap function $\Delta_{j'j''}$ replaces 
the corresponding pairing operator. Indeed within mean-field theory finite-temperature long-range orders 
may occur in the 2D system. Although this is excluded by the exact Mermin-Wagner-Berezinskii Theorem 
\cite{MW,Mer,Bere-70,Bere}, mean-field theory refers to an additional effective way to indirectly account for the 
effects of the very small spin-singlet electron pair Josephson tunneling through nearest-neighboring planes.
In turn, in the pseudogap state considered below the small $t_{\perp}$-dependent terms are behind very small one-electron
transfer between first-neigbhboring planes. For each square-lattice plane such
excitations correspond to excited states of the general spectrum (\ref{DE-3proc}) with a
finite energy gap. Those refer to creation or annihilation of a single electron. As discussed
in Ref. \cite{cuprates0}, such excitations are behind the energy gap of the normal-state conductivity in the direction 
perpendicular to the planes. 

Before introduction in Section IV of the general energy functional corresponding to the VEP quantum-liquid  
microscopic Hamiltonian, in the following we express the creation and annihilation rotated-electron operators of the terms of
such a Hamiltonian that control the thermal and quantum fluctuations in terms of $c$ and $s1$ fermion 
operators. Analysis of the form of the obtained effective Hamiltonian terms reveals that the spin bonds and the 
charge-2e sector of Ref. \cite{duality} correspond to the spin-singlet two-spinon $s1$ fermions and $c$ fermion pairs, 
respectively. For the hole concentration range $x\in (x_c,x_*)$ of the VEP quantum liquid both
the effects of 3D uniaxial anisotropy and intrinsic disorder are small. Hence for such a $x$ range
the interactions of these objects are within our description residual. This results from the residual character
of such interactions within the starting square-lattice quantum liquid of Ref. \cite{companion2}. That property greatly simplifies 
the derivation of the one-electron scattering rate carried out in Ref. \cite{cuprates}. 

The $d$-wave-like structure of the Fermi-line one- and two-electron
energy spectrum given in Eq. (\ref{DE-3proc}) is only a
necessary condition for at zero spin density $m=0$ and finite hole concentrations
in the range $x\in (x_0,x_*)$ for which $\vert\Delta\vert >0$ the ground state of the 
present quantum problem to be that of a $d$-wave superconductor. 
That energy spectrum refers to $\varepsilon^2=m_c^*/M\rightarrow 0$.
It also applies here provided that the 3D anisotropic parameter $\varepsilon^2=m_c^*/M$ 
is very small and $\gamma_d\approx 1$. In the remaining of this section we access the $x$ dependence of the $T=0$ parameters
associated with the VEP quantum-liquid thermal and quantum fluctuations for the approximate range $U/4t\in (u_0,u_{\pi})$.
Indeed some of our expressions are not valid for $U/4t>u_{\pi}$.

In the following evidence from the study of pairing phase fluctuations is found 
that for zero spin density $m=0$ and hole concentrations in the range $x\in (x_c,x_*)$
the ground state of the VEP quantum liquid has phase coherence, which is associated
with a long-range superconducting order.
While our results refer to $U/4t\in (u_0,u_{\pi})$, this may hold for $U/4t>u_{\pi}$
as well. Our results also indicate that for vanishing spin density $m=0$ and finite hole 
concentrations in the range $x\in (x_0,x_c)$ there is
short-range spin order and pairing correlations yet the lack of phase coherence 
prevents long-range superconducting order.
Finally, it is expected that alike for the system of Ref. \cite{duality},
for $m=0$ and $x\in (0,x_0)$ monopole-antimonopole pairs of
the type considered in that reference unbind and proliferate, 
leading to long-range antiferromagnetic order, consistently with the
analysis of Appendix B. 

\subsubsection{Hamiltonian terms that control the
quantum fluctuations of the phases associated with competing orders}

Out of the selected terms of the VEP quantum-liquid microscopic Hamiltonian, here we 
consider those that control the quantum fluctuations of the phases associated with the competing orders. 
It is assumed that the $c$ and $s1$ fermion occupancy configurations that generate the
energy eigenstates of the $t_{\perp}/t=0$ and $\gamma_d=1$ problem generate states
close to energy eigenstates for the VEP quantum liquid at very small $t_{\perp}/t\ll 1$ values 
and $\gamma_d\approx 1$. Such terms are given by,
\begin{equation}
{\hat H}^{bonds} = \sum_{j=1}^{N/2}
\sum_{j',j''[j-const]}\Delta_{j'j''}[{\tilde{c}}^{\dag}_{\vec{r}_{j'},\uparrow}\,
{\tilde{c}}^{\dag}_{\vec{r}_{j''},\downarrow} - {\tilde{c}}^{\dag}_{\vec{r}_{j'},\downarrow}\,
{\tilde{c}}^{\dag}_{\vec{r}_{j''},\uparrow}] + {(\rm h. c.)} \, ,
\label{H-r-el}
\end{equation}
consistently with the maximum number of in-plane independent bonds 
being $N/2$. Indeed, for hole concentrations below $x_*$ and 
zero spin density $m=0$ the ground states are spin-singlet states \cite{companion2}. 
For $x<x_0$ and $x\in (x_0,x_*)$ such states have long-range and short range
spin order, respectively. They contain $N_{s1}=N/2$ two-spinon $s1$ bond
particles \cite{companion2,companion1b}. The two spinons of such spin-neutral objects describe the spin
degrees of freedom of rotated electrons that singly occupy sites in the ground-state configurations. 
Therefore, in the one- and two-electron subspace without 
rotated-electron doubly occupancy where such ground states are
contained there is an energetic 
preference for the formation of spin-singlet rotated-electron bonds. 

The summation $\sum_{j',j''[j-const]}$ on the right-hand side of
Eq. (\ref{H-r-el}) is over all in-plane $N/2$ spin-singlet rotated-electron bonds 
considered in Ref. \cite{companion1b}. Alike in Ref. \cite{duality}, to arrive to the Hamiltonian terms given
here we used the usual mean-field approximation within which the complex gap function $\Delta_{j'j''}$ 
has replaced the corresponding pairing operator. The weak effects of the
very small 3D uniaxial anisotropy perturbation $\hat{H}_{\perp}$ of Eq. (\ref{M-t}) occur on the quantum 
liquid in-plane Hamiltonian terms (\ref{H-r-el}) through the phase of the complex gap function 
$\Delta_{j'j''}$, as discussed below. Each spin-singlet rotated-electron bond
is centered at a real-space point $[\vec{r}_{j'}+\vec{r}_{j''}]/2$,
near that of coordinate $\vec{r}_{j}=[\vec{r}_{j'}+\vec{r}_{j''}]/2-l\,[a_s/2]\,{\vec{e}}_{x_d}$.
Here the two-site bond indices $l\pm 1$ and $d=1,2$
are those used within the notation of Ref. \cite{companion1b} 
discussed below and $a_s=a/\sqrt{1-x}$ is the spacing of the spin effective lattice.  
(The bond index $d=1,2$ is unrelated
to the doublicity $d=\pm 1$ considered in Section II and Appendix A.)

The rotated-electron creation and annihilation operators of the Hamiltonian terms (\ref{H-r-el}) 
act onto the one- and two-electron subspace with zero rotated-electron
double occupancy \cite{companion2}. 
For finite hole concentrations in the range $x\in (x_0,x_*)$ such terms
refer to energy scales below and around the energy $\Delta_0$
of Eq. (\ref{Delta-0-gen}). The complex gap function is defined on the two-site
bonds in the spin effective lattice studied in Ref.
\cite{companion1b}. The energy scale $2\Delta_0$ 
equals the absolute maximum excitation energy below 
which the short-range spin order survives. Therefore,
$\Delta_0$ is the absolute maximum magnitude of the pairing 
energy per spinon of the two-site and two-spinon bond
associated with the complex function $\Delta_{j'j''}$ \cite{companion2}.
The amplitude of $\Delta_{j'j''}$ is for finite hole concentrations in the range $x\in (x_0,x_*)$ frozen below the 
energy $\Delta_0/\sqrt{2}$. Here $1/\sqrt{2}$ is a suitable normalization factor, which is 
absorbed in the expressions of the two-site and two-spinon operators
of the $s1$ bond-particle operators considered below. 
What remains are the fluctuations of the phases $\theta_{j'j''}$ of rotated-electron pairs. 
The complex gap function reads,
\begin{eqnarray}
\Delta_{j'j''} & = & e^{i\theta_{j'j''}} {(-1)^{d-1}\over\sqrt{2}}\Delta_0 
\, ; \hspace{0.35cm} x_0 = {\vert\Delta_{j'j''}\vert\over t}\,C_0
\, ; \hspace{0.35cm} C_0 = \left({\sqrt{2}\,x_*\over (2r_c)^2 + \Delta_0/t}\right)
\approx {1\over 2^{3/2}\pi r_c} \, , \hspace{0.15cm} x_0<x<x_* \, ,
\nonumber \\
\Delta_{j'j''}  & = & e^{i\theta_{j'j''}} {(-1)^{d-1}\over\sqrt{2}}{\mu^0\over 2}
\, , \hspace{0.15cm} 0 \leq x < x_0 \, .
\label{D-jj}
\end{eqnarray}
The hole concentration $x_0$ of Eq. (\ref{x-h}) is
here expressed in terms of $\vert\Delta_{j'j''}\vert=\Delta_0/\sqrt{2}$.
It corresponds to the hole concentration also called $x_0$
in Ref. \cite{duality}. The studies of that reference estimated it
to be proportional to $\vert\Delta_{j'j''}\vert/t$ and such that
$x_0 = [\vert\Delta_{j'j''}\vert/t] C_0\ll 1$. Its expressions 
given in Eqs. (\ref{x-h}) and (\ref{D-jj}) are consistent with such an estimation.

Following the $d$-wave character of the spinon pairing of the $s1$ bond particles,
one has in the expressions of Eq. (\ref{D-jj}) that $d=1$ and $d=2$ for the families of spin-singlet two-site bonds
whose primary bonds considered in Ref. \cite{companion1b}
are horizontal and vertical, respectively.
The different magnitudes $\Delta_0$ and $\mu^0/2$ 
that the energy scale $\vert\Delta_{j'j''}\vert$ has for finite hole concentrations 
in the ranges $x\in (x_0,x_*)$ and $x\in (0,x_0)$, respectively, 
are due to the sharp quantum phase transition 
occurring at $x=x_0$. (We recall that $\mu^0$ is
the energy scale given in Eq. (\ref{DMH}) of Appendix A.)

Usually one restricts the in-plane summations $\sum_{j',j''}$ of Eq. (\ref{H-r-el})
to nearest neighboring sites. In turn, the terms given in that equation
involve contributions from all possible in-plane bonds
whose two rotated electrons are located at arbitrarily distant sites.
In order to capture the physics of the quantum problem studied here one
must start by taking into account all such contributions. After
some algebra one then arrives to an effective Hamiltonian, which
is equivalent to restricting the in-plane summations $\sum_{j',j''}$ of Eq. (\ref{H-r-el})
to nearest neighboring sites. 

The importance of the selected Hamiltonian terms (\ref{H-r-el}) 
is that they contain the phases $\theta_{j'j''}$ whose fluctuations 
control the physics of the VEP quantum liquid. 
It is useful to express  (\ref{H-r-el})  in terms of $c$ fermion operators and
two-site and two-spinon bond operators. This is straightforwardly achieved by 
direct use of the expressions of the rotated-electron creation and annihilation operators in terms of 
$c$ fermion operators and two-site and two-spinon bond operators \cite{companion2},
with the result,
\begin{equation}
\Delta_{j'j''}
[{\tilde{c}}^{\dag}_{\vec{r}_{j'},\uparrow}\,
{\tilde{c}}^{\dag}_{\vec{r}_{j''},\downarrow} -
{\tilde{c}}^{\dag}_{\vec{r}_{j'},\downarrow}\,
{\tilde{c}}^{\dag}_{\vec{r}_{j''},\uparrow}] 
=  (-1)^{1-d}\,\sqrt{2}\,\Delta_{j'j''}\,f_{\vec{r}_{j'},c}^{\dag}\,f_{\vec{r}_{j''},c}^{\dag}
\,b_{\vec{r}_{j'j''},s1,d,l,g}^{\dag} \, .
\label{singlet-conf-simpl-0}
\end{equation}
Here $\vec{r}_{j'j''}=[\vec{r}_{j'}+\vec{r}_{j''}]/2= \vec{r}_{j} +l\,[a_s/2]\,{\vec{e}}_{x_d}$, $a_s=a/\sqrt{1-x}$ for
$x<x_*$, $b_{\vec{r},s1,d,l,g}^{\dag}$ is the in-plane two-site bond operator defined in Ref. \cite{companion1b},
and the index $g=0,1,...,[N_{s1}/4-1]$ refers to the link or bond type also defined in that reference. (The index $g$ appearing here is not 
the amplitude $g$ introduced below in Section III-D.) 

A $s1$ fermion operator of real-space coordinate $\vec{r}_{j}$ is defined
in terms of a superposition of such bond operators,
\begin{equation}
f_{\vec{r}_{j},s1}^{\dag} = e^{i\phi_{j,s1}}\,\sum_{g=0}^{N_{s1}/4-1}\sum_{d=1}^{2}\sum_{l=\pm1}h_{g}^*\,
b^{\dag}_{\vec{r}_{j}+{\vec{r}_{d,l}}^{\,0},s1,d,l,g} \, ,
\label{g-s1+general-gb}
\end{equation}
where $\phi_{j,s1}$ is the operator phase of Eq. (\ref{JW-f+}).
The absolute value $\vert h_{g}\vert=\sqrt{h^*_{g}h_{g}}$ of the coefficients appearing in this 
expression decreases for increasing magnitude of the two-site bond length 
$\xi_{g} \equiv \vert 2\vec{r}_{d,l}^{\,g}\vert$. The minimum and maximum values of
the length $\xi_{g}$ are
$\xi_{0} = a_s$ and $\xi_{g}=\sqrt{2}\,a_s\,(N_{s1}/4-1)$ for $g=[N_{s1}/4-1]$, respectively. These coefficients 
obey the normalization sum-rule
$\sum_{g=0}^{[N_{s1}/4-1]} \vert h_{g}\vert^2 = 1/4$.
Each of the $N_{s1}$ spin-singlet two-spinon bonds of a $s1$ bond particle of real-space coordinate
${\vec{r}}_{j}$ involves two sites of coordinates 
$\vec{r}-\vec{r}_{d,l}^{\,g}$ and $\vec{r}+
\vec{r}_{d,l}^{\,g}$, respectively, where ${\vec{r}}_{j}=\vec{r}-\vec{r}_{d,l}^{\,0}$
and $\vec{r}_{d,l}^{\,0}=l\,[a_s/2]\,{\vec{e}}_{x_d}$. It follows
that the two-site bond centre $\vec{r}\equiv\vec{r}_{j}+{\vec{r}_{d,l}}^{\,0}$ is 
the middle point located half-way between the two sites.
There are four families of bonds labeled by the numbers $d=1,2$
and $l=\pm 1$. For each family there are $N_{s1}/4$ link vectors
$\vec{r}_{d,l}^{\,g}$ of different link type $g=0,1,...,[N_{s1}/4-1]$.

For simplicity, in Eq. (\ref{singlet-conf-simpl-0}) we call $\vec{r}_{j'}$
and $\vec{r}_{j''}$ the two real-space coordinates of the sites of a two-spinon
bond. According to that equation, those are also the real-space coordinates
of the two $c$ fermions, respectively, involved in the corresponding
spin-singlet rotated-electron pair. 
However, we recall that the two real-space coordinates of such a bond and
corresponding $c$ fermion pair refer to well-defined indices
$d$, $l$, and $g$. The $g=0$ primary bonds and primary $c$ fermion pairs
have most of the corresponding rotated-electron pair spectral weight. For each of the $N/2$ in-plane real-space
coordinates $\vec{r}_j=[\vec{r}_{j'}+\vec{r}_{j''}]/2 -{\vec{r}_{d,l}}^{\,0}$ there are
four primary $c$ fermion pairs corresponding to $d=1$ horizontal
and $d=2$ vertical bonds and $l=-1$ left or lowest and $l=+1$
right or upper bonds. 

The phase factor $e^{i\theta_{j'j''}}$ of the complex gap function $\Delta_{j'j''}$ 
appearing in Eq. (\ref{singlet-conf-simpl-0}) can be written as,
\begin{equation}
e^{i\theta_{j'j''}}  = e^{i\theta_{j'j'',0}}\,e^{i\theta_{j'j'',1}} \, .
\label{theta-jj}
\end{equation}
Here $\theta_{j'j'',0}$ corresponds to the overall
centre-of-mass phase. In turn, $\theta_{j'j'',1}$ is the part of $\theta_{j'j''}$,
which cannot be reduced to arbitrary configurations
of site phases and regulates the relative motion
of a pair. Hence the phases $\theta_{j'j'',1}$ are related to the internal
pairing degrees of freedom. Within
charge excitations the phases $\theta_{j'j''}$ are associated
with the $c$ fermion pair of Eq. (\ref{singlet-conf-simpl-0}).
For one-electron and spin excitations they are associated with
the virtual-electron pair defined below. 

Use of expression (\ref{D-jj}) in Eq. (\ref{singlet-conf-simpl-0})
leads to,
\begin{equation}
\Delta_{j'j''}
[{\tilde{c}}^{\dag}_{\vec{r}_{j'},\uparrow}\,
{\tilde{c}}^{\dag}_{\vec{r}_{j''},\downarrow} -
{\tilde{c}}^{\dag}_{\vec{r}_{j'},\downarrow}\,
{\tilde{c}}^{\dag}_{\vec{r}_{j''},\uparrow}]  = e^{i\theta_{j'j''}}\,\Delta_0
\,f_{\vec{r}_{j'},c}^{\dag}\,f_{\vec{r}_{j''},c}^{\dag}
\,b_{\vec{r}_{j'j''},s1,d}^{\dag} \, .
\label{singlet-conf-simpl}
\end{equation}
Hence one finds from use of this expression in Eq. (\ref{H-r-el}),
\begin{equation}
{\hat H}^{bonds} = 
\sum_{j=1}^{N_{s1}} \sum_{j',j''[j-const]} e^{i\theta_{j'j''}}\,\Delta_0\,f_{\vec{r}_{j'},c}^{\dag}\,f_{\vec{r}_{j''},c}^{\dag}
\,b_{\vec{r}_{j'j''},s1,d,l,g}^{\dag} + {(\rm h. c.)} \, .
\label{H-bonds-c-s1-b}
\end{equation}

The $c$ effective lattice is identical to the original lattice \cite{companion2}. 
For the particular set of rotated-electron pairs and
corresponding $c$ fermion pairs of
real-space coordinates $\vec{r}_{j'}$ and $\vec{r}_{j''}$
associated with the same point of coordinate
$\vec{r}_{j} =[\vec{r}_{j'}+\vec{r}_{j''}]/2 -{\vec{r}_{d,l}}^{\,0}$
the phases $\theta_{j'j''}$ of Eq. (\ref{H-bonds-c-s1-b}) are nearly equal. The reason is that such a
set of pairs interact with the two-site and two-spinon bonds
of the same $s1$ fermion of real-space coordinate
$\vec{r}_{j}$ whose operator is given in Eq.
(\ref{g-s1+general-gb}). Thus we introduce the following
notation for the phase factors,
\begin{equation}
e^{i\theta_{j}} = e^{i\theta_{j,0}}\,e^{i\theta_{j,1}} \, .
\label{theta-j}
\end{equation}
The phases $\theta_{j,0}$ and $\theta_{j,1}$
are identical to the phases $\theta_{j'j'',0}$ and $\theta_{j'j'',1}$,
respectively, of the primary $c$ fermion pairs associated with the
four primary two-site bonds referring
to nearest-neighboring sites of the in-plane spin effective lattice
whose centre of mass is near the real-space point of
coordinate $\vec{r}_{j} =[\vec{r}_{j'}+\vec{r}_{j''}]/2 -{\vec{r}_{d,l}}^{\,0}$.

In order to describe the physics of the quantum liquid in terms
of $c$ and $s1$ fermions, in Appendix C it is shown that the 
Hamiltonian terms (\ref{H-bonds-c-s1-b}) are approximately equivalent to,
\begin{equation}
{\hat H}^{bonds}_{eff} = 
\sum_{j=1}^{N_{s1}} \sum_{\langle j',j''\rangle} 
{e^{i\theta_{cp}}\over 4\vert h_0\vert}\Delta_0
f_{\vec{r}_{j'},c}^{\dag}\,f_{\vec{r}_{j''},c}^{\dag}
\,f_{\vec{r}_{j},s1}^{\dag} + {(\rm h. c.)} \, .
\label{H-bonds-c-s1-f}
\end{equation}
Here the summation $\sum_{\langle j',j''\rangle}$ runs over $c$ fermion pairs
associated with rotated-electron pairs on singly occupied sites. In the spin effective 
lattice the centre $[\vec{r}_{j'}+\vec{r}_{j''}]/2$ of the corresponding in-plane 
nearest-neighboring real-space coordinates $\vec{r}_{j'}$ and $\vec{r}_{j''}$ 
of the spins of such rotated electrons is near the real-space 
coordinate  $\vec{r}_{j}=[\vec{r}_{j'}+\vec{r}_{j''}]/2-l\,[a_{s}/2]\,{\vec{e}}_{x_d}$
of the $s1$ fermion. In the present $N_a^2\rightarrow\infty$ thermodynamic limit
it coincides with it. Moreover,
\begin{equation}
e^{i\theta_{cp}} = e^{i\theta_{cp} (\vec{r}_j)} = e^{i[\theta_{j}-\phi^0_{j,s1}]} = e^{i[\theta_{j,0}+\theta_{j,1}-\phi^0_{j,s1}]} 
\, ; \hspace{0.35cm}  \phi^0_{j,s1} = \sum_{j'\neq j} \phi_{j',j,s1} \, .
\label{theta-cp}
\end{equation}
Here $\phi_{j',j,s1}$ are the phases defined in Eq. (\ref{JW-f+}). It follows
from the derivation of Appendix C that the primary two-site
bonds, {\it i.e.} those whose length $\xi_0=a_s=a/\sqrt{1-x}$ is minimum,
have most of the spectral weight of a $s1$ bond particle \cite{companion1b}
and corresponding $s1$ fermion. The coefficient $1/\vert h_0\vert$ of
expression (\ref{H-bonds-c-s1-f}) compensates the weight associated with bonds of
larger length. 

The phases $\theta_{cp}=\theta_{cp} (\vec{r}_j)$ appearing in the Hamiltonian terms 
(\ref{H-bonds-c-s1-f}) are associated with the rotated-electron pair and
corresponding $c$ fermion pair whose centre of mass 
$[\vec{r}_{j'}+\vec{r}_{j''}]/2$ is approximately at the real-space coordinate
$\vec{r}_{j}=[\vec{r}_{j'}+\vec{r}_{j''}]/2-l\,[a_{s}/2]\,{\vec{e}}_{x_d}$
of the $s1$ fermion of operator $f_{\vec{r}_{j},s1}^{\dag}$. Such Hamiltonian terms 
describe the interaction of the $c$ fermion pair with the $s1$ fermion. That pair feels the latter object
through such a phase. Hence the real-space coordinate $\vec{r}_j$ in the argument of 
the phases $\theta_{cp}=\theta_{cp} (\vec{r}_j)$ corresponds both approximately to the 
centre of mass of the $c$ fermion pair and to the real-space coordinate of the $s1$ 
fermion that the two $c$ fermions interact with. For the construction of an effective action for these
important phases the contributions from the Hamiltonian terms (\ref{H-bonds-c-s1-f}) 
involving nearest-neighboring sites at $\vec{r}_{j'}$ and $\vec{r}_{j''}$ are sufficient. 
In turn, concerning some other aspects of the virtual-electron pairing mechanism
the contributions from rotated-electron pairs and corresponding $c$ fermion
pairs at larger distances cannot be ignored.    

Since $\theta_{cp} =[\theta_{j}-\phi^0_{j,s1}]$,
the $c$ fermion pairs of Eq. (\ref{H-bonds-c-s1-f})
feel the $s1$ fermion effective vector potential (\ref{A-j-s1}) 
through the phases $\phi^0_{j,s1}$. Those are obtained from the operator phase 
$\phi_{j,s1}$ also given in that equation
by replacing $f^{\dag}_{{\vec{r}}_{j'},s1}f_{{\vec{r}}_{j'},s1}$
by its average $\langle f^{\dag}_{{\vec{r}}_{j'},s1}f_{{\vec{r}}_{j'},s1}\rangle\approx 1$.
Indeed, for the $s1$ fermion occupancies of the states that span the 
one- and two-electron subspace there are none, one, or two
unoccupied sites in the $s1$ effective lattice. In addition, the total number of such sites 
$N_{a_{s1}}^2\approx N/2=(1-x)N_a^2/2$ is of the order of $N_a^2$. Therefore, within 
the present thermodynamic limit such a replacement is a good approximation 
\cite{companion2}. Furthermore, for the states that span such a subspace
the fluctuations of the phases $\phi^0_{j,s1}$ are very small. Hence the fluctuations of the phases 
$\theta_{cp}$ of Eq. (\ref{theta-cp}) are fully controlled by those of the phases 
$\theta_{j,0}$ and $\theta_{j,1}$ of Eq. (\ref{theta-j}).

The effective Hamiltonian terms (\ref{H-bonds-c-s1-f}) are consistent with considering 
only the Hamiltonian terms,
\begin{equation}
{\hat H}^{bonds} = \sum_{j=1}^{N/2}
\sum_{\langle j',j''\rangle}\Delta_{j'j''}[{\tilde{c}}^{\dag}_{\vec{r}_{j'},\uparrow}\,
{\tilde{c}}^{\dag}_{\vec{r}_{j''},\downarrow} - {\tilde{c}}^{\dag}_{\vec{r}_{j'},\downarrow}\,
{\tilde{c}}^{\dag}_{\vec{r}_{j''},\uparrow}] + {(\rm h. c.)} \, .
\label{H-r-el-nn}
\end{equation}
Those correspond to the spin-singlet rotated-electron pairs of Eq. (\ref{H-r-el})
involving nearest-neighboring sites. Indeed, the spin-singlet rotated-electron pairs
of Eq. (\ref{H-r-el-nn}) involve the same sites as the $c$ fermion pairs of
Eq. (\ref{H-bonds-c-s1-f}).

\subsubsection{The fluctuations of the phases $\theta_{j,0}$ and $\theta_{j,1}$}

A next step is the construction of an effective action for the phases $\theta_{cp}$. The exact calculation
of some of the physical quantities involved in that derivation is an involved open problem. Nevertheless, 
one can construct by means of several approximations
an effective action whose key features faithfully reflect general 
properties of the VEP quantum-liquid microscopic Hamiltonian. 
To follow the fluctuations in $\theta_{j,0}$ and $\theta_{j,1}$, we should
integrate the $c$ and $s1$ fermions in the expression of Eq. (\ref{H-bonds-c-s1-f}) about a suitable
saddle point. Such a procedure can as well be fulfilled by integrating
the rotated-electrons in the expression of Eq. (\ref{H-r-el-nn}). 
Those are however extremely complex problems.

Fortunately, since the VEP quantum-liquid microscopic Hamiltonian 
has the same general form as the microscopic Hamiltonian (1) of Ref. \cite{duality}, the corresponding 
effective action for the phases $\theta_{cp}$ has also basic similarities to that constructed
in that reference. Given the nearly equivalence between the two actions, we omit here the details 
that are common to both investigations. Those can be found in Refs. \cite{duality,fluctua}.
The main results are summarized in the following. They are used
and further developed in Section III-D and following sections. 
Alike in the problem of Refs. \cite{duality,fluctua}, one arrives to an effective continuum Lagrangian. 
In spite of being a simplification of the selected rotated-electron Hamiltonian
terms, such a Lagrangian is expected to faithfully reflect the general properties of the microscopic problem
under consideration. 

First one finds that our phases $\theta_{j,0}$ and $\theta_{j,1}$ correspond
to the phases $\theta_{j'j''}^{CM}$ and $\theta_{j'j''}^{r}$, respectively, of Refs. \cite{duality,fluctua}, 
where the site indices $j'j''$ are called $jk$. Except that the rotated electrons play here the
role plaid by the electrons in that reference, the physics is very similar. 
For instance, the charge$-2e$ sector of Ref. \cite{duality}
refers to the charge$-2e$ $c$ fermion pairs. Indeed, upon expression of the rotated-electron
operators in terms of $c$ fermion and two-site and
two-spinon bond operators one finds that the Cooper
pairs of that reference are mapped onto the virtual-electron
pairs considered below. The charge degrees of freedom of such virtual-electron
pairs refer to the $c$ fermion pairs.
Within charge excitations the latter pairs behave independently of
the $s1$ fermion that contains the two-spinon spin-singlet
configuration of each virtual-electron pair.  

Second it is found that the fluctuations in the phases $\theta_{j,0}$ and
$\theta_{j,1}$ proliferate in different regions of
the phase diagram. For finite hole concentrations in the range $x\in (x_0,x_*)$ and vanishing temperature $T=0$
the fluctuations of the phases $\theta_{j,0}$ increase for $x\rightarrow x_c$ and remain large for
finite hole concentrations in the range $x\in (x_0,x_c)$. Those of $\theta_{j,1}$ increase for $x\rightarrow x_*$.
It follows that the fluctuations of the phases $\theta_{j}$ increase both for $x\rightarrow x_c$
and $x\rightarrow x_*$. Otherwise, the zero-temperture fluctuations
of the phases $\theta_{j,0}$ and $\theta_{j,1}$ remain small for finite hole concentrations 
in the ranges $x\in (x_c,x_*)$ and $x\in (x_0,x_*)$, respectively.
Here $x_*$ is the critical hole concentration of Eq. (\ref{xc-range}) of Appendix A
introduced in Ref. \cite{companion2}. Above it there is no short-range
spin order at zero temperature. As further discussed below, the critical hole concentration 
$x_c$ that emerges from such studies is directly related to the Ginzburg number 
\cite{vortices-RMP} and critical hole concentration $x_0$, $x_c\approx {\rm Gi} + x_0$. The expression of ${\rm Gi}$ 
suitable to the present quantum problem is provided below in Section III-D. 
That relation of $x_c$ to the Ginzburg number and $x_0$ is valid for very small values of the 3D 
uniaxial anisotropy parameter $\varepsilon^2 =m_c^*/M$ and $\gamma_d\approx 1$. For
those $x_c\approx 0.05$ and we find below that ${\rm Gi}\approx 0.037$ for LSCO and
${\rm Gi}\approx 0.026$ for the remaining four representative hole-doped cuprates.

Third one finds that at zero temperature and hole 
concentrations in the range $x\in (x_c,x_*)$ and for temperatures $T$ lower than a
critical temperature $T_c<T^*$ and a smaller temperature-dependent hole
concentration range centered at the optimal hole concentration $x=x_{op}\approx (x_c+x_*)/2$
the phases $\theta_{j}$ and thus $\theta_{cp}$ all line up into a phase-coherent
 superconductor.
For these $x$ and $T$ ranges both the $c$ fermion kinetic energy and the energy order 
parameter $2\vert\Delta\vert$ of the short-range spin correlations are finite.
Combination of the results on the fluctuations of the phases $\theta_{j,0}$ and
$\theta_{j,1}$ with those of Section III-A on the general $d$-wave spectrum
(\ref{DE-3proc}) then provides strong evidence that for 
the VEP quantum liquid there is a $d$-wave long-range superconducting order for 
such hole concentration and temperature ranges. In that phase the fluctuations of
both the phases $\theta_{j,0}$ and $\theta_{j,1}$ are small. This ensures that
the average phase factors $\langle e^{i\theta_{j,0}}\rangle\neq 0$ and
$\langle e^{i\theta_{j,1}}\rangle\neq 0$ are non vanishing. It then follows that
the average phase factor $\langle e^{i\theta_{cp}}\rangle\neq 0$ is also non vanishing. 
This is consistent with the phases $\theta_{cp}$ all lining up into a phase-coherent
 superconductor.

Fourth it is found that for finite hole concentrations in the range 
$x\in (x_0, x_c)$ at zero temperature and a temperature-dependent hole 
concentration range for temperatures belonging to the range $T\in (T_c,T^*)$ 
where $T_c=0$ for $x<x_c$ and $T_c>0$ for $x\in (x_c,x_*)$ and
the pseudogap temperature $T^*$ is approximately given by Eq. (\ref{Dp-T*}) the fluctuations 
of the phases $\theta_{j,1}$ are small yet those of the phases $\theta_{j,0}$ are large. 
As a result, one finds that the average phase factor $\langle e^{i\theta_{j,0}}\rangle= 0$ vanishes and
the average phase factor $\langle e^{i\theta_{j,1}}\rangle\neq 0$ remains non vanishing. 
This then implies that the average phase factor $\langle e^{i\theta_{cp}}\rangle= 0$ vanishes. In the corresponding
pseudogap state there is no phase coherence and thus no long-range superconducting order.
Nonetheless there remain strong pairing correlations. Indeed,
virtual-electron pairs remain existing in such a normal
state yet their phases $\theta_{cp}$ do dot line up. 
Consistently, the virtual electron pair -- VEP quantum liquid refers
both to the normal and superconducting states. In addition the short-range spin 
order prevails and infinite
vorticity loops proliferate. Since the fluctuations of the phases $\theta_{j,1}$ 
remain small, the monopole-antimonopole pairs 
of the type considered in Ref.
\cite{duality} remain bound
and the vorticity is conserved in the low-energy limit. 

Finally one finds that for the hole concentration range $x\in (0,x_0)$ at
zero temperature $T=0$ and a small temperature
dependent hole concentration range for finite temperatures the fluctuations 
of the phases $\theta_{j,0}$ and $\theta_{j,1}$ are large. 
It follows that the average phase factors $\langle e^{i\theta_{j,0}}\rangle= 0$,
$\langle e^{i\theta_{j,1}}\rangle= 0$, and
$\langle e^{i\theta_{cp}}\rangle= 0$ all vanish. As a result
vorticity is not properly conserved. Then the above
monopole-antimonopole pairs unbind and proliferate. At $T=0$ this is a Mott-Hubbard
insulator quantum phase. The corresponding ground state has long-range antiferromagnetic order. 

Alike in the problem of Ref. \cite{duality},
a sharp distinction is drawn between the microscopic properties
of the quantum liquids referring to hole concentrations in the range $x\in (0,x_0)$
and finite hole concentrations $x\in (x_0,x_*)$, respectively, in terms
of the global symmetry of the effective action for the
phases $\theta_{j}$. For the latter hole concentrations it is a global $U(1)$ symmetry.
It is commonly understood that Hamiltonian symmetries 
by themselves are not sufficient to prove that a particular symmetry is broken in the
ground state. However, for vanishing spin density $m=0$ and finite hole concentrations 
in the range $x\in (x_0,x_*)$
the global symmetry of that action is related to the hidden global $U (1)$ symmetry of the Hubbard model on the
square lattice found recently \cite{bipartite}. The latter 
$U(1)$ symmetry is contained in the
model global $SO(3)\times SO(3)\times U(1)=[SO(4)\times U(1)]/Z_2$
symmetry, beyond its previously well-known global $SO(4)$ symmetry \cite{Zhang}.
Indeed the representations of the model hidden global $U(1)$ symmetry correspond
to the occupancy configurations of the $c$ fermions in their $c$ momentum
band \cite{companion2,bipartite}. 

Consistently, for the present quantum problem in the normal
state at zero temperature and finite hole concentrations in the range 
$x\in (x_0,x_c)$ and in the superconducting
state at zero temperature and hole concentrations in the range $x\in (x_c,x_*)$ the $c$ fermions
configurations are organized in terms of zero-momentum $c$ fermion pairs. Those 
play a key role in the non-coherent and coherent 
virtual-electron pairing, respectively, of these two states. In turn, at $m=0$ 
and $x=0$ the $c$ band is for the absolute ground state of the Hubbard model on a square 
lattice full so that there is a single occupancy
configuration. Consistently, the symmetry of the effective action for the
phases $\theta_{j}$ is instead a local compact gauge symmetry. Here
that holds for $x\in (0,x_0)$ as well.

Our results are inconclusive on whether the ground state
of the $t_{\perp}/t=0$ square-lattice Hubbard model (\ref{H}) is superconducting
for intermediate $U/4t$ values and hole concentrations in the range
$x\in (x_c,x_*)$. If it is, the zero-temperature lower hole concentration $x_c$ must
have an expression different from that found here, $x_c\approx {\rm Gi}+x_0$.
Indeed, this expression gives $x_c\rightarrow\infty$ for $t_{\perp}/t\rightarrow 0$
and thus $M/m_c^*\rightarrow\infty$ for $U/4t>0$. However the
expression $x_c\approx {\rm Gi}+x_0$ applies provided that
$x_c$ remains small. It indicates though that the superconducting dome hole-concentration
width $(x_*-x_c)$ decreases upon decreasing $t_{\perp}/t$ in the regime where
$(x_c-x_0) ={\rm Gi} =G\,[M/m_c^*]+x_0$ remains small.

For small hole concentrations obeying the inequality $0<(x-x_0)\ll1$ the motion of 
the $c$ fermion pairs and the associated kinetic energy
play the role of symmetry-breaking Higgs terms. The presence
of such terms supresses free monopoles and is behind
the replacement at $T=0$ of the long-range antiferromagnetic order
for $x\in (0,x_0)$ by the short-range incommensurate-spiral spin
order for $0<(x-x_0)\ll1$ \cite{companion2}. 
Indeed, independently of the occurrence of phase coherence virtual-electron pairs and corresponding
$c$ fermion pairs prevail for $x>x_0$ and only for $x\in (0,x_0)$ 
are replaced by the monopoles and antimonopoles.
The Higgs terms associated with the motion of the $c$ fermion pairs 
leaves behind the global $U(1)$ symmetry of the
short-range spin ordered phase. This is alike in the related problem of Ref. \cite{duality},
where the motion of the $c$ fermion pairs corresponds to  
that of the $-2e$ charges. 

\subsection{Relation of virtual-electron pairing to spinon and $c$ fermion pairings
and the superconducting order parameter}

In this and following sections further evidence is provided that for intermediate (and probably large) $U/4t$
values a ground-state long-range superconducting order occurs in the VEP quantum liquid at 
zero spin density $m=0$ and hole concentrations in the range $x\in (x_c,x_*)$, as a by-product of the 
short-range spin order.

\subsubsection{Small suppression effects and relation of virtual-electron pairing to spinon and $c$ fermion pairings}

For the hole-concentration range $x\in (x_c,x_*)$ at zero temperature and
a smaller temperature dependent $x$ range centered at $x=x_{op}\approx (x_c +x_*)/2$
for finite temperatures below the critical temperature $T_c$ given below the fluctuations of the phases 
$\theta_{j,0}$ and $\theta_{j,1}$ remain small. Hence the amplitudes,
\begin{equation}
g= \vert\langle e^{i\theta_j}\rangle\vert=\vert\langle e^{i[\theta_{j,0}+\theta_{j,1}]}\rangle\vert = g_0\,g_1
\, ; \hspace{0.35cm}
g_0 = \vert\langle e^{i\theta_{j,0}}\rangle\vert
\, ; \hspace{0.35cm}
g_1 = \vert\langle e^{i\theta_{j,1}}\rangle\vert  \, ,
\label{amplitudes-g0-g1}
\end{equation}
remain finite. Such amplitudes play an important role in the physics of the VEP quantum liquid.

According to our scheme, the effects of the cuprates intrinsic disorder or superfluid density anisotropy 
\cite{chains,YBCO-lam}, called here suppression effects, are behind the lessening of the experimental 
critical temperature relative to its magnitude predicted by the $\gamma_d=1$ VEP quantum liquid scheme.  
Within our oversimplified description of such effects, they are accounted for by a single suppression coefficient,
\begin{eqnarray}
\gamma_d & = & \left(1-\alpha_d{4{\breve{g}}\over\gamma_c}\right) = {T_c\over T_c\vert_ {\alpha_d=0}} 
\, ; \hspace{0.25cm}
\gamma_c=\left(1-{x_c\over x_*}\right) 
\, ; \hspace{0.25cm} 
\gamma_d^{min} = \gamma_d\vert_{x=x_{op}} = (1-\alpha_d) \in (\gamma_c,1) \, ,
\nonumber \\
\alpha_d & = & (1-\gamma_d^{min})  
\in\left(0,{x_c\over x_*}\right) \hspace{0.10cm}{\rm for}\hspace{0.10cm}
{x_c\over x_*}\in \left(0,{1\over 4}\right)
\, ; \hspace{0.25cm} 
\alpha_d \in\left({x_c\over x_*}-{1\over 4},{x_c\over x_*}\right) \hspace{0.10cm}{\rm for}\hspace{0.10cm}
{x_c\over x_*}\in \left({1\over 4},{1\over 2}\right) \, .
\label{gamma-d}
\end{eqnarray}
Here ${\breve{g}}$ denotes the amplitude $g$ of Eq. (\ref{amplitudes-g0-g1}) at $T=0$, 
whose maximum magnitude ${\breve{g}}^{max}=\gamma_c/4$
is found below to be reached at $x=x_{op}=(x_c+x_*)/2$, $T_c\vert_ {\alpha_d=0}$ denotes the 
critical temperature in the absence of suppression effects introduced in the
following, and $T_c$ is its actual magnitude as obtained from experiments. 

For ${\breve{g}}\rightarrow 0$ the very strong phase fluctuations are the main mechanism 
for depressing $T_c$, so that $\gamma_d \rightarrow 1$. This is found below to occur
both for $0<(x-x_c)\ll 1$ and $0<(x_*-x)\ll 1$. In turn, $\gamma_d$ reaches
its minimum magnitude $\gamma_d^{min}=(1-\alpha_d)$ at the optimal
hole concentration $x\approx x_{op}=(x_c+x_*)/2$, when such fluctuations are smallest. 
For the five representative systems $x_c/x_*<1/4$ so that the strength of their suppression effects 
is measured by the magnitude of $\alpha_d \in (0,x_c/x_*)$ and thus of $\gamma_d^{min}\in (\gamma_c,1)$.
The $\alpha_d=(1-\gamma_d^{min})$ ranges given in Eq. (\ref{gamma-d}) are justified below.

The validity of our scheme requires that the experimental $T_c^{max}$ magnitude
is smaller than $T_c^{max}\vert_{\alpha_d=0}$ and larger than $\gamma_c\,T_c^{max}\vert_{\alpha_d=0}$
where $\gamma_c\approx 0.81$ and thus $x_c/x_*\approx 0.19<1/4$ for $x_c=0.05$ and $x_*=0.27$. 
In Ref. \cite{cuprates0} it is found that $\gamma_d^{min}\in (0.94,0.98)$ and thus
$\alpha_d\in (0.02,0.06)$ for the four representative systems other than LSCO.
The corresponding weakness of the suppression effects is consistent 
with the experimental results of Refs. \cite{disorder-0,Sugimoto}. In turn, it is found
that $\gamma_d^{min}=0.82$ and thus $\alpha_d =0.18$ for LSCO. The inequalities 
$\gamma_c\,T_c^{max}\vert_{\alpha_d=0}<T_c^{max}<T_c^{max}\vert_{\alpha_d=0}$,
$\gamma_d^{min}>\gamma_c$, and $\alpha_d<x_c/x_*$ for $x_c/x_*\approx 0.19<1/4$ 
are then met for all representative systems.

For temperatures below the pseudogap temperature $T^*$ the short-range spin order parameter $2\vert\Delta\vert$ 
is identified with $2\vert\Delta\vert = g_1\,2\Delta_0$. For a large range of finite temperatures below $T^*$ 
such that $(T^*-T)>0$ is not too small the amplitude 
$g_1 = \vert\langle e^{i\theta_{j,1}}\rangle\vert$ is expected to be approximately given by 
$g_1\approx {\breve{g}}_1$ where ${\breve{g}}_1$ is its
zero-temperature value. It follows that provided that $(T^*-T)>0$ is not too
small the energy scale $2\vert\Delta\vert$ is for the range $U/4t\in (u_0, u_{\pi})$
for which $x_*\in (0.23, 0.32)$ approximately given by,
\begin{equation}
2\vert\Delta\vert = g_1\,2\Delta_0 \approx {\breve{g}}_1\,2\Delta_0 \approx 2\Delta_0\left(1-{x\over x_*}\right) \, .
\label{2Delta}
\end{equation}

The results of Ref. \cite{companion2} about the dependence of the energy scale $\vert\Delta\vert$ 
on the hole concentration $x$ then indicate that for the square-lattice quantum liquid
the amplitude ${\breve{g}}_1$ reads ${\breve{g}}_1\approx (1-x/x_*)$ for small hole concentrations 
$0<x\ll1$. This is expected to hold for the VEP quantum liquid for $0<(x-x_0)\ll 1$. Interestingly, this 
expression has for the whole range $x\in (x_0,x_*)$ the correct behavior 
${\breve{g}}_1<1$, which is imposed by the inequality $g_1= \vert\langle e^{i\theta_{j,1}}\rangle\vert\leq 1$.
Indeed its maximum magnitude reached for $(x-x_0)\rightarrow 0$ reads ${\breve{g}}_1=\gamma_0 \equiv (1-x_0/x_*)<1$.
Provided that $U/4t\in (u_0, u_{\pi})$ and thus the expression $2\vert\Delta\vert\approx  2\Delta_0 (1-x/x_*)$ 
is valid \cite{companion2}, we then consider that ${\breve{g}}_1\approx (1-x/x_*)$ for hole concentrations in 
the range $x\in (x_0,x_*)$. 

Concerning properties such as the fluctuations of the
phases $\theta_{j,0}$ and $\theta_{j,1}$, the Hamiltonian terms (\ref{H-r-el-nn})
involving spin-singlet rotated-electron pairs at nearest-neighboring 
sites contain the relevant information. In turn, in order 
to capture the extra information needed for the further study of the
coherent virtual-electron pairing and clarification of its relation to $c$ fermion
pairing and $s1$ fermion spinon pairing one must take into account the more
general Hamiltonian terms provided in Eq. (\ref{H-r-el}). When expressed in terms
of $c$ fermion operators and two-site and two-spinon
bond operators these Hamiltonian terms are of the form given in Eq. (\ref{H-bonds-c-s1-b}). 

The amplitude fluctuations near a given real-space
point can be neglected. Nevertheless the $x$ dependence of the average over the whole system 
$\vert\langle e^{i\theta_{j'j''}}\rangle\vert\,\Delta_0$
of the quantity $e^{i\theta_{j'j''}}\,\Delta_0$
appearing in the Hamiltonian terms provided in Eq. (\ref{H-bonds-c-s1-b}) plays
an important role. It follows from Eqs. (\ref{amplitudes-g0-g1}) and (\ref{2Delta}) that 
$\vert\langle e^{i\theta_{j'j''}}\rangle\vert\,\Delta_0$
can at zero temperature be written as ${\breve{g}}_0\,\vert\Delta\vert={\breve{g}}\,\Delta_0$.
Here ${\breve{g}}_0$ and ${\breve{g}}$ denote the $T=0$ values of the amplitudes $g_0$
and $g$, respectively. The physical meaning of such terms is that
the two-site and two-spinon bonds of the $s1$
fermion of Eq. (\ref{g-s1+general-gb}) provide through
the residual interactions with the $c$ fermion pairs the
energy needed for effective pairing coupling between the $c$ fermions.
The source of such an energy transfer is the energy scale $\vert\Delta\vert$, which is one half the order
parameter associated with the short-range spin correlations. 
Hence within the VEP quantum-liquid scheme the short-range spin correlations provide the energy
needed for the $c$ fermion effective coupling 
involved in the pairing mechanism behind
the long-range superconducting order. And this is consistent with the
short-range spin order parameter $2\vert\Delta\vert$ 
having the form $2\vert\Delta\vert = g_1\,2\Delta_0 \approx {\breve{g}}_1\,2\Delta_0$, as given in Eq. (\ref{2Delta}).

The expression in Eq. (\ref{H-bonds-c-s1-b}) of the VEP quantum-liquid effective microscopic 
Hamiltonian terms of Eq. (\ref{H-r-el}) in terms of the local $c$ fermion operators and the two-site 
and two-spinon bond operators contains important physical information. Indeed it reveals 
that the set of $c$ fermion pairs whose centre of mass $[\vec{r}_{j'}+\vec{r}_{j''}]/2$ nearly coincides with the real-space coordinate 
$\vec{r}_{j} =[\vec{r}_{j'}+\vec{r}_{j''}]/2 -{\vec{r}_{d,l}}^{\,0}\approx [\vec{r}_{j'}+\vec{r}_{j''}]/2$
of a given local $s1$ fermion interact with the two-site and two-spinon
bonds contained in that $s1$ fermion, as given in Eq. (\ref{g-s1+general-gb}). An important concept is 
then that of a local virtual-electron pair of real-space coordinate $\vec{r}_{j}$. 
Such an object has the same real-space coordinate as a given local $s1$ fermion. 
Indeed there is a local virtual-electron pair for each local $s1$ fermion.
In addition to that $s1$ fermion, a local virtual-electron pair of real-space coordinate $\vec{r}_{j}$
involves a superposition of quantum configurations of all $c$ fermion pairs whose centre of mass 
$[\vec{r}_{j'}+\vec{r}_{j''}]/2$ nearly coincides with its real-space coordinate 
$\vec{r}_{j} =[\vec{r}_{j'}+\vec{r}_{j''}]/2 -{\vec{r}_{d,l}}^{\,0}\approx [\vec{r}_{j'}+\vec{r}_{j''}]/2$.
Each of its configurations includes the charge $-2e$ of a $c$ fermion pair of
centre of mass $[\vec{r}_{j'}+\vec{r}_{j''}]/2$
and the spin-singlet configuration of two spin-$1/2$ spinons of the local $s1$ fermion of real-space
coordinate $\vec{r}_{j} =[\vec{r}_{j'}+\vec{r}_{j''}]/2 -{\vec{r}_{d,l}}^{\,0}\approx [\vec{r}_{j'}+\vec{r}_{j''}]/2$.
In each of these configurations the local $s1$ fermion of real-space
coordinate $\vec{r}_{j} =[\vec{r}_{j'}+\vec{r}_{j''}]/2 -{\vec{r}_{d,l}}^{\,0}\approx [\vec{r}_{j'}+\vec{r}_{j''}]/2$
interacts with the $c$ fermions of a pair of centre of mass $[\vec{r}_{j'}+\vec{r}_{j''}]/2$.
As discussed below, the $c$ fermion effective coupling results from the energy supplied to each pair
by the $s1$ fermion under consideration. 

In the pseudogap state whose $x$ and $T$ range was mentioned above the amplitude $g$ vanishes. 
However, there remain $c$ fermion pairing correlations associated with the amplitude $g_1$, 
which remains finite. Indeed, the fluctuations of the phases $\theta_1$ remain small. Therefore, 
the concepts of a virtual-electron pair and a VEP quantum liquid hold for the pseudogap state as well, yet for it 
such a pairing has no phase coherence. 

The $c$ fermion discrete momenta ${\vec{q}}_j=[{\vec{q}}^{\,h}_j-\vec{\pi}]$ 
where $j=1,...,N_a^2$ and $s1$ fermion discrete
momenta ${\vec{q}}_j$ where $j=1,...,N_{a_{s1}}^2$
are the conjugate of the real-space coordinates of
the $c$ and $s1$ effective lattices of such objects, respectively.
Such momenta are good quantum numbers for the square-lattice quantum liquid of Ref. \cite{companion2}.
They are close to good quantum numbers for the VEP quantum liquid at hole 
concentrations in the range $x\in (x_c,x_*)$. Alike a local virtual-electron pair has the 
same spatial coordinate $\vec{r}_{j}$ as the corresponding local $s1$ fermion,
a virtual-electron pair configuration of momentum ${\vec{q}}_j$ involves one
$s1$ fermion carrying that momentum. Therefore, the virtual-electron pair configuration 
momenta belong to the $s1$ band. Such a virtual-electon pair configuration consists
of a zero-momentum pair of $c$ fermions of hole momenta ${\vec{q}}^{\,h}$ and $-{\vec{q}}^{\,h}$, 
which interact with a $s1$ fermion of momentum ${\vec{q}}$.
Within a local virtual-electron pair the well-defined set of $c$ fermion pairs 
with the same centre of mass $[\vec{r}_{j'}+\vec{r}_{j''}]/2$
interact with the $s1$ fermion of real-space coordinate 
$\vec{r}_{j} =[\vec{r}_{j'}+\vec{r}_{j''}]/2 -{\vec{r}_{d,l}}^{\,0}\approx [\vec{r}_{j'}+\vec{r}_{j''}]/2$. 
Consistently, the effective pairing coupling between $c$ fermions of hole momenta 
${\vec{q}}^{\,h}$ and $-{\vec{q}}^{\,h}$ results from elementary processes
where such objects have interactions with a $s1$ fermion of 
well-defined momentum ${\vec{q}}$. 

We emphasize that a pair of $c$ fermions of hole momenta 
${\vec{q}}^{\,h}$ and $-{\vec{q}}^{\,h}$ is a superposition of local $c$ fermion pairs
whose centre of mass $[\vec{r}_{j'}+\vec{r}_{j''}]/2$ is different 
and thus interact with different $s1$ fermions. Therefore, in the
occupancy configurations of the superconducting 
ground state there is no one-to-one correspondence between a coherent $c$ fermion pair 
and a given $s1$ fermion. The occupancies of such a ground state 
involve superpositions of phase-coherent
 virtual-electron pair configurations of
different momentum. The same pair of $c$ fermions of hole momenta 
${\vec{q}}^{\,h}$ and $-{\vec{q}}^{\,h}$ participates in different virtual-electron
pair configurations where it interacts with a $s1$ fermion
of different momentum ${\vec{q}}$. And vice versa, each $s1$ fermion
of momentum ${\vec{q}}$ participates in different virtual-electron
pair configurations where it interacts with different pairs of $c$ fermions. 

The short-range spin correlations provide through the $c$ - $s1$ fermion interactions the 
energy needed for the effective coupling between the two $c$ fermions of a pair.
These zero-momentum $c$ fermion pairs couple to charge probes independently of 
the $s1$ fermions. The superconducting macroscopic condensate refers
to such pairs. In turn, the virtual-electron pair configurations are coupled 
to spin probes through the $s1$ fermions.
It follows that concerning charge excitations the zero-momentum
coherent $c$ fermion pairs behave as independent objects relative to the virtual-electron
pair configurations. In turn, one-electron and spin excitations break phase-coherent
 virtual-electron pairs.
Indeed, as discussed below virtual-electron pairs exist in intermediate virtual states
generated by pair breaking processes upon such excitations. 
A virtual electron pair configuration involves the charge $-2e$ of its $c$ fermion pair and the spin-singlet configuration
of the two spin-$1/2$ spinons of its composite $s1$ fermion. 

A virtual-electron pairing configuration involves two types 
of pairing: (i) The zero-momentum $c$ fermion pairing 
whose effective coupling between the $c$ fermions results from residual 
interactions with the $s1$ fermion; (ii) The spin-singlet spinon 
pairing of the composite two-spinon $s1$ fermion. 
As reported below in Section III-E, at zero temperature and hole concentrations in the range $x\in (x_c,x_*)$ one may reach the pseudogap state 
by applying a magnetic field $H$ aligned perpendicular to the square-lattice planes.
The occupancies of the $g_1>0$ and $g=0$ normal ground state 
involve again superpositions of virtual-electron pair configurations of different
momentum ${\vec{q}}$. However such pair configurations and corresponding $c$ fermion 
pairs have no phase coherence. The same applies to
the normal ground state for vanishing magnetic field $H=0$ and finite hole concentrations 
in the range $x\in (x_0,x_c)$.

Strong effective coupling between $c$ fermions of a pair is defined below 
as that whose breaking under one-electron excitations leads to sharp spectral features in
the corresponding $(\vec{k},\omega)$-space distribution. One then classifies
the $c$ - $s1$ interactions into two classes: Those that lead and do not lead
to strong effective coupling, respectively. Importantly, only $c$ fermion pairs
with strong effective coupling can participate in phase-coherent virtual electron
pair configurations. Indeed, $c$ fermion strong effective coupling is a necessary but
not sufficient condition for coherent pairing.  

According to the analysis presented below in Section IV-C, the momentum ${\vec{q}}$ of
strongly coupled virtual-electron pairs may belong to different $s1-sc$ lines
labelled by the line normal momentum absolute value $q_{arc}^N\in (q_{ec}^N,q_{Bs1}^N)$. 
Here $sc$ stands for {\it strong coupling}. The $s1-sc$ line of the momentum of a
$s1$ fermion contributing to the strong effective coupling of two $c$ fermions is
determined by the absolute value $q^h$ of their hole momenta ${\vec{q}}^{\,h}$ and 
$-{\vec{q}}^{\,h}$. Indeed, there is an one-to-one correspondence between each 
$q_{arc}^N\in (q_{ec}^N,q_{Bs1}^N)$ value labeling a given $s1-sc$ line
and $q^h=\vert{\vec{q}}^{\,h}\vert\in (q_{Fc}^h,q_{ec}^h)$. 
The set of $c$ fermions with the same hole momentum 
absolute value $q^h=\vert{\vec{q}}^{\,h}\vert\in (q_{Fc}^h,q_{ec}^h)$
are involved in virtual-electron pairs whose momentum 
${\vec{q}}$ belongs to the same $s1-sc$ line. 
Hence the $s1$ fermions whose momentum belongs to a given $s1-sc$ line 
interact with the $c$ fermions whose hole momenta 
${\vec{q}}^{\,h}$ and $-{\vec{q}}^{\,h}$ belong to an approximately circular
$c-sc$ line of radius $q^h\in (q_{Fc}^h,q_{ec}^h)$ centered at $-\vec{\pi}=-[\pi,\pi]$.
And vice versa. $q_{ec}^N$ and $q_{ec}^h$ are the minimum momentum
and corresponding maximum hole momentum, respectively, for which there is
$c$ fermion strong effective coupling. (Such properties refer only to $c$ - $s1$ fermion interactions 
leading to $c$ fermion strong effective coupling.)

Combination of the expression of the number of holes in the $s1$ band
$N_{s1}^h $ given in Eq. (\ref{deltaNcs1}) with that provided in Eq. (\ref{Nas1-Nhs1}) of 
its deviation $\delta N_{s1}$ due to a ground-state - excited-state transition 
gives $N^h_{s1} = [\delta N - 2\delta N_{s1} - 2N_{s2}]$. Here $N_{s2}=0,1$
is the number of four-spinon $s2$ fermions \cite{companion2}. This is the form that
the exact relation between the
excited-states number of holes in the $s1$ band and the corresponding
deviation $\delta N$ from the ground-state electron-number value has
for the one- and two-electron subspace. It implies that  
the matrix elements between one-electron (and two-electron)
excitations and excited states with an even (and an odd) 
number of $s1$ fermion holes vanish. Such an exact
selection rule applies when the initial ground state refers to
vanishing spin density $m=0$. It reveals that excited states with a single
hole in the $s1$ band are generated by application onto that
ground state of one-electron operators. In turn, those with none
or two holes in that band are generated by application of
two-electron operators. Furthermore, the transformation laws under
the electron - rotated-electron transformation of the quantum
objects whose occupancy configurations generate the energy
eigenstates of the Hubbard model can be used to show that nearly 
the whole one-electron (and two-electron)  
spectral weight corresponds to excited states with a single $s1$ 
fermion hole (and none or two $s1$ fermion holes) \cite{companion2}. 
The same is expected to hold for the VEP quantum liquid.

In conventional superconductivity the objects that pair are 
Fermi-liquid quasiparticles \cite{Pines,Schrieffer}.
Here virtual-electron pairing involves a pair of charge $c$
fermions and one spin-neutral two-spinon $s1$ fermion.
The occurrence of $s1$ fermion spin-singlet spinon 
pairing is needed for coherent virtual-electron pairing. In contrast, the 
former pairing can occur independently of the latter. Indeed,   
the $s1$ fermion  spin-singlet spinon pairing
occurs both in the superconducting and pseudogap states. 
Consistently, coherent virtual-electron pairing is a by-product of
the $s1$ fermion spin-singlet spinon pairing. It occurs provided that the amplitude 
$g=\vert\langle e^{i\theta_{j}}\rangle\vert$ is finite, owing to a macroscopic quantum 
phase-coherent virtual-electron pair state.

The coherent virtual-electron pairing involves phase coherence. The
$s1$ fermion spinon pairing does not. Concerning charge
excitations the coherent $c$ fermion pairs of charge $-2e$ behave as independent
entities, which carry the superfluid current. The flux quantization is $-hc/2e$. 
Hence in the superconducting state in the presence 
of a vector potential, the $c$ band dispersion whose expression
is provided in Eqs. (\ref{c-band}) and (\ref{bands}) of Appendix A is shifted and 
approximately given by,
\begin{equation}
\epsilon_{c}^{\vec{A}} ({\vec{q}}^{\,h})\approx 
\epsilon_{c} ({\vec{q}}^{\,h}) + 
\left({1\over 2e}\vec{\nabla}\theta_{cp} - \vec{A}\right)\cdot {\vec{j}}_c ({\vec{q}}^{\,h}) \, .
\label{Ec-A}
\end{equation}
Here $\theta_{cp}$ is the phase of Eq. (\ref{theta-cp}) 
and ${\vec{j}}_c ({\vec{q}}^{\,h})$ the $c$ fermion
current of Eq. (\ref{jc}) of Appendix A. It reads 
$j_{c} ({\vec{q}}^{\,h})\approx {\vec{q}}^{\,h}_{Fc}/m^{\rho}_c$
for hole momenta ${\vec{q}}^{\,h}$ at or near the $c$
Fermi line. Its absolute value is for $x\in (x_c,x_*)$ and 
$U/4t\in (u_0,u_{\pi})$ approximately 
given by $j_{c} ({\vec{q}}^{\,h})\approx \sqrt{x\pi}\,2/m_c^{\rho}$. 
The renormalized transport mass $m_c^{\rho}$ is provided in Eq. (\ref{jc}) of Appendix A. In the units
used here it reads $m_c^{\rho}=r_c/2t$.

\subsubsection{Virtual-electron pairs in virtual intermediate states of
one- and two-electron excitations}

As discussed in Section III-A,
removal of two electrons in a spin-singlet configuration leads
to an excited state with no holes in the $s1$ band.
Upon such an excitation, before the two involved $c$ fermions
recombine with one $s1$ fermion giving rise to the 
removed electron pair under consideration, the system is driven into an
intermediate virtual state. In it the $c$ fermion 
pair under consideration stops interacting with all $s1$ fermions other
than that $s1$ fermion. Hence, in that intermediate virtual state
both the $c$ fermion pair and the $s1$ fermion participate
in a single virtual-electron pair. In it such a pair exists as an
individual object. The occurrence of that intermediate virtual state is behind the
designation virtual-electron pair. This holds both for the superconducting and pseudogap states.
Such a two-electron process corresponds to a gapless excitation branch.
It refers to the last excitation energy $\delta E = 0$ given in Eq. (\ref{DE-3proc}).

In turn, under one-electron removal and spin excitations there emerges one and two 
holes in the $s1$ band, respectively. Also within such excitations the system
is driven into an intermediate virtual-electron
state. In it one $c$ fermion pair and one $s1$ fermion
contribute to a single virtual electron pair: that
broken under the excitation. Such a discontinuous 
change is brought about by application onto the
ground state of the corresponding one-electron
and spin two-electron operator, respectively. Within these processes
the $c$ fermion pair ceases to interact with $s1$
fermions other than that under consideration. Furthermore,
that $s1$ fermion ceases to interact with all
remaining $c$ fermion pairs. The virtual
electron pair and the corresponding $c$ fermion
pair and $s1$ fermion spinon pair are then broken under
the one-electron removal or spin excitation. Finally, under the former
excitation one $c$ fermion and one spinon are removed along with
the electron and one hole emerges in the $s1$ band.
In turn, under a spin-triplet excitation a spinon spin-flip occurs.
It generates two independent spinons. Under a spin-singlet excitation 
the two spinons recombine with those of another $s1$ fermion
giving rise to a spin-singlet four-spinon $s2$ fermion \cite{companion2}.
Such spin two-electron processes are followed by the emergence of two holes in the $s1$ 
band. In addition, the $c$ fermion pair left over under these spin excitations restarts interacting within virtual-electron pairs 
with other $s1$ fermions. This is how one-electron and spin quantum measurements destroy 
the ground-state wavefunction and create a new wavefunction.

\subsubsection{Critical temperature, superconducting order parameter, and coherent length}

The order parameter associated with the short-range spin correlations is 
the maximum magnitude of the $s1$
fermion spinon-pairing energy $2\vert\Delta\vert$. It is expected that the absolute value
$2\vert\Omega\vert$ of the superconducting order parameter 
is the maximum magnitude of the coherent virtual-electron pairing 
energy. According to the pairing mechanism emerging from
the above analysis, the short-range spin correlations supply the latter pairing
energy. It is then expected that for hole concentrations in the range $x\in (x_c,x_*)$ and zero temperature both
the inequality $2\vert\Omega\vert<2\vert\Delta\vert$ and the relation 
$2\vert\Omega\vert\propto 2\vert\Delta\vert$ hold.
The pseudogap temperature (\ref{Dp-T*}) is given by
$T^*\approx {\breve{g}}_1\,\Delta_0/k_B$. It is controlled by
the zero-temperature amplitude ${\breve{g}}_1$, which is finite
for hole concentrations in the range $x\in (x_0,x_*)$ where there is short-range spin order.
Within the present approach the superconductivity critical temperature 
reads $T_c\approx \gamma_d\,{\breve{g}}\,\Delta_0/2k_B$, so that
$T_c\vert_ {\alpha_d=0}\approx {\breve{g}}\,\Delta_0/2k_B$ in Eq. (\ref{gamma-d}). It is controlled by the 
amplitude ${\breve{g}}$, which is finite for hole concentrations in the range 
$x\in (x_c,x_*)$. For it a long-range superconducting order coexists with the short-range spin order, as its
by-product. 

Physically, the energy scales (i) $2\vert\Omega\vert\vert_{T=0}$ and 
(ii) $4k_B\,T_c$ are the maximum magnitude of the coherent $c$ 
fermion pairing energy within pair breaking (i) at zero-temperature under spin-triplet
excitations and (ii) upon increasing the temperature. 
The relation $2\vert\Delta\vert\vert_{T=0}\approx 2k_B\,T^*$ holds. The corresponding 
expected relation $2\vert\Omega\vert\vert_{T=0}\approx 4k_B\,T_c$ holds both in the limits
$0<(x-x_c)\ll 1$ and $0<(x_*-x)\ll 1$. Hence in these limits when the phase fluctuations
are strong the superconducting energy scales $2\vert\Omega\vert\vert_{T=0}$ and 
$4k_B\,T_c$ have the same magnitude. In turn, at $x=x_{op}=(x_c+x_*)/2$ such fluctuations
are smallest at zero temperature and these two energy scales have slightly different magnitudes. Due to
correlations the energy scales considered in the following are not additive, so
that our discussion is merely qualitative. The maximum energy 
that the short-range spin correlations associated with the energy scale $2\vert\Delta\vert (x_{op})$
can supply to coherent pairing is at $x=x_{op}$ approximately given by
$\approx 2\vert\Delta\vert (x_{op})\vert_{T=0}/[2\gamma_c]=\Delta_0/2$. In the absence of suppression
effects the above pairing energies read
$4k_B\,T_c (x_{op})\approx {\breve{g}}_0\,2\vert\Delta\vert (x_{op})\vert_{T=0}=\gamma_c\,\Delta_0/2$
and $2\vert\Omega\vert (x_{op})\vert_{T=0}\approx \Delta_0/2$. For pair breaking
under spin excitations the pairing energy $2\vert\Omega\vert (x_{op})\vert_{T=0}$ 
equals that delivered by the short-range spin correlations $\approx \Delta_0/2$. In contrast, in the case
of $4k_B\,T_c (x_{op})\approx (1-x_c/x_*)\,\Delta_0/2$ only a fraction of the latter
energy is supplied to pairing, the energy $\approx (x_c/x_*)\,\Delta_0/2$ left over being lost 
due to phase and thermal fluctuations. If one accounts for the suppression
effects the above two related pairing energies then approximately read
$4k_B\,T_c (x_{op})\approx \gamma_d^{min}\,{\breve{g}}_0\,2\vert\Delta\vert (x_{op})\vert_{T=0}=\gamma_d^{min}\,\gamma_c\,\Delta_0/2$
and $2\vert\Omega\vert (x_{op})\vert_{T=0}\approx\gamma_d^{min}\,\Delta_0/2$. Hence the relation,
\begin{equation}
{4k_B\,T_c (x_{op})\over 2\vert\Omega\vert (x_{op})\vert_{T=0}} = \gamma_c
\, ; \hspace{0.35cm}
x_{op} = {1\over 2}(x_*+x_c) \, ,
\label{x-op}
\end{equation} 
holds. The following expression satisfies such limiting behaviors,
\begin{equation}
2\vert\Omega\vert\vert_{T=0}
\approx {4k_B\,T_c\over {\breve{\beta}}_c} =  {\gamma_d\,{\breve{g}}\,2\Delta_0\over{\breve{\beta}}_c}
\, ; \hspace{0.35cm}
{\breve{\beta}}_c  \equiv  \left(1-{x_c\over x_*}{T_c\over T^{max}_c}\right) \, .
\label{Omega-Tc}
\end{equation}
The inequalities $\alpha_d<x_c/x_*$ and $x_c/x_*<1/2$ in the range of the parameter $\alpha_d=(1-\gamma_d^{min})$
of Eq. (\ref{gamma-d}) are justified by the physical range and inequality $2\vert\Omega\vert\vert_{T=0}^{max}\in ({\breve{g}}\,2\Delta_0,\Delta_0/2)$
and $2\vert\Delta\vert\vert_{T=0}/[2\gamma_c]=\Delta_0/2<2\vert\Delta\vert\vert_{T=0}$, respectively,
corresponding to $x=x_{op}$. The latter follows from the physical requirement that the maximum energy supplied by the short-range
spin correlations being smaller than their own energy scale $2\vert\Delta\vert$. 
In turn, concerning the above range only a fraction $(1-\alpha_d)\Delta_0/2$
of the available maximum energy $\Delta_0/2$ is supplied, whereas the energy $\alpha_d\,\Delta_0/2$ left over is lost due to the
suppression effects. Our oversimplified scheme in terms of such effects is valid provided that
$2\vert\Omega\vert\vert_{T=0}^{max}=(1-\alpha_d)\Delta_0/2\in ({\breve{g}}\,2\Delta_0,\Delta_0/2)$.
At $\alpha_d=x_c/x_*$ the equality $\gamma_d ={\breve{\beta}}_c$ holds,
so that $2\vert\Omega\vert\vert_{T=0}^{max}={\breve{g}}\,2\Delta_0$. Otherwise, $\gamma_d <{\breve{\beta}}_c$
and ${\breve{g}}\,2\Delta_0<2\vert\Omega\vert\vert_{T=0}^{max}<\Delta_0/2$ for $\alpha_d<x_c/x_*$ and $\gamma_d^{min}>\gamma_c$.
Below it is confirmed that $T^{max}_c=T_c (x_{op})$ and $2\vert\Omega\vert\vert_{T=0}^{max}=2\vert\Omega\vert (x_{op})\vert_{T=0}$. 
For the critical hole concentration values $x_c\approx 0.05$ and $x_*\approx 0.27$ of the five representative systems 
the optimal hole concentration of Eq. (\ref{x-op}) is given by $x_{op}\approx 0.16$, in agreement with experiments.

It follows from the above analysis that the order parameter $2\Omega$ and the energy scale $4k_B\,T_c$ read,
\begin{equation}
2\Omega = e^{i\theta_{cp}} 2\vert\Omega\vert 
\, ; \hspace{0.35cm} 
\theta_{cp} = \theta_j -\phi^0_{j,s1}
\, ; \hspace{0.35cm} 
\theta_j =\theta_{j,0}  +\theta_{j,1}
\, ; \hspace{0.35cm}
4k_B\,T_c \approx \gamma_d\,{\breve{g}}\,2\Delta_0 \, .
\label{Omega}
\end{equation}
Here $2\vert\Omega\vert$ is for $T=0$ provided in Eq. (\ref{Omega-Tc}),
$\theta_{cp}=\theta_{cp} (\vec{r}_j)$ are the phases of Eq. (\ref{theta-cp}), and
${\breve{g}} = {\breve{g}}_0\,{\breve{g}}_1$ is the zero-temperature
value of the overall amplitude given in Eq. (\ref{amplitudes-g0-g1}).

Generalization to finite temperature $T>0$ of the zero-temperature $2\vert\Omega\vert $ 
expression provided in Eq. (\ref{Omega-Tc}) leads for a temperature-dependent $x$ range 
centered at $x=x_{op}$ to,
\begin{equation}
2\vert\Omega\vert
= {\gamma_d\,g\,2\Delta_0\over \beta_c}\, ; \hspace{0.35cm}
\beta_c = \left(1-{x_c\over x_*}{4g\over \gamma_c}\right) \, .
\label{Omega-gen}
\end{equation}
Here $\gamma_c$ is the parameter given in Eq. (\ref{gamma-d}). Note that 
${\beta}_c\vert_{T=0} ={\breve{\beta}}_c$ where ${\breve{\beta}}_c$ is defined in Eq. (\ref{Omega-Tc}).

The $2\Omega$ expression of Eq. (\ref{Omega}) is such that 
$2\Omega\propto e^{i\theta_{cp}}\,\vert\langle e^{i\theta_{cp}}\rangle\vert\,2\Delta_0\approx e^{i\theta_{cp}}\,g\,2\Delta_0$.
However, the amplitude fluctuations are not accounted for twice.
The point is that $\vert\langle e^{i\theta_{cp}}\rangle\vert$
refers to an average over the whole system. Hence the amplitude
$\vert\langle e^{i\theta_{cp}}\rangle\vert$ is independent
of the spatial coordinate $\vec{r}_j$. In turn, the $\vec{r}_j$ dependence of the
phase factor $e^{i\theta_{cp}}$
corresponds to a small real-space region around
$\vec{r}_j$ where the phases $\theta_{cp}$ change little and smoothly.
Then amplitude fluctuations can be neglected in that
small region around $\vec{r}_j$ and averaging $e^{i\theta_{cp}}$ 
over the local virtual-electron pairs contained in it gives
$\vert\langle e^{i\theta_{cp}}\rangle_{near-j}\vert\approx 1$.
This justifies why such fluctuations are not accounted for twice.

That local normalization is not fulfilled 
when the phase fluctuations become large. Nevertheless, then $g\rightarrow 0$
and thus $2\Omega\rightarrow 0$. Hence the expressions 
provided in Eqs. (\ref{Omega}) and (\ref{Omega-gen}) 
remain valid. One then concludes that
the range of validity of the expression $2\Omega = e^{i\theta_{cp}} 2\vert\Omega\vert$
whose $\vec{r}_j$ dependence occurs through
the phases $\theta_{cp}=[\theta_{j}-\phi^0_{j,s1}]$
refers to a small real-space region around $\vec{r}_j$.
We recall that the fluctuations of the phases $\theta_{cp}$ 
result mostly from those of the phases $\theta_{j}=[\theta_{j,0}+\theta_{j,1}]$.
Indeed, the fluctuations of the phases $\phi^0_{j,s1}$ are
very small for the whole hole-concentration range $x\in (x_0,x_ *)$.

At $\alpha_d=0$ and thus $\gamma_d=1$ the superconducting energy scale $4k_B\,T_c$
of Eq. (\ref{Omega}) equals the zero-temperature 
average over the whole system $\vert\langle e^{i\theta_{j'j''}}\rangle\vert\,\Delta_0$
of the basic quantity $e^{i\theta_{j'j''}}\,\Delta_0$
appearing in the Hamiltonian terms (\ref{H-bonds-c-s1-b}).
The expressions provided in Eqs. (\ref{2Delta}),
(\ref{Omega-Tc}), (\ref{Omega}), and (\ref{Omega-gen})
confirm that the short-range spin order parameter
$2\vert\Delta\vert$, energy scale $4k_B\,T_c$, and superconducting
order parameter $2\Omega$ are closely related.
This is consistent with the short-range spin order and long-range
superconducting order being closely related as well.
The presence of the amplitude $g_1=\vert\langle e^{i\theta_{j,1}}\rangle\vert$
within the overall amplitude $g=g_0\,g_1$ of the superconducting order parameter 
(\ref{Omega-gen}) is consistent with the short-range
spin correlations providing the energy needed for the
coherent virtual-electron pairing. Simultaneously, such a supplying of
energy by the short-range spin correlations
suppresses them through the increasing fluctuations of the phases
$\theta_{j,1}$. The latter phases refer to the internal
degrees of freedom of the virtual-electron pairs. 

The phases $\theta_{j,0}$ and $\theta_{j,1}$ are such that ${\breve{g}}_0=\vert\langle e^{i\theta_{j,0}}\rangle\vert_{T=0}= 0$
and ${\breve{g}}_1=\vert\langle e^{i\theta_{j,1}}\rangle\vert_{T=0}= \gamma_c$
for $0<(x-x_c)\rightarrow 0$ and ${\breve{g}}_0=\vert\langle e^{i\theta_{j,0}}\rangle\vert_{T=0}= 1$ and
${\breve{g}}_1=\vert\langle e^{i\theta_{j,1}}\rangle\vert_{T=0}= 0$ for $0<(x_*-x)\rightarrow 0$. 
These behaviors are consistent with the critical temperature $T_c$ being small and given by
$T_c\approx {\breve{g}}_0\,\Delta_0/2k_B$ and
$T_c\approx {\breve{g}}_1\,\Delta_0/2k_B$ for $0<(x-x_c)\ll 1$ and $0<(x_*-x)\ll 1$, respectively. 
In these limits ${\breve{g}}={\breve{g}}_0\,{\breve{g}}_1\rightarrow 0$, so that $\gamma_d\rightarrow 1$,
the suppression effects play no role, and the physics is
controlled by the very strong phase fluctuations.
Symmetry arguments associated with the physics specific to these
limits for which the critical temperature $T_c$ is controlled only by strong phase fluctuations  
imply that ${\breve{g}}_0\approx (x/x_*)^{z\nu}$ and ${\breve{g}}_1\approx ([x_*-x]/x_*)^{z\nu}$
are for $0<(x-x_c)\ll 1$ and $0<(x_*-x)\ll 1$, respectively, 
controlled by the same dynamical exponent
$z=1$ and unknown exponent $\nu$. 

Treatments involving the use of the effective action for the phases $\theta_j$
without incorporating the Berry phase \cite{duality} lead to $\nu\approx 2/3$. More detailed treatments,
incorporating the latter phase, lead to different values for that exponent \cite{fluctua}. 
Symmetry arguments suggest that ${\breve{g}}_1\approx ([x_*-x]/x_*)^{z\nu}=(1-x/x_*)^{z\nu}$
has the same overall exponent $z\nu$ for hole concentrations obeying both the
inequalities $0<(x_*-x)\ll 1$ and $0<(x-x_0)\ll 1$. On combining
the expression $2\Delta = g_1\,2\Delta_0$ of Eq. (\ref{2Delta}) with the square-lattice quantum-liquid results of
Ref. \cite{companion2} for $0<x\ll 1$, which in our case hold for $0<(x-x_c)\ll 1$, 
we then find $z\nu=1$ so that $\nu=1$. Within our scheme the suppression effects 
lessen the critical temperature and related energy scales, yet do not affect
the phases and corresponding amplitude $g=g_0\,g_1$.
We then consistently consider that for approximately the range $U/4t\in (u_0,u_{\pi})$ 
and hole concentrations $x\in (x_0,x_*)$  
the zero-temperature amplitude ${\breve{g}}_1$ is given by ${\breve{g}}_1\approx (1-x/x_*)$.
Furthermore, the expression ${\breve{g}}_0 \approx (x-x_c)/(x_*-x_c)$ obeys 
for $x\in (x_c,x_*)$ the inequality ${\breve{g}}_0= \vert\langle e^{i\theta_{j,0}}\rangle\vert_{T=0}\leq 1$.
It also obeys the expected boundary condition ${\breve{g}}_0\rightarrow 1$ as
$x\rightarrow x_*$. Therefore, we consider that ${\breve{g}}_0 \approx (x-x_c)/(x_*-x_c)$
for hole concentrations in the range $x\in (x_c,x_*)$. Hence within the present approach the 
expressions of the amplitudes $g_0$ and $g_1$ of Eq. (\ref{amplitudes-g0-g1}) 
are for intermediate interaction values $U/4t\in (u_0,u_{\pi})$ and zero temperature approximately given by,
\begin{eqnarray}
{\breve{g}} & = & \vert\langle e^{i\theta_{j}}\rangle\vert_{T=0} = {\breve{g}}_0\,{\breve{g}}_1 \approx
{(x-x_c)\over (x_*-x_c)}\,\left(1-{x\over x_*}\right) \, ; \hspace{0.35cm}
{\breve{g}}_0  = \vert\langle e^{i\theta_{j,0}}\rangle\vert_{T=0} \approx {(x-x_c)\over (x_*-x_c)} 
\hspace{0.15cm}{\rm for} \hspace{0.10cm} x \in (x_c,x_*) \, ,
\nonumber \\
{\breve{g}}_1 & = & \vert\langle e^{i\theta_{j,1}}\rangle\vert_{T=0} \approx \left(1-{x\over x_*}\right) 
\hspace{0.15cm}{\rm for} \hspace{0.10cm} x \in (x_0,x_*) \, ; \hspace{0.35cm}
{\rm max}\{{\breve{g}}_1\} = \gamma_0=\left(1-{x_0\over x_*}\right)\hspace{0.10cm}{\rm for}\hspace{0.10cm} 
0<(x-x_0)\ll 1 \, ,
\nonumber \\
{\breve{g}}_0 & = & {\breve{g}} = 0  
\hspace{0.15cm}{\rm for}
\hspace{0.10cm} 0\leq x\leq x_c \hspace{0.10cm}{\rm and}
\hspace{0.10cm} x\in (x_*,1) \, ; \hspace{0.35cm}
{\breve{g}}_0 = {\breve{g}}_1 = {\breve{g}} = 0  
\hspace{0.15cm}{\rm for}
\hspace{0.10cm} x\in (0, x_0)
\hspace{0.10cm}{\rm and}
\hspace{0.10cm} x\in (x_*,1) \, .
\label{g0-g1}
\end{eqnarray}
Due to the hole-trapping effects reported in Appendix B, the fluctuations of the
phases $\theta_{j,1}$ are very strong for $x\in (0,x_0)$ and the zero-temperature 
amplitude ${\breve{g}}_1=\vert\langle e^{i\theta_{j,1}}\rangle\vert_{T=0}$ vanishes.
For $x\in (x_0,x_c)$ the hole-trapping effects remain 
active yet are weaker than for $x\in (0,x_0)$. In contrast to the range $x\in (0,x_0)$, 
the short-range spiral incommensurate spin order of the related square-lattice
quantum liquid of Ref. \cite{companion2} survives for
hole concentrations $x\in (x_0,x_c)$, coexisting with 
the Anderson insulating behavior brought about by the hole-trapping effects.
  
The behaviors of the zero-temperature amplitudes ${\breve{g}}_0$ and ${\breve{g}}_1$ follow from the 
fluctuations of the corresponding phases $\theta_0$ and $\theta_1$ of Eqs. (\ref{theta-j}) and (\ref{theta-cp})
becoming large for $0<(x-x_c)\ll 1$ and $0<(x_*-x)\ll 1$, respectively.
The singular behavior ${\breve{g}}_1\rightarrow\gamma_0$ for $0<(x-x_0)\rightarrow 0$ 
and ${\breve{g}}_1=0$ at $(x-x_0)=0$ is due to a sharp quantum phase transition.
It marks the onset of the long-range antiferromagnetic order for $x<x_0$.
In turn, the singular behavior ${\breve{g}}_0\rightarrow 1$ for $0<(x_*-x)\rightarrow 0$
and ${\breve{g}}_0=0$ at $(x_*-x)=0$ is also due to a sharp quantum phase transition marking the
onset to a disordered state without short-range spin order for $x>x_*$. 
Due to such sharp quantum-phase transitions, the phases 
$\theta_1$ and $\theta_0$ have also a singular behavior at $x=x_0$ and $x=x_*$, respectively. The fluctuations 
of these phases are small for $0< (x-x_0)\ll 1$ and $0<(x_*-x)\ll 1$ and large for $x<x_0$ and
$x>x_*$, respectively. 

Both in the Anderson insulator phase with short-range 
spiral incommensurate spin order occurring in the hole concentration range $x\in (x_0,x_c)$ due to the 
hole-trapping effects and in the phase corresponding to the range $x\in (x_c,x_*)$ 
for which the short-range spin order coexists with a long-range superconducting order the symmetry of the 
phases $\theta=\theta_0 +\theta_1$ action is a global $U(1)$. 
Such a global $U(1)$ symmetry is related to the Hubbard-model
global $[SO(4)\times U(1)]/Z_2$ symmetry found recently beyond its known 
$SO(4)$ symmetry \cite{bipartite}. 
 
On combining Eqs. (\ref{Omega}) and (\ref{g0-g1}) one finds the
following expression for the critical temperature,  
\begin{equation}
T_c \approx \gamma_d\,T_c\vert_ {\alpha_d=0} = 
\breve{g}\left(1-\alpha_d {4\breve{g}\over\gamma_c}\right){\Delta_0\over 2k_B} 
\, ; \hspace{0.35cm}
T_c\vert_ {\alpha_d=0} \approx \breve{g}{\Delta_0\over 2k_B} =
{(x-x_c)\over (x_*-x_c)}\left(1-{x\over x_*}\right){\Delta_0\over 2k_B} \, .
\label{Om-Dp-Tc}
\end{equation} 
The related energy scale $2\vert\Omega\vert\vert_{T=0}$ is then given by
expression (\ref{Omega-Tc}) with $T_c$ as provided here. Hence the
energy scales $2\vert\Delta\vert\vert_{T=0}$ of Eq. (\ref{2Delta}) at zero temperature
and $2\vert\Omega\vert\vert_{T=0}$ associated with the short-range spin order and 
long-range superconducting order read,
\begin{equation}
2\vert\Delta\vert\vert_{T=0} = {\breve{g}}_1\,2\Delta_0\,\theta (x-x_0) \, ;
\hspace{0.35cm}
2\vert\Omega\vert\vert_{T=0}
\approx {4k_B\,T_c\over {\breve{\beta}}_c} = {\gamma_d\,{\breve{g}}\,2\Delta_0 \over {\breve{\beta}}_c} \, ,
\label{Omega-Tc-di}
\end{equation}
respectively. Analysis of these expressions reveals that the physical requirement that 
$2\vert\Omega\vert<2\vert\Delta\vert$ is met for the whole hole concentration range 
$x\in (x_c,x_*)$ provided that the $\alpha_d=(1-\gamma_d^{min})$ ranges given in 
Eq. (\ref{gamma-d}) are fulfilled. 

It follows from the expressions of the energy scale 
$2\vert\Omega\vert\vert_{T=0}$ and critical temperature $T_c$
provided in Eqs. (\ref{Omega-Tc}) and (\ref{Om-Dp-Tc}), respectively, 
that the maximum magnitudes of these quantities read,
\begin{equation}
2\vert\Omega\vert\vert^{max}_{T=0} = \gamma_d^{min}\,{\Delta_0\over 2} 
\, ; \hspace{0.35cm}
T^{max}_c =  \gamma_d^{min}\gamma_c\,{\Delta_0\over 8k_B} 
\, ; \hspace{0.35cm} 
\gamma_d^{min}=(1-\alpha_d) \in (\gamma_c,1) \, .
\label{max-Tc}
\end{equation}
Those are achieved at the optimal hole concentration given in Eq. (\ref{x-op}).

Complementarily to the complex parameter $2\Omega$ of Eq. (\ref{Omega}), 
one can consider the order parameter $\phi_{cp}=\sqrt{n_{cp}/2}\,e^{i\theta_{cp}}$.
Here $n_{cp}$ denotes the density of paired $c$ fermions contributing to phase-coherent virtual-electron pair
configurations. The $x$ dependence
of $n_{cp}$ is studied below in Section IV. The complex parameter $\phi_{cp} (\vec{r}) =
\sqrt{n_{cp}/2}\,e^{i\theta_{cp}}$ describes
the macroscopic properties of the zero-momentum $c$ fermion pairs superfluid condensate.
Indeed, $\vert\phi_{cp}\vert^2= n_{cp}/2$ is a measure of the
local superfluid density of coherent $c$ fermion pairs.
The average $c$ fermion distance or length $\xi_1$ of the set of $c$ fermion pairs contributing to a 
local virtual-electron pair is related to the coherent length $\xi$ associated with 
coherent virtual-electron pairing. For the isotropic Fermi-velocity range $x\in (x_{c1},x_{c2})$
and approximately $U/4t\in (u_0,u_{\pi})$ such lengths read, 
\begin{eqnarray}
\xi_1 & = & \sum_{g=0}^{[N_{s1}/4-1]} \vert 2h_{g}\vert^2\,\xi_g \approx{\hbar V_F\over 2\gamma_d\Delta_0}  
\approx {[2\pi x_*\sqrt{x\pi}]\,t\over \gamma_d\Delta_0}\, a \, ,
\nonumber \\
\xi & \approx & {\gamma_c\over  4g}\,\xi_1
\, ; \hspace{0.35cm}
\xi\vert_{T=0} \approx {\gamma_c\over  4{\breve{g}}}\,\xi_1 \approx  {\gamma_c\over  4}{\hbar V_F\over 4k_B T_c} 
= {\gamma_c\over  4\gamma_d}\,{\hbar V_F\over {\breve{g}}2\Delta_0} 
\, ; \hspace{0.35cm} \xi^{min} = \xi\vert_{T=0,x=x_{op}} = \xi_1 \, .
\label{coherent-length}
\end{eqnarray}
In the expression of $\xi_1$ we included explicitly here the lattice spacing $a$ and
the factor $\gamma_c/4g$ appearing in the $\xi $ expression assures that its minimum magnitude $\xi^{min}$
equals $\xi_1$. It is achieved at zero temperature and $x=x_{op}$.
Furthermore, $\xi_{g}$ is the distance between the $c$ fermions of each pair 
contributing to the same local virtual-electron pair. For the present isotropic 
Fermi-velocity range the Fermi velocity $V_F$ equals approximately the
$c$ fermion velocity $V_{Fc} \approx \sqrt{x\pi}\,2/m_c^*$ of Eq. (\ref{pairing-en-v-Delta}). 
The quantity $\vert 2h_{g}\vert=\sqrt{2h^*_{g}2h_{g}}$ 
is the absolute value of the coefficients controlling the ratios
$\langle f_{\vec{r}_{j'''},c}^{\dag}\,f_{\vec{r}_{j''''},c}^{\dag}\rangle_g/\vert\langle f_{\vec{r}_{j'},c}^{\dag}\,f_{\vec{r}_{j''},c}^{\dag}\rangle_0\vert
\approx  [2h^*_g/\vert 2h_0\vert]$ of Eq. (\ref{ff}) of Appendix C. Those are associated with a $c$ fermion pair with
the same real-space coordinates $\vec{r}_{j'}$ and $\vec{r}_{j''}$
as a two-site bond of a given rotated-electron pair.
The relation between such $c$ fermion and rotated-electron pair
is given in Eq. (\ref{singlet-conf-simpl}).
The coefficients $2h_{g}$ obey the sum-rule provided in Eq. (\ref{ff}) of Appendix C.
They are twice those appearing in the expression of the annihilation operator
of the local $s1$ fermion of a local virtual-electron pair. Such an expression is obtained from Eq. (\ref{g-s1+general-gb}). 
The absolute value $\vert 2h_{g}\vert=\sqrt{2h^*_{g}2h_{g}}$ decreases upon increasing the magnitude of the 
length $\xi_{g}$ \cite{companion1b}. $\vert 2h_{g}\vert$ is largest at
$g=0$, which corresponds to $\xi_{0} = a_s=a/\sqrt{1-x}$. It falls rapidly upon
increasing $g$. Therefore, the average length $\xi_1$ is typically very small 
as compared to the penetration depth considered below in Section IV-E.

The overall amplitude $g=g_0\,g_1$
associated with the phases $\theta_{j}=\theta_{j,0}+\theta_{j,1}$
whose dome-like $x$ dependence is for zero temperature given in Eq. (\ref{g0-g1}) 
gives a measure of the strength of the $c$ fermion effective coupling
leading to the superconducting state coherent pairing. Such an effective coupling
also occurs when $g_1>0$ and $g_0=0$ yet for the corresponding pseudogap state
does not lead to coherent virtual-electron pairing. In either state
it is due to the energy provided to the $c$ fermions of such pairs
by the short-range spin correlations. This occurs through the
residual interactions of the two $c$ fermions with the $s1$
fermion within each virtual-electron pair. The energy lost by the spin-subsystem to supply
the $c$ fermions of a pair with the energy needed for their effective coupling 
weakens the short-range spin order. This occurs through the
enhancement of the fluctuations of the phases $\theta_{j,1}$
associated with the amplitude $g_1$. For the $c$ fermions $g_1$
refers to their effective pair coupling. Complementarily, for the
$s1$ fermions it is behind the weakening of the
short-range spin correlations upon increasing $x$.
Consistently, the value of $g_1$ provides a measure of
the ability and power of the short-range spin correlations 
supplying the $c$ fermions with the energy needed for 
the occurrence of virtual-electron pairing.

The physical picture that emerges is that the coherent-pair superconducting
order rather than competing with the short-range spin
order is a by-product of it. It coexists with
the latter order for the hole concentration range $x\in (x_c,x_*)$ at zero
temperature and a smaller $x$ range centered at $x=x_{op}$ 
for finite temperatures below $T_c$. Moreover, the type of $d$-wave long-range superconducting
order considered here {\it cannot occur} without the
simultaneous occurrence of short-range spin correlations.
As the limit $0<(x_*-x)\ll 1$ is reached at zero temperature,
the short-range spin correlations use up all their 
energy. This is consistent with the behaviors ${\breve{g}}_1\rightarrow 0$,
${\breve{g}}={\breve{g}}_0\,{\breve{g}}_1={\breve{g}}_1\rightarrow 0$,
$2\vert\Delta\vert\vert_{T=0}\rightarrow 0$, $2\vert\Omega\vert\vert_{T=0}\rightarrow 0$,
and $2\vert\Omega\vert\vert_{T=0}/2\vert\Delta\vert\vert_{T=0}/\rightarrow 1$ occurring in that limit.
Indeed, $\gamma_d\rightarrow 1$ as ${\breve{g}}\rightarrow 0$ for $0<(x_*-x)\ll 1$ when the physics
is controlled by strong phase fluctuations.
The order parameter of the long-range superconducting order vanishes 
upon the vanishing of that of the short-range spin order. Such
parameters vanish upon the simultaneous disappearance of the
corresponding orders.

Except in the limits $0<(x-x_c)\ll 1$ and $0<(x_*-x)\ll 1$ when the phase fluctuations
are very strong, for $x\in (x_c,x_*)$ the suppression effects lessen both the critical temperature $T_c$ of Eq. (\ref{Om-Dp-Tc}) 
and superconducting energy scale $2\vert\Omega\vert\vert$ of Eq. (\ref{Omega-Tc-di}) and enhance 
the coherent length $\xi$ of Eq. (\ref{coherent-length}), relative to their $\alpha_d=0$ 
magnitudes. In turn, the $c$ fermion energy dispersion $\epsilon_{c} (\vec{q}^{\,h})$ 
given in Eqs. (\ref{c-band}) and (\ref{bands}) of Appendix A, the form of the $s1$ fermion energy
dispersion $\epsilon_{s1} ({\vec{q}})$ given below, and the magnitudes of the pseudogap temperature 
$T^*$ of Eq. (\ref{Dp-T*}) and the critical hole concentration $x_*$ of Eq. (\ref{x-c}) remain unaltered. 
Although here we consider intrinsic disorder, this is consistent with the experimental
results of Refs. \cite{Yamamoto,Xu}, according to which the disorder induced by 
Zn substitution does not lead to a clear change in the magnitude of the pseudogap
temperature $T^*$. 

\subsubsection{The ratios controlling the effects weak 3D uniaxial anisotropy and electronic correlations}

In the present limit of very weak 3D uniaxial anisotropy its effects occur mainly through the dependence 
of the critical hole concentration $x_c$ on the parameter $\varepsilon^2$.
It follows that the amplitudes ${\breve{g}}$ and ${\breve{g}}_0$ of Eq. (\ref{g0-g1}), whose
expressions involve $x_c$, also depend on the latter parameter. For approximately $U/4t>u_0$
the critical hole concentration deviation $(x_c - x_0)$ and the critical hole concentration $x_*$ 
are proportional to two important ratios,
\begin{equation}
(x_c - x_0) \approx {\rm Gi} = {G\over\varepsilon^2} =
G\,{M\over m_c^*} \, ; \hspace{0.35cm}
G = {1\over 8}\left[{T_c^{max}(K)\,\lambda_{ab}^2 (\AA)\vert_{x=x_{op},T=0}\over (1.964 \times 10^8)\,\xi^{min} (\AA)}\right]^2 
 \, ; \hspace{0.35cm}
x_* = {2r_s\over \pi} = {2\over \pi}\,{\Delta_0\over 4W_{s1}^0} \, ; \hspace{0.35cm}
U/4t > u_0 \, .
\label{x-c}
\end{equation}
That $(x_c-x_0)$ equals approximately the Ginzburg number Gi is a result obtained from the study of the effective action of the 
phases $\theta_j$. In the $G$ expression provided here $\lambda_{ab}$ denotes the in-plane penetration depth introduced below
in Section IV-E, $T_c^{max}$ is the maximum critical temperature of Eq. (\ref{max-Tc}) reached at $x=x_{op}$, and
$\xi^{min} $ is the corresponding $x=x_{op}$ minimum magnitude of the coherent length given in Eq. (\ref{coherent-length}). 
The ratio $\varepsilon^2 =m_c^*/M$ appearing in the $(x_c - x_0)$ expression of Eq. (\ref{x-c}) controls the effects of weak 3D uniaxial anisotropy. 
Here the $U/4t$ dependent mass $m^{*}_{c}$ is that of the $c$ fermion energy dispersion 
of Eq. (\ref{bands}) of Appendix A and $M\gg m^{*}_{c}$ is the effective mass of Eq.
(\ref{M-t}) associated with electron hopping between the planes.

The critical hole concentration $x_*$ expression given in Eq. (\ref{x-c}) is derived in Ref. \cite{companion2}.
The spin ratio $r_s$ appearing in that expression and the $x_*$ limiting values are provided in Eqs. (\ref{m*c/mc-UL}) 
and (\ref{xc-range}) of Appendix A, respectively. For the smaller range of intermediate interaction values 
$U/4t \in (u_0, u_1)$ one approximately has that $r_c=m^{\infty}_c/m^*_c\approx 2r_s$.
Hence for such a range the upper critical hole concentration can be expressed as 
$x_*\approx r_c/\pi$, as found in Ref. \cite{companion2}. The ratios $r_s=\Delta_0/4W_{s1}^0$
and $r_c=m^{\infty}_c/m^*_c$ control the effects of electronic correlations. Here $m^{\infty}_c=\lim_{U/4t\rightarrow\infty}m^{*}_{c}=1/2t$
and $\Delta_0$ and $4W_{s1}^0$ are the energy scales of Eq. (\ref{Delta-0-gen}). 

Our scheme is valid for very small values of the 3D uniaxial anisotropy mass ratio $m_c^*/M$.
That the relation $(x_c -x_0)\approx {\rm Gi}$ holds is fully consistent with the related results of Ref. \cite{duality}.
The studies of Ref. \cite{cuprates0} reveal that a critical hole concentration $x_c\approx 0.05$
corresponds to $\varepsilon^2 =m_c^*/M\approx 10^{-4}$ for LSCO and
$\varepsilon^2 =m_c^*/M\approx 10^{-3}-10^{-2}$ for the other four systems.
This follows from $G$ being very small and given by
$G\approx 10^{-5}$ and $G\approx 10^{-4}$ for LSCO and the other
four systems, respectively. The smallness of $G$ allows that ${\rm Gi} \approx (x_c-x_0)\approx 10^{-2}$ is small
in spite of $1/\varepsilon^2 =M/m_c^*$ being large and
$(x_c-x_0)\propto 1/\varepsilon^2 =M/m_c^*$. Hence the value $x_c\approx 0.05$ is set by choosing an appropriate small 
magnitude for the 3D uniaxial anisotropy parameter $\varepsilon^2 =m_c^*/M$. In turn, the value $x_*\approx 0.27$ is set by
choosing $U/4t\approx 1.525$. Often one considers the range $U/4t\in (u_0,u_{\pi})$ 
for which according to Eq. (\ref{xc-range}) of Appendix A the $x_*$ values
belong to the domain $x_*\in (2e^{-1}/\pi,1/\pi)\approx (0.23,0.32)$. Some of our expressions refer to
the smaller range $U/4t\in (u_0, u_1)$ of intermediate interaction values for which 
$r_c\approx 2r_s\approx \pi\,x_*\approx 2e^{-4t\,u_0/U}$ and $x_*\in (0.23,0.28)$.

Our results reveal that for constant values of $U/4t$ the superconducting-dome hole-concentration
width $(x_*-x_c)$ decreases upon further decreasing the ratio mass  
$m^{*}_{c}/M$. That could mean that such a width vanishes 
in the 2D limit $m^{*}_{c}/M\rightarrow 0$ so that the ground state
of the Hubbard model on the square lattice is not superconducting.
However, note that our results do not apply to the 2D limit $\varepsilon^2 =m_c^*/M\rightarrow 0$. Then Gi 
becomes large, so that the expression $x_c\approx ({\rm Gi}+x_0)$
obtained from the effective action of the phases $\theta_j$ may not be valid. 

\subsubsection{The rate equations for suppression of $2\vert\Delta\vert$ and variation of $4k_B\,T_c$}

The dependence on the hole concentration $x$ of the zero-temperature short-range spin order 
parameter $2\vert\Delta\vert\vert_{T=0}$ of Eq. (\ref{2Delta}) and critical temperature $T_c$ 
provided in Eq. (\ref{Om-Dp-Tc}) is described by the two rate differential equations under suitable and physical
boundary conditions given in the following. Those are valid for vanishing spin density $m=0$ and finite
hole concentrations in the ranges $x\in (x_0,x_*)$ and $x\in (x_c,x_*)$, respectively. 

The rate equation for suppression of the order parameter
of the short-range spin correlations $2\vert\Delta\vert (x)\vert_{T=0}$
upon increasing the value of $x$ and its
boundary condition are for the approximate range $U/4t\in (u_0, u_{\pi})$ 
and $x\in (x_0,x_*)$ given by,
\begin{equation}
{\partial 2\vert\Delta\vert (x)\vert_{T=0}\over \partial x} = 
-\pi\,{\Delta_0\over r_s}\theta (x_* -x) 
\, ; \hspace{0.35cm} 2\vert\Delta\vert (x_0)\vert_{T=0}=\gamma_0\,2\Delta_0 
\, ; \hspace{0.35cm} \gamma_0=\left(1-{x_0\over x_*}\right) \, .
\label{rate-D}
\end{equation}
Here $\theta (z)$ is a theta function such that $\theta (z)=1$ for $z>0$ and $\theta (z)=0$ for
$z\leq 0$. In turn, the rate equation for variation of the superconducting energy scale $4k_B\,T_c$
upon increasing $x$ and the corresponding boundary condition read,
\begin{equation}
{\partial 4k_B\,T_c  (x)\over \partial x} = {(2\gamma_d-1)\over\gamma_c}{\pi\over r_s}\,
[2\vert\Delta\vert (x)-2\vert\Delta\vert (x_{op})]\vert_{T=0}\,\theta (x_*-x) \, ; \hspace{0.35cm}
4k_B\,T_c (x_c) = 0 \, .
\label{D-Tc-x}
\end{equation}

That the source term of the rate equation (\ref{D-Tc-x})
is given by $[\pi (2\gamma_d-1)/\gamma_c r_s][2\vert\Delta\vert (x)-2\vert\Delta\vert (x_{op})]\vert_{T=0}$ is
consistent with the superconducting order 
being a by-product of the short-range spin correlations. The source of the
energy provided by such spin correlations to sustain the superconducting energy $4k_B\,T_c$ 
is the pseudogap energy $2\vert\Delta\vert$ in that source term. 
Consistently, the latter energy is the order parameter
associated with short-range spin correlations. For hole concentrations
below and above the optimal hole concentration $x_{op}$ of Eq. (\ref{max-Tc}) 
the short-range spin correlations can be considered strong and weak, respectively. Indeed, for
$x\in (x_c,x_{op})$ and $x\in (x_{op},x_*)$ the above
source term is positive and negative, respectively. 

\subsection{Zero-temperature magnetic-field $c$ fermion diamagnetic orbital coupling and $s1$ fermion Zeeman coupling}

It is useful for the study of the pseudogap state to consider a uniform magnetic field 
$H$ aligned perpendicular to the square-lattice plane, to suppress 
superconductivity. The usual type II superconductor field $H_{c1}$
is very small for the VEP quantum liquid at intermediate
$U/4t\in (u_0, u_{\pi})$ values, so that here we consider that $H_{c1}\approx 0$. 
Our study refers to such a $U/4t$ range. 

It is well known that for the type-II superconductors the Pauli field $H_p$ for 
which the Zeeman splitting spin alignment starts to be profitable is given by
$H_p\approx k_B\,T_c/\mu_B g$. Here $\mu_B$ is the Bohr magneton and 
$g\approx 2$. For conventional type-II superconductors such a field is found 
to lie above the $H_{c2}$ line. In that case the Pauli pair breaking will become at zero temperature an important 
issue at high magnetic fields $H\gg H_p>H_{c2}$. In contrast, we find in the following 
that here $H_{c2}>H_p$. Important properties are then that:
\vspace{0.25cm}

{\bf The diamagnetic orbital coupling} of the magnetic field is to the charge $c$ fermions.
It occurs through the $c$ band hole momenta, $\vec{q}^{\,h}-e\vec{A}$, where $\vec{A}$
is the vector potential associated with the magnetic field $\vec{H}=\vec{\nabla}\times\vec{A}$. Hence diamagnetic pair
breaking refers here to the zero-momentum charge $c$ fermion pairing. 
\vspace{0.25cm}

{\bf The Zeeman coupling} of the magnetic field is to the spin-singlet two-spinon 
$s1$ fermions. Hence Pauli pair breaking refers here to the spin-singlet spinon
pairing of the $s1$ fermions. 
\vspace{0.25cm}

Since virtual-electron pairing involves both zero-momentum charge $c$ fermion pairing
and spin-singlet spinon pairing of the $s1$ fermions, diamagnetic orbital coupling and
Zeeman coupling are direct probes of the two types of pairing involved in the
virtual-electron pairing.

That the energy needed for coherent $c$ fermion pairing is supplied by the short-range 
spin correlations associated with $s1$ fermion spinon pairing implies an interplay between 
Pauli spinon pair breaking and diamagnetic $c$ fermion pair breaking. Below a critical field
$H_0$ the energy supplied to the $c$ fermions by the spin correlations associated 
with $s1$ fermion spinon pairing is at zero temperature enough to prevent that diamagnetic 
$c$ fermion pair breaking stops phase coherence. For $H<H_p\approx k_B\,T_c/\mu_B g$ 
there are no significant effects on the spin-singlet spinon pairing from Zeeman coupling
to the $s1$ fermions and only diamagnetic orbital coupling to the $c$ fermions plays
an active role in the physics. In turn, for $H>H_p$ the $s1$ fermion spinon pairing starts
to be affected by the Zeeman coupling. This lessens the energy supplied by the
spin sub-system to the $c$ fermion strong effective coupling. As a result, the effects
of diamagnetic $c$ fermion pair breaking become stronger for $H>H_p$ and the critical field
$H_0>H_p$ above which there is no $c$ fermion pairing phase coherence is at
zero temperature proportional to $H_p$. However, for $H\in (H_p,H_0)$ the diamagnetic 
orbital coupling to the $c$ fermions remains much stronger than the Zeeman coupling
to the $s1$ fermions. Within our scheme the critical field $H_0$
is that above which both the diamagnetic orbital coupling to the $c$ fermions and the 
Zeeman coupling to the $s1$ fermions play an active role. It is
approximately given by $H_0\approx 2H_p$.
Since $H_p\approx k_B\,T_c/\mu_B g$ where $g\approx 2$ one then finds that
at zero temperature $H_0 \approx k_B\,T_c/\mu_B$.        

Hence within our picture, at zero temperature the fluctuations of the phases $\theta_{j}$ remain small
and long-range superconducting order prevails provided that the uniform
magnetic field $H$ is below a critical field $H_0 \approx k_B\,T_c/\mu_B$. Above that
critical field, diamagnetic $c$ fermion pair breaking prevents phase coherence yet
there remain $c$ fermion pairing correlations. At zero temperature such a critical field expression 
approximately reads,
\begin{equation}
H_0 \approx k_B\,T_c/\mu_B \approx 
\gamma_d\,{(x-x_c)\over (x_*-x_c)}\left(1-{x\over x_*}\right){\Delta_0\over 2\mu_B} 
\, , \hspace{0.15cm} x\in (x_c,x_*) \, ; \hspace{0.35cm} 
H_0^{max} = \gamma_d^{min}\gamma_c\,{\Delta_0\over 8\mu_B} \, , \hspace{0.15cm} x = x_{op} \, .
\label{HH-x}
\end{equation}
Here the hole concentration $x_{op}$
is given in Eq. (\ref{x-op}). Interestingly, 
the expressions of the energy scale $\vert 2\Omega\vert\vert_{T=0}$
given in Eq. (\ref{Omega-Tc}), critical temperature $T_c$ 
in Eq. (\ref{Om-Dp-Tc}), and magnetic field $H_0$ in Eq. (\ref{HH-x})
have the dome-like $x$ dependence
observed in the five representative hole-doped cuprates 
\cite{two-gaps,k-r-spaces,duality,2D-MIT,Basov,ARPES-review,Tsuei,pseudogap-review}. 
Within our scheme such a type of $x$ dependence
follows directly from that of the overall zero-temperature amplitude ${\breve{g}}={\breve{g}}_0\,{\breve{g}}_1$
of Eq. (\ref{g0-g1}). The $x$ dependence of such an amplitude 
is fully controlled by the interplay of the $x$ dependences of the fluctuations of the phases
$\theta_{j,0}$ and $\theta_{j,1}$ such that  $\theta_{j}=\theta_{j,0}+\theta_{j,1}$,
as discussed above in Section III-D.

At zero temperature, magnetic field $H$ in the range $H\in (H_0,H_{c2})$, and
finite hole concentrations in a $H$-dependent range below a hole concentration
$x=x_{c2}$ the phases $\theta_{j,0}$ have large fluctuations. This refers
to the pseudogap state for which both the diamagnetic orbital coupling to the $c$ fermions and the 
Zeeman coupling to the $s1$ fermions play an active role. In it the phase-factor
averages $\langle e^{i\theta_{j,0}}\rangle$ and $\langle e^{i\theta_{cp}}\rangle$
and the amplitudes $g_0$ and $g$ vanish. However, the
fluctuations of the phases $\theta_{j,1}$ remain small. As a result, the amplitude 
$g_1=\vert\langle e^{i\theta_{j,1}}\rangle\vert$ remains finite 
and the short-range spin order prevails in the pseudogap state. The same applies to the
pseudogap state reached at zero temperature, magnetic field in 
the range $H\in (H_0,H^*)$, and a $H$-dependent hole concentration range above 
$x=x_{c2}$ and below $x=x_*$. Here $H_{c2}$ and $H^*$ are the fields above which the phases 
$\theta_{j,1}$ have large fluctuations, so that the amplitude $g_1=\vert\langle e^{i\theta_{j,1}}\rangle\vert$ 
vanishes and there is no short-range spin order. 

While the phase transition occurring
at $H=H_0$ results mostly from $c$ fermion diamagnetic pair breaking, the disappearance 
of short-range spin order occurring at the $H_{c2}$ and $H^*$ lines results both from
$c$ fermion diamagnetic and Zeeman spin-singlet spinon $s1$ fermion pairs breaking.
Consistently, at zero temperature the fields $H=H_{c2}$ for $x\in (x_0,x_{c2})$ and $H=H^*$ for $x\in (x_{c2},x_*)$ 
are those at which Zeeman spin-singlet spinon $s1$ fermion pair breaking fully suppresses the short-range
spin order and thus the diamagnetic $c$ fermion pair breaking fully suppresses the
corresponding incoherent $c$ fermion pairing correlations.

For hole concentrations in the range $x\in (x_0,x_{c1})$ we use the method of Ref.
\cite{duality} to derive the $x$ dependence of the magnitude $H_{c2}$ 
of the uniform magnetic field $H$ aligned perpendicular to the square-lattice plane. 
This involves expanding the quadratic terms of the continuum Lagrangian representing the charge sector
associated with the $c$ fermion pairs. The quadratic terms reflect the centre of mass
motion of such pairs. They are expanded in charge $-2e$ Landau levels resulting
from the diamagnetic orbital coupling of the magnetic field to the charge $c$ fermions.
In turn, we did not derive an expression valid for hole concentrations in the range $x\in (x_{c1},x_{c2})$.
For that $x$ range we could only access an inequality obeyed by $H_{c2}$ and the hole concentration $x_{c2}$,
respectively. In turn, the $x$ dependence of the upper field $H^{*}$ is a simpler problem. 
Alike $H_0 \approx k_B\,T_c/\mu_B$ is proportional to the superconducting critical temperature 
$T_c$, for $x\in (x_{c2},x_*)$ it is proportional to the pseudogap temperature $T^*$ of Eq. (\ref{Dp-T*}). 
For the interaction range $U/4t\in (u_0,u_{\pi})$ our results for the fields $H_{c2}$ and $H^*$ are,
\begin{eqnarray}
H_{c2} & \approx & {\cal{F}} 
= {(x-x_0)\over (x_*\gamma_0)}{\Delta_0\over 3\mu_B} 
\, , \hspace{0.15cm} x \in (x_0,x_{c1}) \, ; \hspace{0.35cm}
{{\breve{g}}_1 (x)\over {\breve{g}}_1 (x_{c1})}\,{\cal{F}} \leq H_{c2} \leq {\cal{F}} \, , \hspace{0.15cm}
x \in (x_{c1},x_{c2}) \, ,
\nonumber \\
H^* & \approx & {k_B\,T_*\over \mu_B}  \approx  {\breve{g}}_1{\Delta_0\over \mu_B} 
\, , \hspace{0.15cm}
x \in (x_{c2},x_*) \, .
\label{Hc2-H*}
\end{eqnarray}
Here the amplitude ${\breve{g}}_1={\breve{g}}_1 (x)$ is given in Eq. (\ref{g0-g1}).
We call $H^{*}_{c2}$ the field magnitude reached at the hole $x=x_{c2}$ at which 
$H_{c2}=H^{*}$. Hence $H^{*}_{c2}=H_{c2}(x_{c2})=H^{*}(x_{c2})$.
The inequalities obeyed by $H^{*}_{c2}$ and $x_{c2}$ are then given by,
\begin{equation}
0 \leq H^{*}_{c2} \leq {\gamma_0\,\Delta_0\over (3\gamma_0 +1) \mu_B}
\, ; \hspace{0.35cm}
x^{min}_{c2} \leq x_{c2} \leq x_*   \, ; \hspace{0.35cm}
x^{min}_{c2} = \left({2\gamma_0 +1\over 3\gamma_0 +1}\right)\,x_* \, ,
\label{H*c2-xc2}
\end{equation}
respectively. Here $H^{*}_{c2}=\gamma_0\Delta_0/[(3\gamma_0 +1)\mu_B]$ and $H^{*}_{c2}=0$
for $x=x^{min}_{c2}$ and $x=x^{max}_{c2}=x_*$, respectively, and $\gamma_0$
is the parameter given in Eq. (\ref{rate-D}). It is expected that $H^{*}_{c2}$ is closer
to $\approx \gamma_0\Delta_0/[(3\gamma_0 +1)\mu_B]$ than to zero,
so that $x_{c2}\approx x^{min}_{c2}$ and $H_{c2} \approx {\cal{F}}$
both for $x\in (x_0,x_{c1})$ and $x\in (x_{c1},x_{c2})$.

That the expression $H^{*} \approx {\breve{g}}_1[\Delta_0/\mu_B]$  
involves the zero-temperature and zero-field value of the amplitude 
$g_1=\vert\langle e^{i\theta_{j,1}}\rangle\vert$ is consistent
with it marking the full suppression of the short-range spin correlations.
Indeed, for $H>H^{*}$ and hole concentrations in the range $x\in (x_{c2},x_*)$
the system is driven into a disordered state without short-range spin order. 
In turn, for hole concentrations $x\in (x_0,x_{c1})$ the field $H_{c2}$
grows linearly with $x$. If such a behavior is a good approximation for
$x\in (x_{c1},x_{c2})$ as well, for $x_*= 0.27$ one finds 
$x_{c2}\approx x^{min}_{c2}\approx 0.20$ for the five representative systems.
For $x\in (x_1,x_{c2})$ the actual $H_{c2}$ $x$ dependence may slightly deviate 
to below the linear-$x$ behavior. If so, the hole concentration $x_{c2}\approx 0.20$ 
may increase to $\approx 0.21-0.22$. 

The pseudogap state occurring for fields below $H_{c2}$ and $H^{*}$ and above $H_0$
for a $H$-dependent hole-concentration range centered at $x=x_{c2}$ 
is a quantum vortex liquid. It shows strong vortex fluctuations and enhanced diamagnetism. 
Along the line $H=H_0$, this quantum vortex liquid freezes into a solid. Then any small deviation 
$-\delta H<0$ of the field $H=[H_0-\delta H]$ near $H_0$ leads to superconductivity. For magnetic fields 
above the field $H^{*}$ corresponding to the range $x\in (x_{c2},x_*)$ the system is driven into a disordered spin state.
It is similar to that reached at zero field and zero temperature for hole concentrations larger than $x_*$. 
Upon lowering $H$ from above $H^*$, the field $H^{*}$ marks the onset of the
short-range spin order. Therefore, it marks a crossover rather than a sharp transition. The same applies to 
the pseudogap temperature $T^*$ of Eq. (\ref{Dp-T*}) for vanishing magnetic field $H=0$ and finite hole 
concentrations in the range $x\in (x_0,x_*)$. Within a mean-field treatment, the upper field $H_{c2}$ 
refers to a crossover as well. For finite fields $H$ and vanishing temperature the true quantum transition 
to the superconducting phase takes place at the line $H_0=H_0 (x)$. Along it the quantum vortex liquid
freezes into a solid. Whether beyond mean-field theory the line associated with $H_{c2}(x)$ marks a sharp 
quantum phase transition and the onset of a new order for fields $H>H_{c2}$ and finite hole concentrations 
in the range $x\in (x_0,x_{c2})$ remains an open question. 

\section{The quantum-liquid energy functional, virtual-electron strong effective 
coupling pairing mechanism, and superfluid density}

Here we study the general energy functional associated with the VEP quantum-liquid
microscopic Hamiltonian considered in Section III-C. It refers to that Hamiltonian in normal order
relative to the $m=0$ initial ground state. The 2D energy functional considered here
accounts for the weak 3D uniaxial anisotropy effects through the dependence of
the critical hole concentration $x_c$ and pairing phases on the parameter
$\varepsilon^2 =m_c^*/M\ll 1$. 

Such a functional is obtained from that constructed in Ref. \cite{companion2} on introducing
in it the effects of the coherent virtual-electron pairing phase fluctuations, accounting for the 
small suppression effects. For $x\in (x_c,x_*)$ this refers to the occurrence in the VEP quantum 
liquid of a long-range superconducting order. That program involves the evaluation of the 
general superconducting virtual-electron pairing energy,
which emerges in the spectrum of such an energy functional. The derivation
of such a pairing energy requires the study of the virtual-electron strong effective 
coupling pairing mechanism. This involves as well investigations on the energy 
range of the $c$ - $s1$ fermion interactions behind it and signatures of
virtual-electron pairs in the one-electron removal spectral-weight
distribution. In addition, in this section we study the hole-concentration 
dependence of the zero-temperature superfluid density.

\subsection{The quantum-liquid energy functional}

The phase and amplitude of the complex gap function (\ref{D-jj}) appearing 
in the VEP quantum-liquid Hamiltonian terms (\ref{H-r-el}) show up in the superconducting order parameter $2\Omega$
of Eq. (\ref{Omega}). The absolute value of this order parameter is the maximum magnitude of
the superconducting virtual-electron pairing energy
 $2\vert\Omega_{s1} ({\vec{q}})\vert$
evaluated below in Section IV-B for $s1$ band momenta ${\vec{q}}\approx {\vec{q}}^{\,d}_{Bs1}$
at or near the $s1$ boundary line and in Section IV-F for momenta belonging to
the general coherent $s1-sc$ lines introduced in that section. (Here the index $sc$ stands for {\it strong coupling}.)
For temperatures below $T_c$ and a $T$ dependent $x$ range centered at $x=x_{op}$, which
is largest and given by $x\in (x_c,x_*)$ at $T=0$, the superconducting pairing energy per electron 
$\vert\Omega_{s1} ({\vec{q}})\vert$ derived in these sections emerges in the $s1$ fermion energy dispersion 
$\epsilon_{s1} (\vec{q})$. This dispersion appears in the general energy functional introduced in the 
following. That functional involves the ground-state normal-ordered $c$ and $s1$ fermion hole 
momentum distribution function deviations \cite{companion2},
\begin{equation}
\delta N^h_{c}({\vec{q}}^{\,h}_{j}) = [N^h_{c}({\vec{q}}^{\,h}_{j}) -
N^{h,0}_{c}({\vec{q}}^{\,h}_{j})] \, ;
\hspace{0.35cm}
\delta N^h_{s1}({\vec{q}}_{j}) = [N^h_{s1}({\vec{q}}_{j}) -
N^{h,0}_{s1}({\vec{q}}_{j})] \, .
\label{DNq}
\end{equation}
Here $N^{h,0}_{c}({\vec{q}}^{\,h}_{j})$ and $N^{h,0}_{s1}({\vec{q}}_{j})$
are the corresponding initial-ground-state values. According to the results of Ref. \cite{companion2}, 
the $c$ and $s1$ fermion discrete hole momenta ${\vec{q}}^{\,h}_{j}$ and 
momenta ${\vec{q}}_{j}$, respectively, are good quantum numbers for the $\varepsilon^2=0$, $\alpha_d=0$, and
$x_0=0$ square-lattice quantum liquid of Ref. \cite{companion2}. Under the very weak 3D 
uniaxial anisotropy effects and suppression effects they are for $x\in (x_c,x_*)$ close to 
good quantum numbers of the VEP quantum liquid. This is in 
contrast to the range $x\in (0,x_c)$, in which the strong hole tapping effects reported in Appendix B
change qualitatively the square-lattice quantum liquid physics of Ref. \cite{companion2}. 

It follows that alike for that quantum liquid, for the the VEP quantum liquid considered here 
the hole momentum distribution functions $N^h_{c}({\vec{q}}^{\,h}_{j})$ and $N^h_{s1}({\vec{q}}_{j})$ 
approximately read $1$ and $0$ for unfilled and filled, respectively, discrete-hole-momentum 
values ${\vec{q}}^{\,h}_{j}$ or discrete-momentum values ${\vec{q}}_{j}$. Furthermore, the first-order energy terms of
the energy functional given in the following correspond to the dominant contributions.
The $c-c$ fermion interactions vanish or are extremely small and can be ignored,
whereas the $s1-s1$ fermion interactions do not lead to inelastic scattering \cite{companion2,cuprates}. 
In turn, the effects of the $c-s1$ fermion interactions of interest for the
studies of this paper are implicitly accounted for in
the pairing energy involved in the $s1$ band energy dispersion $\epsilon_{s1} ({\vec{q}})$
appearing in the first-order terms of the energy functional. This justifies why its
second-order terms in the deviations (\ref{DNq}) are not studied here. An approximate
expression of the matrix element $\lim_{{\vec{p}}\rightarrow 0}W_{c,s1} ({\vec{q}}^{\,h},{\vec{q}};{\vec{p}})$ of the 
$c$ - $s1$ fermion effective interaction between the initial and final states needed 
for the derivation of the one-electron inverse lifetime and related to the second-order
forward-scattering $f_{c,s1} ({\vec{q}}^{\,h},{\vec{q}})$ function is for transfer momentum 
${\vec{p}}\rightarrow 0$ evaluated in Ref. \cite{cuprates}.

For the interaction range $U/4t\in (u_0, u_{\pi})$, hole concentrations 
$x\in (x_c,x_*)$, and vanishing spin density $m=0$ 
the VEP quantum liquid general energy functional is to first order 
in the deviations (\ref{DNq}) given by, 
\begin{equation} 
\delta E = -\sum_{\vec{q}^{\,h}}\epsilon_{c} (\vec{q}^{\,h})\delta N^h_{c}({\vec{q}}^{\,h}) 
- \sum_{{\vec{q}}} \epsilon_{s1} (\vec{q})\delta N^h_{s1}({\vec{q}}) \, .
\label{DeltaE-plus}
\end{equation}
The energy dispersion $\epsilon_{c} (\vec{q}^{\,h}_j)$ appearing
here is given in Eqs. (\ref{c-band}) and (\ref{bands}) of Appendix A. 
The dispersion $\epsilon_{s1} ({\vec{q}}_{j})$ reads,
\begin{equation} 
\epsilon_{s1} ({\vec{q}}) = -\sqrt{\vert\epsilon^0_{s1} ({\vec{q}})\vert^2
+ \vert\Delta_{v-el} ({\vec{q}})\vert^2} \, ;
\hspace{0.35cm}
\vert\Delta_{v-el} ({\vec{q}})\vert = \vert\Delta_{s1}({\vec{q}})\vert
+  \vert\Omega_{s1} ({\vec{q}})\vert \, .
\label{band-s1}
\end{equation}
Here the energy dispersion $\epsilon^0_{s1} ({\vec{q}})$ and pairing energy per spinon 
$\vert\Delta_{s1} ({\vec{q}})\vert$ are given in Eq. (\ref{bands-bipt}) of Appendix A, 
$\vert\Delta_{v-el} ({\vec{q}})\vert$ is the virtual-electron pairing energy per electron, 
and $\vert\Omega_{s1} ({\vec{q}})\vert$ is its coherent part introduced below in 
Sections IV-B and IV-F. The short-range spin order and its order parameter $2\vert\Delta\vert$ 
refer to the range $x\in (x_0,x_*)$. However, due to the effects of the Anderson insulator behavior
coexisting with that order for $x\in (x_0,x_c)$, the $s1$ fermion energy dispersion 
$\epsilon_{s1} ({\vec{q}})$ of Eq. (\ref{band-s1}) and the corresponding momentum-dependent
energies $\vert\Delta_{s1}({\vec{q}})\vert$ and $\vert\Omega_{s1} ({\vec{q}})\vert$ are
well defined only for the range $x\in (x_c,x_*)$.

As discussed in Section III-D, removal of two electrons in a spin-singlet configuration leads
to an excited state with no holes in the $s1$ band. Hence it follows from the form
of the energy functional (\ref{DeltaE-plus}) that no virtual-electron
pair is broken under that excitation. Instead, such a pair becomes the
removed electron pair. In turn, one-electron and spin excitations lead to the emergence
of one and two holes in the $s1$ band, respectively. Thus their energy spectrum
involves a finite pairing energy associated with virtual-electron pair breaking.

We recall that the virtual-electron pairs 
carry the same momentum ${\vec{q}}$ as the corresponding $s1$ fermion. Indeed, their two $c$ fermions have hole 
momentum ${\vec{q}}^{\,h}$ and $-{\vec{q}}^{\,h}$, respectively. Therefore, the $c$ fermion pair net momentum
contribution vanishes. This justifies why the energy scale $\vert\Delta_{v-el} ({\vec{q}})\vert$ 
given in Eq. (\ref{band-s1}) appears in the ${\vec{q}}$
dependent energy dispersion $\epsilon_{s1} ({\vec{q}}) = -\sqrt{\vert\epsilon^0_{s1} ({\vec{q}})\vert^2
+ \vert\Delta_{v-el} ({\vec{q}})\vert^2}$. Indeed, that it and thus
$\vert\Omega_{s1} ({\vec{q}})\vert$ appear in such a dispersion follows from the virtual-electron pair
associated with these pairing energies per electron carrying momentum ${\vec{q}}$.
As discussed in Section III-D, such virtual-electron pairs of momentum $\vec{q}$ exist as individual objects only in 
intermediate virtual states of one-electron and spin excitations. Nevertheless, the virtual-electron pair energy 
is related to the spectrum of one-electron excitations and has physical significance, as confirmed below
in Section IV-D. It reads,
\begin{equation}
E_{v-el} ({\vec{q}}) = 2\vert\epsilon_c (\pm\vec{q}^{\,h})\vert + \vert\epsilon_{s1} ({\vec{q}})\vert \, .
\label{E-v-el}
\end{equation}
Note that such an energy is not in general a pairing energy and thus is different from the virtual-electron pairing energy per 
electron $\vert\Delta_{v-el} ({\vec{q}})\vert$ of Eq. (\ref{band-s1}). It is found below that for 
momenta $\vec{q}\approx {\vec{q}}^{\,d}_{Bs1}$ at or near the $s1$ boundary line coherent pairing
is associated with $c$ fermions of hole momenta ${\vec{q}}^{\,h}$ and $-{\vec{q}}^{\,h}$ near
the $c$ Fermi line. In this case $E_{v-el} ({\vec{q}}^{\,d}_{Bs1}) 
\approx \vert\Delta_{v-el} ({\vec{q}}^{\,d}_{Bs1})\vert$, so that the virtual-electron pair energy
becomes the virtual-electron pairing energy per electron. 

At zero temperature the energy $\vert\Omega_{s1} ({\vec{q}})\vert$ is finite for 
$s1$ band momenta ${\vec{q}}$ corresponding to $s1$ fermions that mediate 
coherent $c$ fermion pairing. These momenta belong to the 
range ${\vec{q}} \in Q^{s1}_{cp}$ defined in Section IV-F. It refers to a set of
coherent $s1-sc$ lines centered at zero $s1$ band momentum.
Moreover, $s1$ fermions whose momentum belongs to a given 
coherent $s1-sc$ line only interact with $c$ fermions whose hole momenta belong to 
a related nearly circular coherent $c-sc$ line centered at the $c$ band momentum $-\vec{\pi}=-[\pi,\pi]$.
There is a one-to-one correspondence between such $s1-sc$ lines and $c-sc$ lines.

The non-coherent $\vert\Delta_{s1}({\vec{q}})\vert$ and coherent $\vert\Omega_{s1} ({\vec{q}})\vert$
parts of the virtual-electron pairing energy $\vert\Delta_{v-el} ({\vec{q}})\vert$ of Eq.
(\ref{band-s1}) are associated with $s1$ fermion spin-singlet spinon pairing
and coherent zero-momentum $c$ fermion pairing, respectively. That they appear in the 
spin $s1$ fermion spectrum and their maximum magnitudes $\vert\Delta\vert$ 
and $\vert\Omega\vert=\gamma_d\,g_0\,\vert\Delta\vert$ are closely related  
and as confirmed in Ref. \cite{cuprates0} correspond to the superconducting- and
normal-state maximum one-electron gap $\vert\Delta\vert$ and superconducting-state
low-temperature magnetic resonance energy $2\vert\Omega\vert$, respectively, 
is consistent with the interplay between magnetic fluctuations and unconventional 
superconductivity discussed in Ref. \cite{Yu-09}.

The corresponding general excitation momentum functional
is linear in the deviations (\ref{DNq}) and reads,
\begin{equation}
\delta {\vec{P}} = \delta {\vec{q}}_{c}^{\,0}
-\sum_{\vec{q}^{\,h}}[\vec{q}^{\,h}-\vec{\pi}]\,\delta N^h_{c}({\vec{q}}^{\,h}) -
\sum_{\vec{q}}
{\vec{q}}\,\delta N^h_{s1}({\vec{q}}) \, .
\label{DP}
\end{equation}
Here $\delta {\vec{q}}_{c}^{\,0}$ is the subspace-dependent small momentum deviation 
considered in Ref. \cite{companion2}.

Except for the superconducting pairing energy
 per electron $\vert\Omega_{s1} ({\vec{q}})\vert$
appearing in the expression of the dispersion $\epsilon_{s1} ({\vec{q}})$ of Eq. (\ref{band-s1}),
the first-order terms of the energy functional (\ref{DeltaE-plus}) are those derived in Ref.
\cite{companion2} for approximately the range $U/4t> u_0\approx 1.302$. Indeed and 
as found in that reference, for $U/4t<u_0$ the energy scale $2\Delta_0$ whose limiting 
behaviors are given in Eq. (\ref{Delta-0}) of Appendix A becomes very small. It follows that the amplitude fluctuations of the order 
parameter of the short-range spin correlations cannot be ignored. As discussed in Ref.
\cite{companion2}, for small $U/t$ values the $s1$ fermion energy dispersion
$\epsilon_{s1} ({\vec{q}}_{j})$ is then not expected to have the form given in Eq. (\ref{band-s1}).

\subsection{$d$-wave virtual-electron pairing mechanism near the Fermi line}

Here we evaluate the superconducting pairing energy
 per electron $\vert\Omega_{s1} ({\vec{q}})\vert$
in the $s1$ band energy spectrum of Eq. (\ref{band-s1}) for 
momenta ${\vec{q}}\approx {\vec{q}}^{\,d}_{Bs1}$ at or near the $s1$ boundary line.
As given in Eq. (\ref{D-x}) of Appendix A, the order parameter $2\vert\Delta\vert$ 
associated with the short-range spin correlations is the maximum magnitude of the
$s1$ fermion spinon-pairing energy $2\vert\Delta_{s1}({\vec{q}})\vert$. Similarly, the
absolute value $2\vert\Omega\vert\vert_{T=0}$ of the order parameter 
associated with phase-coherent-pair superconducting order 
is the maximum magnitude of the superconducting virtual-electron pairing energy
. The maximum magnitude of 
the $s1$ fermion spinon-pairing energy $2\vert\Delta_{s1}({\vec{q}})\vert$
corresponds according to Eq. (\ref{D-x}) of Appendix A
to $s1$ band momentum values ${\vec{q}}={\vec{q}}^{\,d\,AN}_{Bs1}$.
Those belong to the $s1$ boundary line and their corresponding auxiliary
momentum values ${\vec{q}}_0={\vec{q}}^{\,AN}_{Bs1}$
given in Eq. (\ref{qFc-qBs1}) of that Appendix point
in the anti-nodal directions. (The equality ${\vec{q}}^{\,d\,AN}_{Bs1}={\vec{q}}^{\,AN}_{Bs1}$ 
holds for ground states at vanishing spin density $m=0$ and their two-electron excited states. In turn,
one has that ${\vec{q}}^{\,d\,AN}_{Bs1}\neq {\vec{q}}^{\,AN}_{Bs1}$
for their one-electron excited states \cite{companion2}.)

The VEP quantum-liquid $d$-wave long-range superconducting order occurring at zero temperature
for the $x$ range $x\in (x_c,x_*)$ is a by-product of the short-range spin correlations. 
For small $0<(x-x_c)\ll 1$ this shows up in the relation of the zero temperature 
energy scale $\vert\Omega_{s1} ({\vec{q}}^{\,d}_{Bs1})\vert$ to the  $s1$ fermion pairing 
energy per spinon associated with such short-range correlations. Indeed for $0<(x-x_c)\ll 1$
the energy scale $\vert\Omega_{s1} ({\vec{q}})\vert$
is the deviation of the $s1$ fermion pairing 
energy per spinon $\vert\Delta_{s1}({\vec{q}}^{\,d}_{Bs1}-{\vec{q}}_{{\breve{g}}_0})\vert$
from its value $\vert\Delta_{s1}({\vec{q}}^{\,d}_{Bs1})\vert$
at ${\vec{q}}_{{\breve{g}}_0}=0$. Here ${\vec{q}}_{{\breve{g}}_0}\equiv [{\breve{g}}_0/\sqrt{2}]\,{\vec{e}}_{\phi^{\Delta}_{s1}}$.
(The physical meaning of the vector ${\vec{q}}_{{\breve{g}}_0}$ is discussed below.) Expanding
$\vert\Delta_{s1}({\vec{q}}^{\,d}_{Bs1}-{\vec{q}}_{{\breve{g}}_0})\vert$ gives,
\begin{equation}
\vert\Delta_{s1}({\vec{q}}^{\,d}_{Bs1}-{\vec{q}}_{{\breve{g}}_0})\vert
\approx \vert\Delta_{s1}({\vec{q}}^{\,d}_{Bs1})\vert
+  {\vec{V}}^{\Delta}_{s1} ({\vec{q}}^{\,d}_{Bs1})\cdot {\vec{q}}_{{\breve{g}}_0} = \vert\Delta_{s1}({\vec{q}}^{\,d}_{Bs1})\vert
+ \vert\Omega_{s1} ({\vec{q}}^{\,d}_{Bs1})\vert 
\, , \hspace{0.15cm} T = 0 \, .
\label{2D-2Omega}
\end{equation}
Hence for $0<(x-x_c)\ll 1$ the zero-temperature superconducting virtual-electron pairing energy
 per electron reads,
\begin{equation}
\vert\Omega_{s1} ({\vec{q}}^{\,d}_{Bs1})\vert =
{\vec{V}}^{\Delta}_{s1} ({\vec{q}}^{\,d}_{Bs1})\cdot {\vec{q}}_{{\breve{g}}_0} 
= {\breve{g}}_0\,\sqrt{2}\,V^{\Delta}_{Bs1}  = 
\vert\Omega\vert\vert_{T=0}\vert\sin 2\phi\vert \, ,
\hspace{0.15cm} T = 0 \, .
\label{Omega-c}
\end{equation}
Here the $s1$ fermion velocity ${\vec{V}}^{\Delta}_{s1} ({\vec{q}}^{\,d}_{Bs1})$
is given in Eq. (\ref{g-velocities-def}) of Appendix A, $V^{\Delta}_{Bs1}$ denotes its absolute
value of Eq. (\ref{pairing-en-v-Delta}), and $\vert\Omega\vert\vert_{T=0}$ is the 
superconducting energy scale of Eq. (\ref{Omega-Tc}). 

The expansion of Eq. (\ref{2D-2Omega}) refers to small 
$0<(x-x_c)\ll 1$ and thus to small absolute values ${\breve{g}}_0\,[1/\sqrt{2}]=[(x-x_c)/(x_*-x_c)]\,[1/\sqrt{2}]$
of the vector ${\vec{q}}_{{\breve{g}}_0}= [{\breve{g}}_0/\sqrt{2}]\,{\vec{e}}_{\phi^{\Delta}_{s1}}$. 
As mentioned above, that for such a range of $x$ values near and above the critical hole concentration $x_c$ the coherent virtual-electron
pairing energy per electron $\vert\Omega_{s1} ({\vec{q}}^{\,d}_{Bs1})\vert$
associated with the phase-coherent-pair superconducting order is the energy 
deviation ${\vec{V}}^{\Delta}_{s1} ({\vec{q}}_{Bs1})\cdot {\vec{q}}_{{\breve{g}}_0}$
of the expansion (\ref{2D-2Omega}) is consistent with such an order being a by-product of the
short-range spin correlations. Indeed, the latter are associated with the $s1$ fermion
pairing energy per spinon $\vert\Delta_{s1}({\vec{q}}^{\,d}_{Bs1})\vert$,
which corresponds to the zeroth-order term of that expansion. 

For larger hole concentration values in the range $x\in (x_c,x_*)$ the expansion of Eq. (\ref{2D-2Omega})
is not valid anymore. However, within our scheme the expression for the zero-temperature coherent 
virtual-electron pairing energy per electron $\vert\Omega_{s1} ({\vec{q}}^{\,d}_{Bs1})\vert\approx
\gamma_d\,{\breve{g}}_0\,\sqrt{2}\,V^{\Delta}_{Bs1}$ of Eq. (\ref{Omega-c}) multiplied by
the suppression coefficient $\gamma_d$ of Eq. (\ref{gamma-d}) is. Indeed, that equation refers to 
the limit $0<(x-x_c)\ll 1$ for which $\gamma_d=1$.
For virtual-electron pairs of momentum ${\vec{q}}\approx {\vec{q}}^{\,d}_{Bs1}$
the full virtual-electron pairing energy per electron has the following form,
\begin{equation}
\vert\Delta_{v-el} ({\vec{q}}^{\,d}_{Bs1})\vert =
\vert\Delta_{s1}({\vec{q}}^{\,d}_{Bs1})\vert
+  \vert\Omega_{s1} ({\vec{q}}^{\,d}_{Bs1})\vert 
\approx \vert\Delta\vert\vert\cos 2\phi\vert + \vert\Omega\vert\vert\sin 2\phi\vert \, .
\label{full-v-el-en}
\end{equation}
For small $0<(x-x_c)\ll 1$ this energy reads
$\vert\Delta_{s1}({\vec{q}}^{\,d}_{Bs1}-{\vec{q}}_{{\breve{g}}_0})\vert$,
as given in Eq. (\ref{2D-2Omega}). The form of the pairing energy (\ref{full-v-el-en})
is consistent with virtual-electron pairing having a $d$-wave character and
involving both spin-singlet spinon $s1$ fermion pairing and
$c$ fermion pairing. For the normal-state ground states at finite magnetic field $H$
considered in Section III-E the virtual-electron pairs exist yet loose phase coherence. Then the
virtual-electron pairing energy per electron equals for virtual-electron
pairs of momentum ${\vec{q}}\approx {\vec{q}}^{\,d}_{Bs1}$
the spin-singlet spinon $s1$ fermion pairing energy for the same momentum,
$\vert\Delta_{v-el} ({\vec{q}}^{\,d}_{Bs1})\vert = \vert\Delta_{s1}({\vec{q}}^{\,d}_{Bs1})\vert$.

It follows from the above analysis that for hole concentrations in the range $x\in (x_c,x_*)$ the zero-temperature 
absolute value $2\vert\Omega\vert\vert_{T=0}$ defined by Eqs. (\ref{Omega-Tc}) and (\ref{Om-Dp-Tc}) 
of the order parameter introduced in Eq. (\ref{Omega})
associated with the phase-coherent-pair superconducting order reads
$2\vert\Omega\vert\vert_{T=0}=2\vert\Omega_{s1} ({\vec{q}}^{\,d,N}_{Bs1})\vert$.
It corresponds to the maximum magnitude of the coherent virtual-electron pairing
energy. Such a maximum magnitude refers to virtual-electron pairs of momentum at the $s1$ boundary line 
and whose auxiliary momentum points in the nodal directions.
We recall that the maximum magnitude of the 
zero-temperature spin-singlet spinon $s1$ fermion pairing energy
${\rm max}\,2\vert\Delta_{s1}({\vec{q}}^{\,d}_{Bs1})\vert=
2\vert\Delta_{s1}({\vec{q}}^{\,d,AN}_{Bs1})\vert=2\vert\Delta\vert$ corresponds to
momenta belonging to the $s1$ boundary
line and whose auxiliary
momenta ${\vec{q}}^{\,AN}_{Bs1}$ point instead
in the anti-nodal directions. 

For hole concentrations in the range $x\in (x_c,x_*)$, momenta ${\vec{q}}\approx {\vec{q}}^{\,d}_{Bs1}$ 
at or near the $s1$ boundary line, and temperatures $T$ smaller than $T_c$ 
the coherent part $\vert\Omega_{s1} ({\vec{q}})\vert$ of the pairing energy per electron reads,
\begin{equation}
\vert\Omega_{s1} ({\vec{q}})\vert = \vert\Omega_{s1} (-{\vec{q}})\vert
\approx \gamma_d\,{\vec{V}}^{\Delta}_{s1} ({\vec{q}})\cdot {\vec{q}}_{g_0} = 
\gamma_d\,g_0\sqrt{2}\,V^{\Delta}_{s1} ({\vec{q}}) \, ; \hspace{0.35cm}
{\vec{q}}_{g_0}  = g_0\,{\vec{e}}_{\phi^{\Delta}_{s1}} 
\, ; \hspace{0.35cm} {\vec{q}} \approx {\vec{q}}^{\,d}_{Bs1} \, .
\label{q-g-0}
\end{equation}
It involves the general amplitude $g_0$ rather than its
$T=0$ magnitude ${\breve{g}}_0=(x-x_c)/(x_*-x_c)$ of Eq. (\ref{g0-g1}). 

The $c$ - $s1$ fermion residual interactions are behind the 
effective pairing coupling of the two $c$ fermions of hole momentum ${\vec{q}}^{\,h}$
and $-{\vec{q}}^{\,h}$ such that ${\vec{q}}^{\,h} \approx {\vec{q}}^{\,h\,d}_{Fc}$ is at or near
the $c$ Fermi line. Such $c$ fermion pairs
are associated with the pairing energy per electron $\vert\Omega_{s1} ({\vec{q}})\vert\approx
\gamma_d\,{\vec{V}}^{\Delta}_{s1} ({\vec{q}})\cdot {\vec{q}}_{g_0}$ of phase-coherent virtual-electron 
pair configurations of momentum ${\vec{q}} \approx {\vec{q}}^{\,d}_{Bs1}$ at or near the $s1$ boundary line.  
The short-range spin correlations provide the energy $\vert\Omega_{s1} ({\vec{q}})\vert$ 
through the interactions of the $s1$ fermion of momentum 
${\vec{q}} \approx {\vec{q}}^{\,d}_{Bs1}$, energy  $\epsilon_{s1} ({\vec{q}}_{Bs1})\approx
-\vert\Delta_{s1} ({\vec{q}}_{Bs1})\vert$, and velocity ${\vec{V}}^{\Delta}_{s1} (\vec{q}) \approx
-{\vec{\nabla}}_{\vec{q}}\vert\Delta_{s1} ({\vec{q}})\vert$ with the two $c$ fermions. 

The vector ${\vec{q}}_{g_0}$ plays the role of coupling constant of the $c$ fermion-pair degrees 
of freedom to the $s1$ fermion of velocity ${\vec{V}}^{\Delta}_{s1} ({\vec{q}})$. It is  through it
that the pairing energy $\vert\Omega_{s1} ({\vec{q}})\vert\approx
\gamma_d\,{\vec{V}}^{\Delta}_{s1} ({\vec{q}})\cdot {\vec{q}}_{{\breve{g}}_0}$
needed for the $c$ fermion effective coupling and corresponding 
coherent virtual-electron pairing is supplied to the two $c$ fermions. The absolute value $g_0/\sqrt{2}$
of the vector ${\vec{q}}_{g_0}$ involves the amplitude $g_0=g/g_1$ of Eq. (\ref{amplitudes-g0-g1}). 
It measures the relative strength of the coherent virtual-electron
pairing and $s1$ fermion spinon pairing. Indeed, the 
magnitudes of the amplitudes $g$ and $g_1$ also given in Eq. (\ref{amplitudes-g0-g1})
provide a measure of the strengths of the coherent virtual-electron pairing and 
$s1$ fermion spinon pairing, respectively.  

As confirmed in Section IV-D, breaking under one-electron excitations of
the virtual-electron pairs considered here refers to removal or addition of
electrons from and to the Fermi line. Therefore, virtual-electrons of $s1$
band momentum at or near the $s1$ boundary line are within the electron
representation at or near the Fermi line, respectively. Consistently,
the superconducting fluctuations lead to an additional contribution 
$\vert\Omega_{s1} ({\vec{q}}^{\,d}_{Bs1})\vert\approx \vert\Omega\vert\vert\sin 2\phi\vert$
to the anisotropic term $\delta E_F =\vert\Delta_{s1} ({\vec{q}}^{\,d}_{Bs1})\vert$ of the Fermi energy  
introduced in Ref. \cite{companion2}, which appears in the spectra given in Eq. (\ref{DE-3proc}).
Hence for the present VEP quantum liquid the Fermi energy reads,
\begin{eqnarray}
E_F & = & \mu +\delta E_F (\phi) \, ; \hspace{0.35cm} 
\mu\approx {\breve{\mu}}^0+W_c^h \, ,
\nonumber \\
\delta E_F (\phi) & = & -\epsilon_{s1} ({\vec{q}}^{\,d}_{Bs1}) = \vert\Delta_{v-el} ({\vec{q}}^{\,d}_{Bs1})\vert =
\vert\Delta_{s1} ({\vec{q}}^{\,d}_{Bs1})\vert+\vert\Omega_{s1} ({\vec{q}}^{\,d}_{Bs1})\vert \approx
\vert\Delta\vert\vert\cos 2\phi\vert +\vert\Omega\vert\vert\sin 2\phi\vert \, ,
\label{EF}
\end{eqnarray}
where the superconducting energy scale $\vert\Omega\vert$ is provided in Eq. (\ref{Omega-gen}) and
$W_c^h$ is the ground-state $c$ fermion-hole energy bandwidth given in Eq. (\ref{bands}) of Appendix A.
For intermediate $U/4t$ values approximately in the range $U/4t\in (u_0, u_1)$ the $W_c^h$ expression 
provided in that equation is a very good approximation for hole concentrations $x\in (x_c,x_{c1})$ and a 
reasonably good approximation for the $x$ range $x\in (x_{c1},x_*)$. 
Rather than the square-lattice quantum-liquid expression $\mu\approx \mu_0 +W_c^h$, the expression 
$\mu\approx {\breve{\mu}}^0+W_c^h$ given here for the chemical potential refers to the VEP quantum liquid.
In it ${\breve{\mu}}^0=\mu^0+\delta \mu$ where $\mu^0 =\lim_{x\rightarrow 0}\mu$ is one half the Mott-Hubbard
gap of Eq. (\ref{DMH}) of Appendix A and the shift $\delta \mu$ generated by the hole trapping effects
is given in Eq. (\ref{mu-x-AL}) of Appendix B. 

As discussed in Ref. \cite{companion2}, in the Fermi energy expression $E_F = \mu +\delta E_F (\phi)$
the chemical potential $\mu$ arises from the isotropic $c$ fermion energy dispersion and 
$\delta E_F =\vert\epsilon_{s1} ({\vec{q}}^{\,d}_{Bs1})\vert=\vert\Delta_{v-el} ({\vec{q}}^{\,d}_{Bs1})\vert$ 
stems from the anisotropic $s1$ fermion energy dispersion.
The extra term $\vert\Omega_{s1} ({\vec{q}}^{\,d}_{Bs1})\vert\approx \vert\Omega\vert\vert\sin 2\phi\vert$ 
appearing for $x\in (x_c,x_*)$ in the expression of the anisotropic Fermi energy term $\delta E_F$ given in Eq. (\ref{EF}) does 
not change the physics discussed in Section III-A. In turn, for $x\in (x_0,x_c)$ and $T=0$ the equality 
$\delta E_F=\vert\Delta_{v-el} ({\vec{q}}^{\,d}_{Bs1})\vert =\vert\Delta_{s1} ({\vec{q}}^{\,d}_{Bs1})\vert$ holds. 
The anisotropic Fermi energy term $\delta E_F$ plays an important role in the VEP quantum-liquid physics. 
Its maximum magnitude ${\rm max}\,\{\delta E_F\}= \vert\Delta\vert$ determines and equals that of the 
anti-nodal one-electron gap $\delta E_F^{AN} = {\rm max}\,\{\delta E_F\}= \vert\Delta\vert$. From the
use of the critical-temperature expressions given in Eq. (\ref{Om-Dp-Tc}) and (\ref{max-Tc}) and
$\delta E_F (\phi)$ expression provided in Eq. (\ref{EF}), $\delta E_F^{AN}$ can be expressed 
as a function of $T_c/T^{max}_c$ as,
\begin{equation}
\delta E_F^{AN} = {\rm max}\,\{\delta E_F\} = \vert\Delta\vert (x) =
\vert\Delta\vert (x_{op})(1-{\rm sgn}\,\{x-x_{op}\}\vert\sqrt{1-T_c (x)/T^{max}_c}\vert) 
\, ; \hspace{0.35cm} \vert\Delta\vert (x_{op}) = \gamma_c\,{\Delta_0\over 2} \, .
\label{Delta-1-kBTc}
\end{equation}

\subsection{The general virtual-electron strong effective coupling pairing mechanism}

According to our approach, $c$ fermion strong effective coupling
is that whose breaking upon one-electron excitations leads to sharp
features in the one-electron spectral function. It is a necessary but
not sufficient condition for the occurrence of coherent pairing in the initial ground state. 
The absolute value $q^h=\vert{\vec{q}}^{\,h}\vert$ of the
hole momenta ${\vec{q}}^{\,h}$ and $-{\vec{q}}^{\,h}$ of two $c$ fermions
with strong effective coupling belongs to a well-defined range $q^h\in (q_{Fc}^h,q_{ec}^h)$.
Here $q_{ec}^h$ is the maximum hole momentum given below for which there is
$c$ fermion strong effective coupling. We denote by $Q^{c}_{ec}$ and $Q^{s1}_{ec}$
the corresponding $c$ and $s1$ band momentum domains for which the $c$ and $s1$ fermions, 
respectively, participate in virtual-electron pairs with strong effective coupling.
$Q^{c}_{ec}$ and $Q^{s1}_{ec}$ correspond to a set of $c-sc$ and $s1-sc$ lines,
respectively.
As further discussed below, within the present scheme:
\vspace{0.25cm}

i) Only strong effective coupling, {\it i.e.} that whose breaking upon one-electron 
excitations leads to sharp features in the one-electron spectral function, can lead 
to coherent virtual-electron pairing.
\vspace{0.25cm}

ii) For each $c$ band hole momentum absolute value $q^h$ in a well-defined 
range $q^h\in (q_{Fc}^h,q_{ec}^h)$ there is a $c$ band approximately circular $c-sc$ line 
of radius $q^h$ centered at $-\vec{\pi}$. Such a range refers to the $c$ fermions of hole momenta 
$\vec{q}^{\,h}$ and $-\vec{q}^{\,h}$ whose effective coupling is strong. 
There is no one-to-one correspondence between the $c$ fermion pairs and
the $s1$ fermions of strongly coupled virtual-electron pairs. 
As justified in this section, the strong effective coupling of $c$ fermions of hole momenta 
$\vec{q}^{\,h}$ and $-\vec{q}^{\,h}$ at or near a $c-sc$ line of 
radius $q^h$ results from interactions with $s1$ fermions of momentum 
$\vec{q}$ at or near a corresponding $s1-sc$ line centered at the $s1$ band zero momentum.
The four nodal momenta belonging to the latter line have the same absolute value $q_{arc}^N =q_{arc}^N (q^h)$, which is 
uniquely determined by the radius magnitude $q^h$ of the corresponding $c-sc$ line. Here $q_{arc}^N\in (q_{ec}^N,q_{Bs1}^N)$ 
where $q_{arc}^N =q_{ec}^N$ and $q_{arc}^N =q_{Bs1}^N$ correspond to $q^h= q_{ec}^h$ 
and $q^h= q_{Fc}^h$, respectively. For $q_{arc}^N<q_{Bs1}^N$ and thus $q^h> q_{Fc}^h$ 
the $s1-sc$ line is constituted by four disconnected line arcs centered at
zero momentum and crossing perpendicularly the four nodal lines, respectively, 
whose auxiliary momenta refer to Fermi angles $\phi=\pi/4, 3\pi/4, 5\pi/4, 7\pi/4$. The 
angular width of these four line arcs is a decreasing function of $q^h$, which vanishes at
$q^h= q_{ec}^h$. As found below, in that limit these four line arcs reduce
to four nodal momenta whose absolute value is $q_{ec}^N$. The
function $q_{arc}^N =q_{arc}^N (q^h)$ and the magnitudes of $q_{ec}^h$
and $q_{ec}^N$ are given below. That the number of $s1$ fermions that
mediate the strong effective coupling of $c$ fermion pairs of hole momenta 
$\vec{q}^{\,h}$ and $-\vec{q}^{\,h}$ decreases upon increasing $q^h$,
vanishing for $q^h> q_{ec}^h$, is consistent with for the range $q^h\in (q_{Fc}^h,q_{ec}^h)$
the strength of such effective coupling being a decreasing function of the $c$ fermion
absolute energy, $\vert\epsilon_c (\vec{q}^{\,h})\vert=([q^h]^2-[q^h_{Fc}]^2)/2m_c^*$.
\vspace{0.25cm}

In turn, weak effective pairing coupling between $c$ fermions
is that whose breaking upon one-electron excitations 
leads to broad flat features. Weak effective pairing
coupling never contributes to coherent virtual-electron pairing. Strong
effective pairing coupling may or may not lead to such a coherent
pairing. 

In the following we justify and supplement the information reported in point (ii).
The ground-state {\it $ec$-pairing line} is the $c-sc$ line of largest radius $q^h= q_{ec}^h$. It
separates in the $c$ momentum band the $c$ fermions
with strong effective pairing coupling from those with only weak effective pairing coupling, respectively. 
(The designation $ec$-pairing line follows from it separating $c$ fermions with
two different types of effective coupling [$ec$].) The hole momentum domain $Q^{c}_{ec}$ 
refers to the set of $c-sc$ lines. It is limited by the $c$ Fermi line and $ec$-pairing line, respectively. 
As confirmed below in Section IV-E, the energy scale,
\begin{equation}
W_{ec} = {[q^h_{ec}]^2-[q^h_{Fc}]^2\over 2m_c^*} = {4\Delta_0\over \gamma_c} \, ,
\label{W-ec}
\end{equation}
is the maximum magnitude of both the energy bandwidth of the superconducting-ground-state 
sea of $c$ fermions contributing to coherent pairing and the energy bandwidth 
corresponding to the hole momentum domain $Q^{c}_{ec}$ considered here.

Alike the smaller $c-sc$ lines enclosed by it, for the interaction range $U/4t\in (u_0,u_1)$
the $ec$-pairing line is nearly circular for hole concentrations $x\in (x_c,x_{c1})$.
It is a fairly good approximation to consider that it remains circular for $x\in (x_{c1},x_*)$. The energy range 
$\vert\epsilon_c (\vec{q}^{\,h})\vert=\vert\epsilon_c (-\vec{q}^{\,h})\vert \in (0,W_{ec})$
of $c$ fermions having strong effective coupling in the ground state 
corresponds to the uniquely defined hole momentum range $q^h\in (q_{Fc}^h,q_{ec}^h)$
reported in (ii). For these $c$ fermions, the hole momenta ${\vec{q}}^{\,h}_{ec}$ belonging to the $ec$-pairing line 
are those of largest absolute value $q^h_{ec}$. 
For the interaction range $U/4t\in (u_0, u_1)$ 
and $x\in (x_c,x_*)$ it is approximately given by,
\begin{equation}
q^{h}_{ec} = \sqrt{1+{W_{ec}\over [q_{Fc}^h]^2r_c t}}\,q_{Fc}^h 
= \sqrt{1+{\Delta_0\over x (x_*-x_c)\pi^2 t}}\,q_{Fc}^h \, .
\label{q-h-Fcp}
\end{equation}

An interesting property is that for $x\in (x_c,x_*)$ the $c$ band
hole momentum area corresponding to $c$ fermions with strong effective coupling
in the ground state is independent of $x$ and given by,
\begin{equation}
S_{ec}^c = \pi ([q^{h}_{ec}]^2-[q_{Fc}^h]^2) = {4\Delta_0\over (x_*-x_c)\,t} \, .
\label{S-ec-c}
\end{equation}
It refers to the hole momentum area of the domain $Q_{ec}^c$. Its expression
involves only the superconducting dome hole concentration width $(x_*-x_c)$
and the basic energy scales $\Delta_0$ and $t$. 

In Appendix D it is found that for $U/4t\in (u_0, u_1)$ 
and hole concentrations in the range $x\in (x_c,x_*)$
the energy bandwidth $\vert\epsilon_c (\vec{q}^{\,h})\vert$ 
of $c$ fermions of momenta $\vec{q}^{\,h}$
and $-\vec{q}^{\,h}$ having strong effective pairing
coupling due to residual interactions with $s1$
fermions of momentum ${\vec{q}}\in Q^{s1}_{ec}$
obeys the inequality,
\begin{equation}
\vert\epsilon_c (\vec{q}^{\,h})\vert  
\leq W_{ec}\left(1-{\vert\Delta_{s1} ({\vec{q}})\vert\over\vert\Delta\vert}\right) 
\approx W_{ec}(1-\vert\cos 2\phi\vert) \, ; \hspace{0.35cm}
\vert\Delta_{s1} ({\vec{q}})\vert \approx \vert\Delta\vert\vert\cos 2\phi\vert
\hspace{0.15cm}{\rm for}\hspace{0.15cm} {\vec{q}}\in Q^{s1}_{ec} \, .
\label{range-2-ec}
\end{equation}
For $x\in (x_c,x_*)$ the inequality $W_{ec}<W_c^p$ holds. 
(Here $W_c^p$ is the energy bandwidth of the ground-state $c$ momentum band
filled by $c$ fermions.) Hence only $s1$ fermions whose auxiliary momenta point in the nodal
directions have interactions with strongly coupled $c$ fermions whose 
energy $\vert\epsilon_c (\vec{q}^{\,h})\vert$ belongs to the whole range
$\vert\epsilon_c (\vec{q}^{\,h})\vert\in (0,W_{ec})$.

The inequality (\ref{range-2-ec}) is obeyed by the following hole momentum range,
\begin{equation}
q^h \in (q_{Fc}^h, q^h_q) \, ; \hspace{0.35cm}
q^h_q = \alpha_q\,q_{Fc}^h \, ; \hspace{0.35cm}
\alpha_q= \sqrt{1 + (\alpha_{ec}^2-1)\left(1-{\vert\Delta_{s1} ({\vec{q}})\vert\over\vert\Delta\vert}\right)}
\approx \sqrt{1 + (\alpha_{ec}^2-1)\left(1-\vert\cos 2\phi\vert\right)} \, ,
\label{alpha-q}
\end{equation}
Equivalent to that inequality are the following inequalities obeyed by the magnitude $\vert\Delta_{s1} ({\vec{q}})\vert$ 
of the $s1$ fermion pairing energy per spinon and Fermi angle $\phi$ defined in Eq. (\ref{phi-F}),
\begin{equation}
\vert\Delta_{s1} ({\vec{q}})\vert \leq \vert\Delta\vert
\left(1-{\vert\epsilon_c (\vec{q}^{\,h})\vert\over W_{ec}}\right)
\, ; \hspace{0.35cm} \vert\cos 2\phi\vert \leq \left(1-{\vert\epsilon_c (\vec{q}^{\,h})\vert\over W_{ec}}\right) \, ,
\label{range-s1-q}
\end{equation}
respectively. The second inequality given here involving the Fermi angle $\phi$ is 
in turn equivalent to restricting the range $\phi\in (0,\pi/2)$ to an arc of angular 
width $2\phi_{arc}$ centered at $\phi =\pi/4$, 
\begin{equation}
\phi \in (\pi/4 -\phi_{arc},\pi/4+\phi_{arc}) \, ; \hspace{0.35cm}
\phi_{arc} = {1\over 2}\arcsin\left({[q^h_{ec}]^2-[q^h]^2
\over [q^h_{ec}]^2-[q^h_{Fc}]^2}\right) =
{1\over 2}\arcsin\left({\epsilon^{ec}_c (\vec{q}^{\,h})\over W_{ec}}\right) 
\in (0,\pi/4) \, .
\label{phi-arc}
\end{equation}
Here $\epsilon_c^{ec} (\vec{q}^{\,h})$ is the following $c$ fermion energy dispersion, 
\begin{equation}
\epsilon_c^{ec} (\vec{q}^{\,h}) = W_{ec}+\epsilon_c (\vec{q}^{\,h})
\, ; \hspace{0.35cm}
0 \leq \epsilon_c^{ec} (\vec{q}^{\,h}) \leq W_{ec} \, .
\label{Fc-q-h-ec}
\end{equation}
It is the energy of a $c$ fermion of hole 
momentum $\vec{q}^{\,h}$ measured from the energy level of 
the $ec$-pairing line. The energy (\ref{Fc-q-h-ec}) is largest at 
the $c$ Fermi line and vanishes at the $ec$-pairing line,
\begin{equation}
\epsilon_c^{ec} ({\vec{q}}^{\,h}_{ec}) = 0
\, ; \hspace{0.35cm} \epsilon_c^{ec} ({\vec{q}}_{Fc}^{\,h\,d}) = W_{ec} \, .
\label{Fc-q-h-limits-ec}
\end{equation}

The shape of the $ec$-pairing line is for $x\in (x_c,x_*)$ fully determined by the 
form of the $c$ band energy dispersion $\epsilon_c^{ec} (\vec{q}^{\,h})$ as follows,
\begin{equation}
{\vec{q}}^{\,h}_{ec} \in ec-{\rm pairing} 
\hspace{0.10cm} {\rm line}
\hspace{0.05cm} \Longleftrightarrow  \hspace{0.05cm}
\epsilon_c^{ec} ({\vec{q}}^{\,h}_{ec}) = 0 \, .
\label{g-Ul}
\end{equation}

For simplicity we limit our analysis to the quadrant referring to the Fermi angle range
$\phi\in (0,\pi/2)$ and corresponding $s1-sc$ line arc whose angular range is that
given in Eq. (\ref{phi-arc}), $\phi \in (\pi/4 -\phi_{arc},\pi/4+\phi_{arc})$. To each 
$c$ band approximately circular $c-sc$ line of radius $q^h$ centered at $-\vec{\pi}$ 
corresponds such a $s1$ band $s1-sc$ line arc whose momenta 
${\vec{q}}^{\,d}_{arc}$ are approximately given by,
\begin{equation}
{\vec{q}}^{\,d}_{arc} (\phi) = A^d_{s1} (\phi)\,{\vec{q}}_{arc} (\phi) 
\, ; \hspace{0.30cm} {\vec{q}}_{arc} (\phi) \approx {q_{arc}^N\over q^N_{Bs1}}\,{\vec{q}}_{Bs1} (\phi)
\, ; \hspace{0.30cm} q_{arc} (\phi) = q_{arc} (\pi/2-\phi) \, ; \hspace{0.25cm}
\phi \in (\pi/4 -\phi_{arc},\pi/4+\phi_{arc}) \, .
\label{s1-sc-line-mom}
\end{equation}
Here $q_{arc} (\phi)\approx [q_{arc}^N/q^N_{Bs1}]\,q_{Bs1} (\phi)$ is the absolute value of the 
auxiliary momentum ${\vec{q}}_{arc} (\phi)$, $A^d_{s1}$ is the matrix of Eq. (\ref{qs1-qhc}), 
and the $s1-sc$ line arc nodal momentum absolute value $q_{arc}^N = q_{arc}^N (q^h)$
is given below. As stated in (ii), for each $c$ band $c-sc$ line of radius $q^h$ in the 
range $q^h\in (q_{Fc}^h,q_{ec}^h)$ there is for $\phi\in (0,\pi/2)$ exactly one of such $s1$ band $s1-sc$ line arcs.
The line arc limiting angular widths $2\phi_{arc}=0$ and $2\phi_{arc} =\pi/2$ refer to $q^h=q_{ec}^h$
and $q^h =q_{Fc}^h$, respectively. The angle range $\phi\in (\pi/4-\phi_{arc},\pi/4+\phi_{arc})$ of 
each $s1-sc$ line arc runs symmetrical around $\phi=\pi/4$. 
The physical importance of the $s1-sc$ line arcs follows from the
corresponding sharp spectral features line arcs studied below, which emerge in the one-electron removal
weight distribution and are observed in experiments on the cuprates \cite{cuprates}. 

We note that for $\phi \in (0,\pi/2)$ the $s1$ band auxiliary momenta belong to the quadrant for which their
two components are negative \cite{companion2}. For instance, for momenta near
the $s1$ boundary line and $x\in (x_{c1},x_{c2})$ this follows from the relation of the
$s1$ band momentum angle $\phi_{s1}=\phi + \pi$ of Eq. (\ref{phiF-c-s1}) to the Fermi angle $\phi$. 
Indeed the range $\phi_{s1}\in (\pi,3\pi/2)$ refers to $\phi \in (0,\pi/2)$. Consistently with Eq. (\ref{s1-sc-line-mom}), 
the $s1$ boundary line is for $\phi \in (0,\pi/2)$ the $s1-sc$ line arc whose nodal momentum absolute value is given by
$q_{arc}^N = q^N_{Bs1}$. It corresponds to the approximately circular $c$ Fermi line whose 
radius reads $q^h =q_{Fc}^h$: One confirms below that $q_{arc}^N = q^N_{Bs1}$ for $q^h =q_{Fc}^h$.
Hence the auxiliary momentum angle $\phi_{s1}=\phi + \pi$ can be generalized to
all $s1-sc$ line arcs. Each of such line arcs then refers to an auxiliary momentum 
angle range $\phi_{s1}\in (5\pi/4 -\phi_{arc},5\pi/4 +\phi_{arc})$ corresponding to 
$\phi\in (\pi/4 -\phi_{arc},\pi/4 +\phi_{arc})$.

As mentioned above, there is no one-to-one correspondence between the $c$ fermion pairs and
corresponding $s1$ fermions participating in virtual-electron pairs with strong effective coupling. 
Concerning general $c-s1$ fermion interactions, there are in average two $c$ fermions for each
$s1$ fermion. However, only part of these $c$ fermions and $s1$ fermions participate in 
strongly coupled virtual-electron pairs. There are four classes of strongly coupled virtual-electron
pairs associated with the four disconnected $s1-sc$ line arcs, respectively. For any strongly 
coupled virtual-electron pair configuration whose auxiliary momentum Cartesian components are
$[q_{arc}\cos \phi_{s1},q_{arc}\sin \phi_{s1}]$ where $\phi_{s1}=\phi +\pi$ there are in the other 
three disconnected $s1-sc$ line arcs three strongly coupled virtual-electron pair configurations, respectively, 
with the same energy and momentum absolute value $q_{arc}$.
The $s1$ band auxiliary momenta of the four strongly coupled virtual electron pair configurations 
under consideration have Cartesian components $\pm [q_{arc}\cos \phi_{s1},q_{arc}\sin \phi_{s1}]$ and 
$\pm [-q_{arc}\sin \phi_{s1},q_{arc}\cos \phi_{s1}]$. The set of four $s1-sc$ line arcs 
that such four momenta belong to are
associated with four sub-domains $Q_{ec}^{\pm,+,s1}$ and $Q_{ec}^{\pm,-,s1}$, respectively,
of the $s1$ band momentum domain $Q_{ec}^{s1}$
of strongly coupled virtual-electron pair configurations. 

The point is that the strongly coupled virtual electron pair configurations 
whose auxiliary momentum has Cartesian components $\pm [q_{arc}\cos \phi_{s1},q_{arc}\sin \phi_{s1}]$ couple
to different $c$ fermion pairs than those of Cartesian components
$\pm [-q_{arc}\sin \phi_{s1},q_{arc}\cos \phi_{s1}]$. Hence one can divide the 
$c$ band domain $Q_{ec}^c$ into two sub-domains that in average have half
of the $c$ fermion pairs each. We call them $Q_{ec}^{+c}$ and $Q_{ec}^{-c}$,
respectively. The strong effective coupling of $c$ fermions of hole momenta belonging  
to $Q_{ec}^{+c}$ and $Q_{ec}^{-c}$ results from interactions with $s1$ fermions
whose momenta belong to $Q_{ec}^{\pm,+,s1}$ and $Q_{ec}^{\pm,-,s1}$, respectively.
Furthermore, symmetry arguments concerning the
relationship of the two strongly coupled virtual-electron pairs with the
same energy, momentum absolute value, and momentum direction but different
senses suggest that the number of $s1$ fermions of each of the two
$s1$ band sub-domains $Q_{ec}^{\pm,+,s1}$ (and $Q_{ec}^{\pm,-,s1}$) is one half 
that of $c$ fermions in the $c$ band domain $Q_{ec}^{+c}$ (and $Q_{ec}^{-c}$). 
Therefore, for each of the four sub-domains $Q_{ec}^{\pm,+,s1}$ and $Q_{ec}^{\pm,-,s1}$
the $s1$ band momentum area $S_{ec}^{s1}$ corresponding to $s1$ fermions of virtual-electron
pairs with strong effective coupling in the ground state is for $x\in (x_c,x_*)$ one fourth that of Eq. (\ref{S-ec-c}),
\begin{equation}
S_{ec}^{s1} = {\Delta_0\over (x_*-x_c)\,t} \, .
\label{S-ec-s1}
\end{equation}
It follows that the momentum area of the whole domain $Q_{ec}^{s1}$ equals that of
$Q_{ec}^{c}$ and reads $4S_{ec}^{s1} = 4\Delta_0/(x_*-x_c)\,t$.
It refers to the four sets of concentric $s1-sc$ line arcs centered at $\vec{0}$. The present results are valid for 
$U/4t\in (u_0,u_1)$. 

The simultaneous validity of the equalities (\ref{S-ec-c})
for the whole $c$ band domain $Q_{ec}^c$ and (\ref{S-ec-s1}) for each of
the four sub-domains of $Q_{ec}^{s1}$ requires the hole concentration to obey
the inequality $x<x_{ec}^{max}=1-[2\Delta_0]/[\pi^2 (x_*-x_c)\,t]$. Hence $x_*$ must be
smaller than or equal to $x_{ec}^{max}$. This requires that
$\Delta_0/t<[\pi^2/2](1-x_*)(x_*-x_c)$. For $x_c=0.5$ and $x_*=0.27$ this
gives approximately $\Delta_0/t<0.79$. The fulfillment of this inequality assures that 
for $x\in (x_c,x_*)$ the total numbers of $c$ fermions and $s1$ fermions are larger than
those of $c$ fermions and $s1$ fermions, respectively, participating in
strongly coupled virtual-electron pairs. For $U/4t=1.525$ and the $t$ magnitude
$t\approx 195$\,meV found in Ref. \cite{cuprates0} to be appropriate to the four representative hole doped cuprates 
other than LSCO for which 
$x_c=0.5$ and $x_*=0.27$ one finds $\Delta_0\approx 84$\,meV, so that
$\Delta_0/t\approx 0.28$. In turn, the LSCO cation-randomness effects considered
in that section lessen the $\Delta_0$ magnitude to $\Delta_0\approx 42$\,meV,
so that $\Delta_0/t\approx 0.14$ for that random alloy. 

That in average the number of $s1$ fermions whose momenta ${\vec{q}}$ belong to
the domain $Q_{ec}^{s1}$ is larger than that of $c$ fermion pairs of hole momenta
${\vec{q}}^{\,h}$ and $-{\vec{q}}^{\,h}$ belonging to $Q_{ec}^c$ is consistent
with the dependences on $q^h$ of the $s1-sc$ and $c-sc$ lines length. The
$c-sc$ line length $\approx 2\pi q^h$ increases upon increasing $q^h$. In contrast, 
that of the $s1-sc$ line,
\begin{equation}
L_{s1} (q_{arc}^N) = 4\,l_{s1} (q_{arc}^N) \, ; \hspace{0.35cm}
l_{s1} (q_{arc}^N) \approx 2(1-\sin 2\phi_{arc})\sqrt{[q^{AN}_{Bs1}]^2-[q^{N}_{Bs1}]^2} 
=  2{(q^N_{Bs1}-q_{arc}^N)\over (q^N_{Bs1}-q^N_{ec})}\sqrt{[q^{AN}_{Bs1}]^2-[q^{N}_{Bs1}]^2} \, ,
\label{L-s1-lines}
\end{equation}
decreases. Here $l_{s1} (q_{arc}^N)$ is the approximate length of each of the four $s1-sc$ line arcs,
$q^{N}_{ec}$ is the minimum absolute value of the nodal $s1-sc$-line-arc momenta
${\vec{q}}^{\,d\,N}_{s1} \equiv {\vec{q}}^{\,d}_{arc} (\pi/4)$ and their auxiliary momenta
${\vec{q}}^{\,N}_{s1} \equiv {\vec{q}}_{arc} (\pi/4)$ 
of Eq. (\ref{s1-sc-line-mom}), and $q^{N}_{Bs1}$ and $q^{AN}_{Bs1}$
are the absolute values of Eq. (\ref{qN-qAN}) of Appendix E of the nodal and 
anti-nodal $s1$ boundary-line momenta ${\vec{q}}^{\,d\,N}_{Bs1}$ and ${\vec{q}}^{\,d\,AN}_{Bs1}$, 
respectively. According to the definition of Ref. \cite{companion2},
their corresponding auxiliary momenta  ${\vec{q}}^{N}_{Bs1}$ and ${\vec{q}}^{AN}_{Bs1}$,
respectively, have for the quadrant for which $\phi \in (0,\pi/2)$ so that $\phi_{s1}\in (\pi,3\pi/2)$ and 
thus $q_{0x_1}\leq 0$ and $q_{0x_2}\leq 0$ 
the Cartesian components given in Eq. (\ref{q-A-q-AN-c-s1}) of Appendix A.
The $x$ dependence of $q^N_{Bs1}$ and $q^{AN}_{Bs1}$ is discussed in Appendix E.

Hence the number of $s1$ fermions
available to supply the energy needed for $c$ fermion strong effective
pairing coupling decreases upon increasing the $c-sc$ line radius $q^h$. For $q^h$ equal or close to
$q^h_{Fc}$ there are plenty $s1$ fermions to supply the energy 
needed for the $c$ fermion strong effective coupling. In the opposite limit of
$q^h$ tending to $q^h_{ec}$ the $c-sc$ line length $2\pi q^h_{ec}$ is
maximum. However, the length $L_{s1} (q_{arc}^N)$ given in Eq. (\ref{L-s1-lines})
tends to zero, as each $s1-sc$ line arc becomes a single discrete momentum
value. Therefore, then only two sets of two $s1$ fermions whose auxiliary momenta
have Cartesian components $\pm [q^N_{ec}/\sqrt{2},q^N_{ec}/\sqrt{2}]$ and 
$\pm [-q^N_{ec}/\sqrt{2},q^N_{ec}/\sqrt{2}]$ are available to supply the energy 
needed for the ground-state strong effective coupling of
$c$ fermions of hole momenta belonging to $Q_{ec}^{+c}$ and $Q_{ec}^{-c}$, respectively. 
Consistently, there is no strong effective  pairing coupling for $q^h>q^h_{ec}$ 
and $q_{arc}^N<q^N_{ec}$. 

The momentum area $S_{ec}^{s1}$ of Eq. (\ref{S-ec-s1}) is approximately given by,
\begin{equation}
S_{ec}^{s1} = \int_{q^N_{ec}}^{q^N_{Bs1}}dq_{arc}^N\,l_{s1} (q_{arc}^N) \approx 
(q^N_{Bs1}-q^N_{ec})\sqrt{[q^{AN}_{Bs1}]^2-[q^{N}_{Bs1}]^2} \, .
\label{S-ec-c-lines}
\end{equation}
Here $l_{s1} (q_{arc}^N)$ is the approximate length of each $s1-sc$ line arc given in 
Eq. (\ref{L-s1-lines}). From the use of the two $S_{ec}^{s1}$ expressions provided
in Eqs. (\ref{S-ec-s1}) and (\ref{S-ec-c-lines}), respectively, one straightforwardly arrives to,
\begin{equation}
q^N_{ec} \approx  q^N_{Bs1} - {\Delta_0\over (x_*-x_c)\sqrt{[q^{AN}_{Bs1}]^2-[q^{N}_{Bs1}]^2}\,t} \, .
\label{q-N-ec-coherent}
\end{equation}
On combining this expression with the auxiliary $s1-sc$-line momentum expression of
Eq. (\ref{s1-sc-line-mom}) one confirms that,
\begin{eqnarray}
{\vec{q}}_{arc} (\phi) & = & {\vec{q}}_{Bs1} (\phi)\hspace{0.35cm}
{\rm for}\hspace{0.15cm} q_{arc}^N = q^N_{Bs1} \, , \hspace{0.35cm}
\phi \in (0,\pi/2) \, ,
\nonumber \\
& = & {\vec{q}}^{\,N}_{ec} (\phi)\hspace{0.35cm}
{\rm for}\hspace{0.15cm} q_{arc}^N = q^N_{ec} \, , \hspace{0.35cm}
\phi = \pi/4 \, .
\label{qq-qq}
\end{eqnarray}
Here ${\vec{q}}^{\,N}_{ec}$ is the auxiliary momentum of the
$s1-sc$-line momentum ${\vec{q}}^{\,d\,N}_{ec}=A^d_{s1}\,{\vec{q}}^{\,N}_{ec}$
of smallest absolute value (\ref{q-N-ec-coherent}).

The absolute value $q_{arc}^N\in (q^{N}_{ec},q^{N}_{Bs1})$ of the nodal momentum belonging to 
a given $s1-sc$ line arc is obtained by replacing  in expression (\ref{q-N-ec-coherent}) 
the energy scale $\Delta_0$ by the energy spectrum $\vert\Delta_{ec} ({\vec{q}}^{\,h})\vert$ defined in Eq.
(\ref{epsilon-c-W-c}) of Appendix D. As given in Eq. (\ref{Deltacp-limits})
of that Appendix, its magnitude varies from $\vert\Delta_{ec} ({\vec{q}})\vert = 0$
for ${\vec{q}}^{\,h}={\vec{q}}_{Fc}^{\,h}$ to $\vert\Delta_{ec} ({\vec{q}}^{\,h})\vert = \Delta_0$
for ${\vec{q}}^{\,h}={\vec{q}}^{\,h}_{ec}$. This is consistent with $c$ fermions
with strong effective coupling and hole momenta at or near the $c$ Fermi line interacting
with $s1$ fermions of momenta at or near the $s1$ boundary line. In turn, 
$c$ fermions with strong effective coupling and hole momenta at or near the $ec$ pairing line interact
with the $s1$ fermion whose nodal momentum has absolute value $q_{ec}^N$
given in Eq. (\ref{q-N-ec-coherent}). In the limit $q_{arc}^N\rightarrow q^N_{ec}$ the $s1-sc$ line arc of minimum absolute
nodal momentum $q^N_{ec}$ has vanishing length and for $\phi \in (0,\pi/2)$ and thus
$\phi_{s1} \in (\pi,3\pi/2)$ reduces to a single discrete nodal momentum 
${\vec{q}}^{\,d\,N}_{ec}=A^d_{s1}\,{\vec{q}}^{\,N}_{ec}$. 

Replacement in expression (\ref{q-N-ec-coherent}) of the energy scale $\Delta_0$ by the energy 
spectrum $\vert\Delta_{ec} ({\vec{q}}^{\,h})\vert$ then leads to,
\begin{equation}
q_{arc}^N \approx q^N_{Bs1} - {\vert\Delta_{ec} ({\vec{q}}^{\,h})\vert\over (x_*-x_c)\sqrt{[q^{AN}_{Bs1}]^2-[q^{N}_{Bs1}]^2}\,t}
= q^N_{Bs1} -{[q^h]^2-[q^h_{Fc}]^2\over 4\sqrt{[q^{AN}_{Bs1}]^2-[q^{N}_{Bs1}]^2}}\,\pi
\, ; \hspace{0.35cm} 
q^h \in (q_{Fc}^h,q_{ec}^h) \, .
\label{q-N-coherent}
\end{equation}
These expressions are equivalent to,
\begin{equation}
q_{arc}^N \approx q^N_{Bs1} - (1-\sin 2\phi_{arc})[q^N_{Bs1}-q^N_{ec}] \, .
\label{q-N-coherent-angles}
\end{equation}
The angle $\phi_{arc}$ of Eq. (\ref{phi-arc}) can be expressed in terms of the nodal momentum $q_{arc}^N$
of Eq. (\ref{q-N-coherent}) as follows,
\begin{equation}
\phi_{arc} = {1\over 2}\arcsin\left({q_{arc}^N -q^N_{ec}\over q^N_{Bs1}-q^N_{ec}}\right)\in (0,\pi/4) \, .
\label{phi-arc-qN}
\end{equation}
As given in Eq. (\ref{q-N-coherent-angles}), the absolute value $q_{arc}^N$ provided in Eq. (\ref{q-N-coherent}) of the
nodal momentum belonging to the $s1-sc$ line arc can be expressed as a function of the 
angle $\phi_{arc}$.

Consistently with the limiting behaviors reported in Eq. (\ref{qq-qq}),
the $s1-sc$ line arc considered here becomes for $q_{arc}^N =q^N_{ec}$ and thus for $q^h =q_{ec}^h$ 
a single discrete nodal momentum ${\vec{q}}^{\,d\,N}_{ec}$ whose 
absolute value is given in Eq. (\ref{q-N-ec-coherent}). Hence ${\vec{q}}^{\,d}_{arc} ={\vec{q}}^{\,d\,N}_{ec}$. 
In the opposite limit reached at $q_{arc}^N =q^N_{Bs1}$
and then $q^h =q_{Fc}^h$ the $s1-sc$ line arc becomes the $s1$ boundary 
line of angular range $\phi\in (0,\pi/2)$, so that ${\vec{q}}^{\,d}_{arc} ={\vec{q}}^{\,d}_{Bs1}$. Between these two limits
the angular range $\phi\in (\pi/4 -\phi_{arc},\pi/4 + \phi_{arc})$ of the $s1-sc$ line arcs
increases from a single angle $\phi=\pi/4$ for $q_{arc}^N =q_{ec}^N$ and ${\vec{q}}^{\,d}_{arc}  ={\vec{q}}^{\,d\,N}_{ec}$
to a maximum angular width $\phi\in (0,\pi/2)$ for $q_{arc}^N =q^N_{Bs1}$ and ${\vec{q}}^{\,d}_{arc} ={\vec{q}}^{\,d}_{Bs1}$.
This corresponds to auxiliary momentum angular ranges increasing from a single
angle $\phi_{s1}=5\pi/4$ to a maximum angular width $\phi_{s1}\in (\pi,3\pi/2)$,
respectively.

The inequality (\ref{range-2-ec}) is equivalent to restricting the
energy dispersions $\epsilon_c^{ec} (\vec{q}^{\,h})$ of Eq. (\ref{Fc-q-h-ec}) and
$\epsilon_c (\vec{q}^{\,h})$ to the following ranges,
\begin{equation}
\epsilon_c^{ec} (\vec{q}^{\,h}) \in (V_{ec}^{eff} ({\vec{q}}),W_{ec}) \, ; \hspace{0.35cm}
\epsilon_c (\vec{q}^{\,h}) \in (-[W_{ec}-V_{ec}^{eff} ({\vec{q}})],0) \, ,
\label{range-2-int-ec}
\end{equation}
respectively. Here the energy $[W_{ec}-V_{ec}^{eff} ({\vec{q}})]$ is positive
or vanishing. For a $c$ fermion of a $c$ strongly coupled pair whose energy $\epsilon_c^{ec} (\vec{q}^{\,h})$
is measured from the $ec$-pairing line the energy scale,
\begin{equation}
V_{ec}^{eff} ({\vec{q}}) = {W_{ec}\over\vert\Delta\vert}\vert\Delta_{s1} ({\vec{q}})\vert 
\, ; \hspace{0.35cm} V_{ec}^{eff} ({\vec{q}}) \in (0,W_{ec}) \, ,
\label{Vc-s1-ec}
\end{equation}
with limiting values,
\begin{equation}
V_{ec}^{eff} ({\vec{q}}^{\,N\,d}_{ec}) = V_{ec}^{eff} ({\vec{q}}^{\,N\,d}_{Bs1}) = 0 
\hspace{0.50cm}
V_{ec}^{eff} ({\vec{q}}^{\,AN\,d}_{Bs1}) = W_{ec} \, ,
\nonumber
\end{equation}
plays the role of an effective potential energy. 

Since the momentum ${\vec{q}}$ of a strongly coupled virtual-electron pair configuration
equals that of the corresponding $s1$ fermion, ${\vec{q}}$ is at or near a $s1-sc$ line arc.
It follows from Eq. (\ref{range-2-ec}) that the 
energy $E_{v-el} ({\vec{q}})$ given in Eq. (\ref{E-v-el}) of 
the corresponding virtual-electron pair obeys the inequality,
\begin{equation}
E_{v-el} ({\vec{q}}) \leq E_1 (\phi) + \vert\epsilon_{s1} ({\vec{q}}^{\,d\,\pm}_{arc})\vert 
\, ; \hspace{0.35cm}  E_1 (\phi) = 2W_{ec}(1-\vert\cos 2\phi\vert) \, .
\label{Evep}
\end{equation}
Here ${\vec{q}}^{\,d\,\pm}_{arc}$ is a $s1-sc$-line arc momentum ${\vec{q}}^{\,d}_{arc}$
pointing in the directions defined by the angle $\phi =\pi/4\pm\phi_{arc}$. Hence breaking virtual-electron pairs of 
momentum ${\vec{q}} = {\vec{q}}^{\,d}_{arc}$ belonging to a $s1-sc$ line arc
under one-electron excitations leads to sharp spectral features
provided that their energy obeys the inequalities of Eq. (\ref{Evep}).
Moreover, only such virtual-electron pairs may have phase coherence. 
Note that the Fermi angle range $\phi\in (\pi/4 -\phi_{arc},\pi/4 + \phi_{arc})$ 
of a $s1-sc$ line arc has been constructed to inherently 
the inequalities of Eqs. (\ref{range-2-ec}) and (\ref{Evep}) and the energy
ranges of Eq. (\ref{range-2-int-ec}) being obeyed. 

For $U/4t\in (u_0,u_1)$ and $x\in (x_A,x_{c2})$ where $x_A\approx x_*/2$ the maximum
magnitude of the energy $\vert\epsilon_{s1} ({\vec{q}}^{\,d\,\pm}_{arc})\vert$ is often below the resolution of 
experiments on hole doped cuprates \cite{cuprates}. In that case the inequality (\ref{Evep})
approximately reads,
\begin{equation}
E_{v-el} ( {\vec{q}}^{\,d}_{arc}) \leq E_1 (\phi) = 2W_{ec}(1-\vert\cos 2\phi\vert)  \, .
\label{Evep-app}
\end{equation}

\subsection{Virtual-electron pairing breaking under electron removal: Signatures of the virtual-electron pairs
in the one-electron spectral-weight distribution}

Taking into account the ranges of Eqs. (\ref{range-2-ec}), (\ref{Evep}), and (\ref{Evep-app}) and using the 
general energy functional of Eq. (\ref{DeltaE-plus}), one finds that the maximum energy of the one-electron removal 
processes that lead to sharp spectral features is reached as follows: Upon breaking a virtual-electron pair, a $c$ fermion pair 
whose $c$ fermions have hole momenta ${\vec{q}}^{\,h}\,'$ and
$-{\vec{q}}^{\,h}\,'$ at or near the $c-sc$ line corresponding to the energy 
$\vert\epsilon_c (\pm\vec{q}^{\,h}\,')\vert = W_{ec}\,(1-\vert\cos 2\phi\vert)$
is broken. Simultaneously, a spinon pair of a $s1$ fermion whose momentum
${\vec{q}} = {\vec{q}}^{\,d}_{arc}$ is at or near the $s1-sc$ line arc whose nodal momentum absolute
value $q_{arc}^N$ is given by Eq. (\ref{q-N-coherent}) with $q^h=q^h\,'\equiv\vert{\vec{q}}^{\,h}\,'\vert$
is broken as well. The maximum energy condition imposes that such a $s1-sc$ line arc 
momentum corresponds to a minimum $\pi/4 -\phi_{arc}$
or maximum $\pi/4 +\phi_{arc}$ Fermi angle $\phi$ magnitude. Furthermore, one of the two involved $c$ fermions 
recombines with one spinon within the removed electron.
The spinon left behind then leads to the emergence of one $s1$ band hole \cite{companion2} whose
momentum ${\vec{q}} = {\vec{q}}^{\,d}_{arc}$ is that of the broken $s1$ fermion. 
Finally, the second $c$ fermion goes over to the $c$ Fermi line. The resulting energy spectrum 
is straightforwardly obtained by use of the hole-momentum-distribution-function deviations,
\begin{equation}
\delta N^h_{c}({\vec{q}}^{\,h})= 
[\delta_{\vec{q}^{\,h},-{\vec{q}}^{\,h}\,'}+\delta_{\vec{q}^{\,h},{\vec{q}}^{\,h}\,'}-\delta_{\vec{q}^{\,h},{\vec{q}}^{\,h\,d}_{Fc}}]
\, ; \hspace{0.35cm} \delta N^h_{s1}({\vec{q}}) = \delta_{\vec{q},{\vec{q}}^{\,d}_{arc}} \, ,
\label{oe-devi}
\end{equation}
in the energy functional of Eq. (\ref{DeltaE-plus}). It is given by the energy scale $E_1 (\phi)$ of Eqs. (\ref{Evep})
and (\ref{Evep-app}). Such an energy scale appears on the right-hand side of the inequality $E_{v-el} ({\vec{q}}^{\,d}_{arc})\leq E_1 (\phi)$ 
also provided in the former equation. It equals the maximum magnitude of the virtual-electron pair energy $E_{v-el} ({\vec{q}}^{\,d}_{arc})$ 
of strongly coupled virtual-electron pairs of momentum $\vec{q}= {\vec{q}}^{\,d}_{arc}$.
It follows that the one-electron spectrum $E_1 (\phi)$ corresponds to a line in the $(\phi,\omega)$ plane associated 
with the boundary separating the one-electron sharp features from broad incoherent features. Such a boundary refers 
to the equality $E_{v-el} ({\vec{q}}^{\,d\,\pm}_{arc})\approx E_1 (\phi)$ associated with virtual-electron pairs
whose $s1-sc$-line arc momentum has $\phi$ limiting magnitudes given by $\pi/4\pm\phi_{arc}$. 
In turn, the above inequality $E_{v-el} ({\vec{q}}^{\,d}_{arc})\leq E_1 (\phi)$ 
of Eq. (\ref{range-2-ec}) corresponds to a $(\phi,\omega)$ plane
domain bounded by it and associated with virtual-electron pairs
whose $s1-sc$-line arc momentum has $\phi$ magnitudes belonging to
the whole corresponding range $\phi\in (\pi/4 -\phi_{arc},\pi/4 +\phi_{arc})$. 
Indeed, breaking virtual-electron pairs of momentum $\vec{q}\approx {\vec{q}}^{\,d}_{arc}$ 
upon one-electron removal excitations leads to sharp spectral features provided that their effective coupling is strong.
This implies that their energy obeys such an inequality. It corresponds to a $(\phi,\omega)$ plane domain
constituted by a set of sharp one-electron removal spectral features lines.

There is an one-to-one correspondence between each sharp one-electron removal spectral 
feature line arc (1-el-sharp-feature line arc) and a virtual-electron-pair $s1$ band $s1-sc$ line arc. Specifically, for $x\in (x_A,x_{c2})$
and $U/4t\in (u_0,u_1)$ the energy scale $\vert\epsilon_{s1} ({\vec{q}}^{\,d}_{arc})\vert$ is small and below the
experimental resolution, so that the energy and hole momentum of the 1-el-sharp-feature line arcs read,
\begin{eqnarray} 
E (\vec{k}^{\,h}) & \approx & {4\over\pi}{(q^N_{Bs1}-q_{arc}^N)\sqrt{[q^{AN}_{Bs1}]^2-[q^{N}_{Bs1}]^2}\over m_c^*}
= {\cal{E}}_{v-el} (2\phi_{arc})
= 2W_{ec} (1-\sin 2\phi_{arc}) \, ,
\nonumber \\
\vec{k}^{\,h} & = & {\vec{q}}^{\,h\,d}_{Fc} - {\vec{q}}^{\,d}_{arc} 
\, , \hspace{0.15cm} \phi\in (\pi/4 -\phi_{arc},\pi/4 +\phi_{arc}) 
\, ; \hspace{0.35cm} \vec{k}^{\,N} \equiv \vec{k}^{\,h\,N} - \vec{\pi} = - {k^N\over q^N_{arc}}{\vec{q}}^{\,N}_{arc}
\approx - {k_F^N\over q^N_{Bs1}}{\vec{q}}^{\,N}_{arc} \, ,
\label{k-phi}
\end{eqnarray}
respectively. Importantly, the energy scale $E (\vec{k})\approx E (q_{arc}^N)$ exactly equals the energy $E_{v-el}$ of the 
strongly coupled virtual-electron pair broken under the one-electron removal
excitation. Except for the small energy scale $\vert\epsilon_{s1} ({\vec{q}}^{\,d}_{arc}\vert$
ignored here, strongly coupled virtual-electron pairs associated with the
same 1-el-sharp-feature line arc have the same
energy. Indeed, $E_{v-el} (q_{arc}^N)$ depends only on the nodal momentum
absolute value $q_{arc}^N$ of the corresponding $s1-sc$ line arc. It may alternatively
be expressed as a function of the angular width $2\phi_{arc}$ of the
line arc under consideration. 

Note that different 1-el-sharp-feature line arcs have different 
energy magnitudes $E (q_{arc}^N) = E_{v-el} (q_{arc}^N)$. Hence they refer
to broken strongly coupled virtual-electron pairs of different energy.
It follows that the energy $E (q_{arc}^N) = E_{v-el} (q_{arc}^N)$ of each
1-el-sharp-feature line arc provides 
a direct experimental signature of the corresponding 
broken strongly coupled virtual-electron pairs. Its energy 
exactly equals that of such pairs. The ${\cal{E}}_{v-el} (2\phi_{arc})$ expression provided in Eq. (\ref{k-phi})
refers to the energy $2\vert\epsilon_{c} ({\vec{q}}^{\,h})\vert$
expressed in terms of the corresponding $s1-sc$-line arc angular width $2\phi_{arc}$.
The coefficient $(1-\sin 2\phi_{arc})$ equals the factor $(1-\vert\cos 2\phi\vert)$ of the $E_1$ expression given in
Eq. (\ref{EF}) for $2\phi =(\pi/2\pm 2\phi_{arc})$ and $2\phi_{arc} \in (0,\pi/2)$.

The 1-el-sharp-feature line arc
excitation hole momentum $\vec{k}^{\,h}=\vec{k}+\vec{\pi}$ also given in Eq. (\ref{k-phi})
is centered at momentum $-\vec{\pi}$. In that expression an expression for
the corresponding momentum $\vec{k}$ valid only for the nodal direction is
also provided. Each of the four arcs of the 1-el-sharp-feature line has
exactly the same angular width $2\phi_{arc}$ as the corresponding
$s1-sc$ line four arcs. Moreover, the Fermi angle $\phi$ range provided in Eq. (\ref{k-phi})
of each 1-el-sharp-feature line arc has been constructed to inherently 
the inequalities of Eqs. (\ref{range-2-ec}), (\ref{Evep}), and (\ref{Evep-app}) being obeyed.
Its expression provided in that equation
can be expressed in terms of the momentum ${\vec{q}}^{\,d}_{arc}$ of the hole that
emerges in the $s1$ band and final hole momentum ${\vec{q}}^{\,h\,d}_{Fc}$ of the $c$ fermion
transferred from the $c$ band hole momentum ${\vec{q}}^{\,h}$ or $-{\vec{q}}^{\,h}$ to the
$c$ Fermi line. The  auxiliary momentum
${\vec{q}}_{arc}=[A^d_{s1}]^{-1}\,{\vec{q}}^{\,d}_{arc}$
has been constructed to inherently pointing in the same
direction as the hole momentum $-\vec{k}^{\,h}$. This implies
that the absolute value $k^N$ of the
excitation nodal momentum $\vec{k}^{\,N}=\vec{k}^{\,h\,N}-\vec{\pi}$ is approximately given
by $k^N\approx k_F^N \,[q_{arc}^N/q_{Bs1}^N]$.
A comparative analysis of the hole Fermi momentum $\vec{k}_F^{\,h} (\phi)$ 
and nodal Fermi momentum $\vec{k}_F^{\,N}$ 
expressions of Eq. (\ref{khF-kF})
and 1-el-sharp-feature line arc hole momentum $\vec{k}^{\,h} (\phi)$ 
and nodal momentum $\vec{k}^{\,N}$ expressions 
of Eq. (\ref{k-phi}) reveals that for $q_{arc}^N =q^N_{Bs1}$ and thus ${\vec{q}}_{arc}={\vec{q}}_{Bs1}$
the sharp one-electron spectral features line centered at $-\vec{\pi}$ is the Fermi line. In turn,
for $q_{arc}^N =q^N_{ec}$ and thus ${\vec{q}}_{arc}={\vec{q}}_{ec}$ where
$\phi=\pi/4$ the arc of such a line considered here becomes a single discrete momentum value.
It corresponds to a nodal momentum that we denote by $\vec{k}^{\,N}_{ec}$ and
reads $\vec{k}^{\,N}_{ec}\approx -[k_{F}^N/q_{Bs1}^N]\,{\vec{q}}^{\,N}_{ec}$. Here 
${\vec{q}}^{\,N}_{ec}$ is the auxiliary nodal $s1-sc$-line momentum of Eq. (\ref{qq-qq})
whose absolute value is given in Eq. (\ref{q-N-ec-coherent}).

The studies of Ref. \cite{cuprates} check the validity of the strongly coupled virtual-electron
pairing mechanism introduced in this paper in experiments on hole-doped cuprates.
As discussed in that reference, the 1-el-sharp-feature line arcs, whose energy
provides a direct signature of the strongly coupled virtual-electron pairs, are observed in
the experiments on LSCO of Ref.  \cite{LSCO-ARPES-peaks}. The studies of Ref. \cite{cuprates}
confirm that the boundary that separates the broad one-electron removal spectral features
from the 1-el-sharp-feature line arcs is indeed the theoretical
line $E_{v-el} ({\vec{q}}^{\,d}_{arc}) \approx 2W_{ec}(1-\vert\cos 2\phi\vert)$
of the virtual-electron energy allowed by the inequality (\ref{Evep-app}). It refers to
the momenta associated with the minimum and maximum Fermi
angles $\pi/4-\phi_{arc}$ and $\pi/4+\phi_{arc}$, respectively, of each
1-el-sharp-feature line arc.
The theoretical expression $E_{v-el} ({\vec{q}}^{\,d}_{arc}) \approx 2W_{ec}(1-\vert\cos 2\phi\vert)$
has exactly the same form and, for the theoretical parameters appropriate to LSCO, the same 
$2W_{ec}$ magnitude as the corresponding experimental empirical formula given
in Eq. (1) of Ref. \cite{LSCO-ARPES-peaks}. That formula refers indeed to 
the boundary separating the one-electron sharp features
from broad incoherent features in that hole-doped cuprate superconductor.
Such an overall agreement seems to confirm that the strongly coupled virtual-electron pairs 
are observed in the hole-doped cuprate superconductors, consistently with
the pairing mechanism proposed here for the VEP quantum liquid.

\subsection{Zero-temperature superfluid density}

Concerning charge excitations and currents associated with the phases 
$\theta_{cp}$ the zero-momentum $c$ fermion pairs participating in
phase-coherent virtual-electron pair configurations behave independently 
of the corresponding $s1$ fermions of momentum $\vec{q}$.  
Hence within such phenomena the phases $\theta_{cp}$
are associated with the zero-momentum $c$ fermion pairs of the
phase-coherent-virtual-electron-pair superconducting state macroscopic condensate. 
In the following we call them {\it coherent $c$ fermion pairs}.
For such a state the energy cost of 
a phase twist is for small $0<(x-x_c)\ll 1$ and zero temperature
approximately given by $\rho_{cp} (0)\vert{\vec{\nabla}}\theta_{cp}\vert^2/8$.
Here $\rho_{cp} (T)$ is the superfluid density at temperature $T$
of coherent $c$ fermion pairs. For $0<(x-x_c)\ll 1$
the fluctuations of the phases $\theta_{j,0}$ and thus of the
$c$ fermion-pair phases $\theta_{cp}$ are strong. Hence the critical temperature
$T_c$ is governed by such fluctuations. As a result for $0<(x-x_c)\ll 1$ the transition 
taking place from the superconducting state to the quantum vortex liquid is 
for the quasi-2D VEP quantum liquid a Berezinskii-Kosterlitz-Thouless like transition \cite{Bere,KT}. 
One finds according to the properties of such a transition that under the suppression
effects the following relation holds,
\begin{equation}
{8\over \pi}\,k_B\,T_c={n_{cp} (T_c)\over m^*_c} = 
\rho_{cp} (T_c) \, .
\label{Yc-n-rho}
\end{equation}
Here $n_{cp} (T_c)$ is the density $\lim_{T\rightarrow T_c}n_{cp} (T)$ of coherently paired $c$ fermions 
for $0<(T_c-T)\ll 1$. (For $0<(T-T_c)\ll 1$ such a density vanishes.) 
$n_{cp} (T_c)/2$ is the corresponding density of the macroscopic condensate coherent $c$ fermion pairs. In units of lattice spacing $a$ one, 
the $c$ fermion mass can be written in terms of the coefficient $r_c$ of Eq. (\ref{m*c/mc-UL}) of Appendix A
as $m^*_c=\hbar^2/2r_c t$. Here $\hbar$ is the Planck constant.
(Often we write such a mass in units of both $a$ and $\hbar$ one, so that it reads
$m^*_c=1/2r_c t$.) From the $T_c$ expression given in Eq. (\ref{Om-Dp-Tc}) one finds 
$T_c={\breve{g}}_0\Delta_0/2k_B\rightarrow 0$ as $(x-x_c)\rightarrow 0$. Indeed, in 
that limit the suppression coefficient of Eq. (\ref{gamma-d}) reads $\gamma_d =1$.
For $0<(x-x_c)\ll 1$ the use of the relations given in Eq. (\ref{Yc-n-rho}) then leads to,
\begin{equation}
\rho_{cp}(0) \approx {\breve{g}}_0\,{4\Delta_0\over \pi\hbar^2} \, .
\label{rho-0}
\end{equation}
Here ${\breve{g}}_0$ is the amplitude given in Eq. (\ref{g0-g1}). So that the relations 
$r_c\approx 2r_s\approx 2e^{-4t\,u_0/U}$ and $x_*=2r_s/\pi\approx r_c/\pi\approx [2/\pi] e^{-4t\,u_0/U}$ 
apply, expression (\ref{rho-0}) and most expressions given below are derived for the approximate
range $U/4t\in (u_0\,u_1)$. It follows from Eq. (\ref{rho-0}) that for $0<(x-x_c)\ll 1$ the superconducting-ground-state
$c$ band momentum area that corresponds to the $c$ fermions contributing to coherent pairing is,
\begin{equation}
S_{cp}^c = (x-x_c)\,{4\pi^2\over r_c (r_c-\pi x_c)} {2\Delta_0\over t}\, .
\label{Acp}
\end{equation} 
The hole-momentum domain $Q^{c}_{cp}$ of the $c$ fermions
that contribute to coherent pairing refers to a set of $c-sc$ lines 
whose radius $q^h$ belongs to the range $q^h\in (q^{h}_{Fc},q^{h}_{cp})$.
Here $q^{h}_{cp}$ is the radius of the 
zero-temperature {\it coherent unpaired $c$ fermion line}
whose hole momenta are denoted by ${\vec{q}}^{\,h}_{cp}$. It is
the {\it coherent $c-sc$ line} of largest radius $q^{h}_{cp}$. For $0<(x-x_c)\ll 1$ it is 
nearly circular and separates in the $c$ momentum band the $c$ fermions
participating in coherent pairing from those that do not participate.
The hole-momentum domain $Q^{c}_{cp}$ is bounded by the $c$ Fermi
line and coherent unpaired $c$ fermion line, respectively.
  
For $0<(x-x_c)\ll 1$ the coherent unpaired $c$ fermion line radius $q^h_{cp}$ 
approximately reads,
\begin{equation}
q^{h}_{cp} \approx \sqrt{1+{W_{cp}\over [q_{Fc}^h]^2r_c t}}\,q_{Fc}^h 
= \sqrt{1+{W_{cp}\over x4\pi r_c t}}\,q_{Fc}^h  \, .
\label{q-h-Fcp-coherent}
\end{equation}
For $U/4t\in (u_0, u_1)$ this expression is a very good approximation for the range $x\in (x_c,x_{c1})$ 
and a reasonably good approximation for the range $x\in (x_{c1},x_*)$.

The energy bandwidth $W_{cp}\approx [(q^{h}_{cp})^2-(q^{h}_{Fc})^2]/2m^*_c$
of the corresponding superconducting-ground-state sea of $c$ fermions 
contributing to coherent pairing reads,
\begin{equation}
W_{cp} = {\breve{g}}_0\,8\Delta_0 
\, ; \hspace{0.35cm}
W_{ec} \equiv {\rm max}\,W_{cp}={4\Delta_0\over \gamma_c} \, .
\label{W-c-qq}
\end{equation}
This confirms the validity of the maximum energy bandwidth expression
provided in Eq. (\ref{W-ec}). The parameter $\gamma_c$ appearing here is given in Eq. (\ref{x-op}).
The expression of $W_{cp}$ provided in this equation is a good approximation for a 
range of hole concentrations $x\in (x_c,x_{cp})$ for which $W_{cp}<W_{ec}$.
The maximum magnitude $W_{ec}$ of the energy bandwidth 
$W_{cp}$ is reached at a hole concentration
$x=x_{cp}$ given below. Only for $x$ below $x_{cp}$ there is formation
of coherent $c$ fermion pairing upon increasing $x$. Such a formation 
occurs while the short-range spin correlations are strong enough to
supply the energy needed for it. For the hole concentration range
$x\in (x_{cp},x_*)$ the energy bandwidth $W_{cp}$ is independent of
$x$ and reads $W_{cp}=W_{ec}$. Furthermore, it vanishes for $x>x_*$. 
The singular behavior occurring at $x=x_*$ marks a sharp quantum phase transition to
a disordered state without short-range spin order and thus
without long-range superconducting order. 

For the approximate range $U/4t\in (u_0, u_1)$
generalization of the zero-temperature superfluid density expression (\ref{rho-0}) 
found above for $0<(x-x_c)\ll 1$ to the hole-concentration range $x\in (x_c,x_*)$ 
for which there is long-range superconducting order leads to, 
\begin{equation}
\rho_{cp} \approx  \left[{\breve{g}}_0 \Theta (x_{cp}-x) + {1\over 2\gamma_c} \theta (x-x_{cp})\right]{4\Delta_0\over \pi \hbar^2} 
\, ; \hspace{0.35cm} x_{cp}= {x_*\over 2} + x_c \, .
\label{rho-gen}
\end{equation}
Here the theta function $\Theta (z)$ is such that $\Theta (z)=1$ for $z\geq 0$ and $\Theta (z)=0$ for $z<0$. 
The superfluid density $\rho_{cp}$ vanishes both for $x<x_c$ and $x>x_*$. Hence it has
a singular behavior at $x=x_*$. It marks the above mentioned sharp quantum phase transition to
a disordered state without short-range spin order and thus without long-range superconducting 
order for $x>x_*$. 

Why does the superfluid density (\ref{rho-gen}) not depend of the suppression coefficient $\gamma_d$?
Since there are no suppression effects both for $0<(x-x_c)\ll 1$ and $0<(x_*-x)\ll 1$ when $\gamma_d=1$ 
and such effects are strongest at $x=x_{op}$, to answer such a question it is useful to consider the ratio,
\begin{equation}
\left[{2\vert\Omega\vert\over 2\vert\Delta\vert}\right]\vert_{T=0,x=x_{op}} =
{\gamma_d^{min}\over 2\gamma_c} = {(1-\alpha_d)\over 2\gamma_c} \, .
\label{value-2-ratios}
\end{equation} 
The energy scale $2\vert\Omega\vert\vert_{T=0}$ is the maximum magnitude of the 
superconducting virtual-electron pairing energy
. The dependence on the suppression coefficient 
$\gamma_d^{min}=(1-\alpha_d)$ of the ratio $[2\vert\Omega\vert/2\vert\Delta\vert]\vert_{T=0,x=x_{op}}$
justifies physically why the zero-temperature superfluid density $\rho_{cp}$ remains unaltered under the suppression effects 
and given by Eqs. (\ref{rho-gen}). If the ratio $[2\vert\Omega\vert/2\vert\Delta\vert]\vert_{T=0,x=x_{op}}$
was independent of $\gamma_d^{min}$, the superfluid density would be suppressed. In that case one 
would expect that $\rho_{cp}\propto \gamma_d^{min}$ at $x=x_{op}$. However, while the energy available to the 
short-range correlations $2\vert\Delta\vert\vert_{T=0}$ to supply pair formation remains unaltered, 
the energy cost $2\vert\Omega\vert\vert_{T=0}$ of that pair formation decreases by a factor given
exactly by the suppression coefficient $\gamma_d^{min}$. This effect cancels the decreasing of $\rho_{cp}$,
whose magnitude remains independent of $\gamma_d^{min}$. Such an analysis is straightforwardly 
generalized to the whole range $x\in (x_c,x_*)$.

It follows from expression (\ref{rho-gen}) that the corresponding 
superconducting-ground-state energy bandwidth of the sea of $c$ fermions whose pairs contribute 
to the superfluid density is for hole concentrations in the range $x\in (x_c,x_*)$, vanishing spin density 
$m=0$, and zero temperature given by,
\begin{equation}
W_{cp} = \left[{\breve{g}}_0\,\Theta (x_{cp}-x) +{1\over 2\gamma_c}\,\theta (x-x_{cp})\right] 8\Delta_0  \, .
\label{Wc-Wcp}
\end{equation} 

The in-plane penetration depth $\lambda_{ab}$ can be expressed in terms of the 
superfluid density of Eq. (\ref{rho-gen}) as follows,
\begin{equation}
\lambda_{ab} = {1\over (-e)}\,\sqrt{d_{\parallel}\over \mu_0\,\rho_{cp}}
\, ; \hspace{0.35cm}  
{1\over \lambda_{ab}^2} = {\mu_0\,e^2\over d_{\parallel}}\,\rho_{cp} \, .
\label{depth}
\end{equation}
Here $d_{\parallel}$ is the plane separation of the system described by the quasi-2D 
Hamiltonian (\ref{M-t}), $-e$ denotes the electronic charge, and $\mu_0$ is the vacuum permeability.
As confirmed in Ref. \cite{cuprates0}, for the parameter values appropriate to the five representative hole-doped 
cuprate superconductors the penetration depth $\lambda_{ab}$ is much larger than the coherence length
$\xi$ of Eq. (\ref{coherent-length}).

Finally, the $x$ dependence of the zero-temperature superfluid density $\rho_{cp}$ of 
Eq. (\ref{rho-gen}) can for approximately $U/4t\in (u_0, u_1)$ be obtained from solution 
of the following rate differential equation of superfluid-density increase upon 
increasing the hole concentration $x$, under suitable physical boundary conditions,
\begin{equation}
{\partial \rho_{cp}\over \partial x} =   
{W_{ec}\over 2\hbar^2 r_s}\,\theta (x_{cp}-x)  \, ; \hspace{0.35cm}
\rho_{cp}\vert_{x_{cp} < x<x_*} = \rho_{cp}\vert_{x=x_{cp}} = {W_{ec}\over \pi\hbar^2} 
 \, ; \hspace{0.35cm}
\rho_{cp}\vert_{x=x_c} = \rho_{cp}\vert_{x=x_*} = 0 \, .
\label{rate-x}
\end{equation}

\subsection{General superconducting virtual-electron pairing energy}

Here we generalize the expression of the superconducting virtual-electron pairing energy per electron
$\vert\Omega_{s1} ({\vec{q}})\vert$ given in Eq. (\ref{q-g-0}) for momenta ${\vec{q}}\approx {\vec{q}}^{\,d}_{Bs1}$ 
at or near the $s1$ boundary line to momenta ${\vec{q}}={\vec{q}}^{\,d}_{arc}$ belonging to any
other phase-coherent virtual-electron pair configuration $s1-sc$ line arc, as defined below. The whole set of such lines
generates the momentum range ${\vec{q}}\in Q^{s1}_{cp}$ of the phase-coherent virtual-electron pair configurations.
The general function $\vert\Omega_{s1} ({\vec{q}})\vert$ 
given here is that involved in the $s1$ band energy dispersion $\epsilon_{s1} ({\vec{q}})$
expression of Eq. (\ref{band-s1}). That dispersion appears in the first-order terms of
the general energy functional provided in Eq. (\ref{DeltaE-plus}).

The general coherent virtual-electron
pairing energy per electron $\vert\Omega_{s1} ({\vec{q}})\vert$
refers to pairs of $c$ fermions whose hole momenta
belong to any $c-sc$ line between the $c$ Fermi line and the 
{\it coherent unpaired} $c$ fermion line. The radius $q^h$
of such a $c-sc$ line belongs to the range $q^h\in (q^{h}_{Fc}, q^{h}_{cp})$.
Only for it is the coherent virtual-electron
pairing energy per electron introduced here finite.
The energy bandwidth $W_{cp}$ 
of such $c$ fermions provided in Eq. (\ref{Wc-Wcp}) plays
the same role for the coherent unpaired $c$ fermion line
as the energy bandwidth $W_{ec}$ given in
Eq. (\ref{W-ec}) for the $ec$-pairing line. As found below,
for $x\in (x_{cp},x_*)$ these two lines are the same line.

For the range of hole concentrations $x\in (x_c,x_*)$ the shape of the coherent unpaired 
$c$ fermion line is at zero temperature defined by the following relation,
\begin{equation}
{\vec{q}}^{\,h}_{cp} \in {\rm coherent} 
\hspace{0.10cm} {\rm unpaired} 
\hspace{0.10cm} c  
\hspace{0.10cm} {\rm line}
\hspace{0.05cm} \Longleftrightarrow  \hspace{0.05cm}
\epsilon_{cp} ({\vec{q}}^{\,h}_{cp})= 0 \, .
\label{g-cp}
\end{equation}
Here the $c$ fermion energy dispersion $\epsilon_{cp} (\vec{q}^{\,h})$,
\begin{equation}
\epsilon_{cp} (\vec{q}^{\,h}) = W_{cp}+\epsilon_c (\vec{q}^{\,h}) \, , \hspace{0.15cm}
0 \leq \epsilon_{cp} (\vec{q}^{\,h}) \leq W_{cp} \, ; \hspace{0.35cm}
\epsilon_{cp} ({\vec{q}}^{\,h}_{cp}) = 0
\, ; \hspace{0.15cm}  \epsilon_{cp} ({\vec{q}}^{\,h\,d}_{Fc}) = W_{cp} \, ,
\label{Fcp-q-h}
\end{equation}
has the vanishing energy level at 
the coherent unpaired $c$ fermion line.
For $U/4t\in (u_0,u_1)$ and the hole-concentration range $x\in (x_c,x_*)$ 
the coherent unpaired $c$ fermion line radius is approximately 
given by Eq. (\ref{q-h-Fcp-coherent}). 

The corresponding $s1$ band momentum domain $Q^{s1}_{cp}$ 
of the phase-coherent virtual electron pair configurations refers to the set of coherent $s1-sc$ line arcs associated with 
coherent $c-sc$ lines contained in the $c$ band hole-momentum domain 
$Q^c_{cp}$. For instance, for the $s1-sc$ line arc corresponding to the Fermi angle range $\phi\in (0,\pi/2)$
whose $s1$ band quadrant has auxiliary momentum
angle range $\phi_{s1}\in (\pi,3\pi/2)$ and components
$q_{0x_1}\leq 0$ and $q_{0x_2}\leq 0$ it is limited by the $s1$ boundary line and the 
coherent $s1-sc$ line arc with minimum nodal momentum absolute value $q^N_{cp}$.
The latter is given by expression (\ref{q-N-coherent}) at $q^h=q^h_{cp}$. This leads to,
\begin{equation}
q^N_{cp} \approx q^N_{Bs1} -{[q^h_{cp}]^2-[q^h_{Fc}]^2\over 4\sqrt{[q^{AN}_{Bs1}]^2-[q^{N}_{Bs1}]^2}}\,\pi =
q^N_{Bs1} - { {\breve{g}}_0\, 2\Delta_0\over r_c \sqrt{[q^{AN}_{Bs1}]^2-[q^{N}_{Bs1}]^2}\,t}\,\pi \, ,
\label{q-N-cp-coherent}
\end{equation}
for $x\in (x_c,x_{cp})$ and $q^N_{cp}=q^N_{ec}$ for $x\in (x_{cp},x_*)$
where $q^N_{ec}$ is provided in Eq. (\ref{q-N-ec-coherent}). The $s1$ boundary momenta 
$q^{N}_{Bs1}$ and $q^{AN}_{Bs1}$ appearing here are given in Eq. (\ref{qN-qAN}) of Appendix E.

For the hole concentration range $x\in (x_c,x_{cp})$ the coherent $s1-sc$ line arc angle 
width $2\phi_{arc}$ belongs to the range $2\phi_{arc}\in (\pi/2-2\phi_{arc}^{cp},\pi/2+2\phi_{arc}^{cp})$. 
Here $\phi_{arc}^{cp}$ is the angle $\phi_{arc}$ of Eq. (\ref{phi-arc-qN}) 
at $q_{arc}^N =q^N_{cp}$ and thus $q^h=q^h_{cp}$. It reads,
\begin{eqnarray}
\phi_{arc}^{cp} & = & {1\over 2}\arcsin\left({q^N_{cp}-q^N_{ec}\over q^N_{Bs1}-q^N_{ec}}\right)
= {1\over 2}\arcsin\left({2(x_{cp}-x)\over x_*}\right)\in (0,\pi/4) \, ; \hspace{0.35cm} x\in (x_c,x_{cp}) \, ,
\nonumber \\
& = & 0 \, ; \hspace{0.35cm} x\in (x_{cp},x_*) \, .
\label{phi-arc-qN-cp}
\end{eqnarray}
Its limiting values are $\phi_{arc}^{cp} =\pi/4$ at $x=x_c$ and $\phi_{arc}^{cp}=0$ 
for $x\in (x_{cp},x_*)$.

The $c$ fermions of hole momenta $\vec{q}^{\,h}$ and $-\vec{q}^{\,h}$ at or near the $c$ Fermi line
participate in phase-coherent virtual-electron pair configurations of momentum 
${\vec{q}}\approx {\vec{q}}^{\,d}_{Bs1}$ at or near the $s1$ boundary line. The strong effective coupling 
of such $c$ fermions results from residual interactions with the corresponding $s1$ fermions of momentum 
${\vec{q}}\approx {\vec{q}}^{\,d}_{Bs1}$ at or near the $s1$ boundary line. In turn, in the opposite limit of
$c$ fermion hole momenta $\vec{q}^{\,h}$ and $-\vec{q}^{\,h}$ belonging to the coherent unpaired $c$ fermion line
such a residual interactions are with $s1$ fermions of momentum ${\vec{q}}$ belonging to the
coherent $s1-sc$ line arc whose nodal momentum has minimum absolute 
value $q^N_{cp}$. For the hole concentration range $x\in (x_c,x_{cp})$ that line has the
minimum angular range $\phi\in (\pi/4-\phi_{arc}^{cp},\pi/4+\phi_{arc}^{cp})$ of the coherent $s1-sc$ line arcs
and its nodal momentum has absolute value $q^N_{cp}$ given in Eq. (\ref{q-N-cp-coherent}).
In turn, for $x\in (x_{cp},x_*)$ the angular width $2\phi_{arc}$ vanishes and $q^N_{cp}$ equals the 
absolute-value nodal momentum $q^N_{ec}$ given in Eq. (\ref{q-N-ec-coherent}). For the latter $x$ range 
the coherent $s1-sc$ line arc of minimum nodal momentum absolute value $q^N_{cp}$
then reduces to the discrete nodal momentum ${\vec{q}}= {\vec{q}}^{\,d\,N}_{ec}$.
Hence for $x\in (x_{cp},x_*)$ the coherent unpaired $c$ fermion line is the $ec$-pairing $c$ fermion line. 
For intermediate $\vec{q}^{\,h}$ and $-\vec{q}^{\,h}$ hole momenta at or near the coherent $c-sc$ lines
between the $c$ Fermi line and the coherent unpaired $c$ fermion line the strong effective coupling 
associated with coherent pairing is due to residual interactions 
of the $c$ fermions under consideration with $s1$ fermions of momentum at or near the 
corresponding coherent $s1-sc$ line arcs whose nodal momentum absolute value belongs to the range
$q_{arc}^N\in (q^N_{cp},q^N_{Bs1})$.

Based on the results of Sections IV-C and IV-E, we are now ready to introduce the
superconducting pairing energy
 spectrum $\vert\Omega_{s1} ({\vec{q}})\vert$ contributing to the $s1$ band energy 
dispersion $\epsilon_{s1} ({\vec{q}})$ expression (\ref{band-s1}).
For a virtual-electron pair of momentum ${\vec{q}}={\vec{q}}^{\,d}_{arc}$ belonging to a given
coherent $s1-sc$ line arc the general superconducting virtual-electron pairing energy
 per electron reads,
\begin{equation}
\vert\Omega_{s1} ({\vec{q}}^{\,d}_{arc})\vert =
\Theta \left(\epsilon_{cp} (\vec{q}^{\,h}) -V_{cp}^{eff} ({\vec{q}}^{\,d}_{arc})\right)
\Theta \left(W_{cp}-\epsilon_{cp} (\vec{q}^{\,h})\right)
\gamma_d\,g_0\sqrt{2}\,V^{\Delta}_{s1} ({\vec{q}}^{\,d}_{arc})\,
{\epsilon_{cp} (\vec{q}^{\,h})
+\epsilon_{cp} (-\vec{q}^{\,h})\over 2W_{cp}} \, .
\label{omega-gene-q-h}
\end{equation}
Here $q^h$ is the radius of the corresponding $c-sc$ line. It is uniquely related to
the nodal momentum absolute value $q_{arc}^N$ of the coherent $s1-sc$ line arc
under consideration by the function $q_{arc}^N = q_{arc}^N (q^h)$ defined in Eq. (\ref{q-N-coherent}). 
Moreover, since coherent pairing requires strong effective coupling, the energy range of
Eq. (\ref{range-2-ec}) together with the expression $\epsilon_{cp} (\vec{q}^{\,h}) = 
W_{cp}+\epsilon_c (\vec{q}^{\,h})$ of Eq. (\ref{Fcp-q-h}) imply that the energy scale
$V_{cp}^{eff} ({\vec{q}})$ appearing in this pairing energy expression reads,
\begin{eqnarray}
V_{cp}^{eff} ({\vec{q}}^{\,d}_{arc}) & = & 0 \, ; \hspace{0.5cm} 
W_{cp} < W_{ec}\left(1-{\vert\Delta_{s1} ({\vec{q}}^{\,d}_{arc})\vert\over\vert\Delta\vert}\right)
\nonumber \\
& = & W_{cp} - W_{ec}\left(1-{\vert\Delta_{s1} ({\vec{q}}^{\,d}_{arc})\vert\over\vert\Delta\vert}\right)
\, ; \hspace{0.5cm} 
W_{cp} > W_{ec}\left(1-{\vert\Delta_{s1} ({\vec{q}}^{\,d}_{arc})\vert\over\vert\Delta\vert}\right) \, .
\label{Vcp-s1}
\end{eqnarray}

The pairing energy given in Eq. (\ref{omega-gene-q-h}) is expressed in terms of the energy 
$\epsilon_{cp} (\vec{q}^{\,h})+\epsilon_{cp} (-\vec{q}^{\,h})$
of the $c$ fermion pairs of hole momentum at or near the coherent 
$c-sc$ line of radius $q^h$. Those interact with the $s1$ fermion 
of momentum ${\vec{q}}={\vec{q}}^{\,d}_{arc}$ at or near the
corresponding coherent $s1-sc$ line arc. This occurs within the phase-coherent virtual-electron pair configuration of 
the same momentum. The use of the function $q_{arc}^N = q_{arc}^N (q^h)$ of Eq. (\ref{q-N-coherent}) 
leads to the following equivalent yet simpler expression for the general pairing energy per electron 
defined by Eqs. (\ref{omega-gene-q-h}) and (\ref{Vcp-s1}),
\begin{equation}
\vert\Omega_{s1} ({\vec{q}}^{\,d}_{arc})\vert =
\gamma_d\,g_0\sqrt{2}\,V^{\Delta}_{s1} ({\vec{q}}^{\,d}_{arc})
\left({q_{arc}^N -q_{cp}^N\over
q_{Bs1}^N -q^N_{cp}}\right) \approx
\gamma_d\,g\left({q_{arc}^N -q_{cp}^N\over
q_{Bs1}^N -q^N_{cp}}\right)\Delta_0\vert\sin 2\phi\vert \, .
\label{omega-gene}
\end{equation}
Here, 
\begin{equation}
V^{\Delta}_{s1} ({\vec{q}}) = 
\vert{\vec{\nabla}}_{\vec{q}}\vert\Delta_{s1} ({\vec{q}})\vert\vert =
{\vert\Delta\vert\over\sqrt{2}}\,G_{s1}({\vec{q}}) 
\, , \hspace{0.25cm} {\vec{q}}={\vec{q}}^{\,d}_{arc}
\, ; \hspace{0.35cm} G_{s1}({\vec{q}}^{\,d}_{arc}) \approx \vert\sin 2\phi\vert \, ,
\label{v-D-Fs-G}
\end{equation}
where the function $G_{s1}({\vec{q}})$ is defined in Ref. \cite{companion2}. 
The value $G_{s1}({\vec{q}}^{\,d}_{arc}) \approx \vert\sin 2\phi\vert$ is a good
approximation for $s1$ band momenta belonging to coherent $s1-sc$ line arcs.

The energy $\vert\Omega_{s1} ({\vec{q}}^{\,d}_{arc})\vert$ has its
maximum magnitude for momenta ${\vec{q}}^{\,d}_{arc}\approx {\vec{q}}^{\,d}_{Bs1}$,
when it involves $c$ fermions of hole momenta at or near the $c$ Fermi
line such that $\epsilon_{cp} (\pm\vec{q}^{\,h})= W_{cp}$. The strong effective
coupling of such fermions results from interactions with $s1$ fermions
of momenta at or near the $s1$ boundary line. It vanishes in the
opposite limit referring to $c$ fermions of hole momenta at or near the coherent unpaired 
$c$ fermion line for which $\epsilon_{cp} (\pm\vec{q}^{\,h})=0$.

Both the superconducting pairing energy
 expressions  (\ref{omega-gene-q-h}) 
and (\ref{omega-gene}) have physical significance only for $s1$
band momenta ${\vec{q}}\approx {\vec{q}}^{\,d}_{arc}$ at or near a coherent $s1-sc$ line arc. 
Interestingly, the energy $\vert\Omega_{s1} ({\vec{q}}^{\,d}_{arc})\vert$ has a
much simpler expression (\ref{omega-gene}) in terms of the
momenta of the virtual-electron pair coherent $s1-sc$ line arc
than as given in Eq.  (\ref{omega-gene-q-h}).
Indeed, the coherent $s1-sc$ line arcs have been constructed to inherently the 
inequalities (\ref{Vcp-s1}) being obeyed. 

On varying the virtual-electron pair coherent $s1-sc$ line arc nodal momentum
absolute value within the range $q_{arc}^N\in (q^{N}_{cp}, q^{N}_{Bs1})$ and thus that of the
$c$ fermions hole momenta within $q^h\in (q^{h}_{Fc}, q^{h}_{cp})$ 
the pairing energy $2\vert\Omega_{s1} ({\vec{q}}^{\,d}_{arc})\vert$ varies within the 
corresponding range
$2\vert\Omega_{s1} ({\vec{q}}^{\,d}_{arc})\vert\in (0,2\vert\Omega\vert\vert\sin 2\phi\vert)$.
It has the following limiting behaviors,
\begin{equation}
{\rm max}\,2\vert\Omega_{s1} ({\vec{q}}^{\,d}_{arc})\vert = 
2\vert\Omega_{s1} ({\vec{q}}^{\,d\,N}_{Bs1})\vert\vert_{T=0}
= 2\vert\Omega\vert\vert_{T=0} \, ; \hspace{0.35cm}
2\vert\Omega_{s1} ({\vec{q}}^{\,d}_{cp})\vert = 0 \, ; \hspace{0.35cm}
2\vert\Omega_{s1} ({\vec{q}}^{\,d}_{Bs1})\vert\approx 2\vert\Omega\vert\vert\sin 2\phi\vert \, .
\label{lim-omega-gene}
\end{equation}

\section{Concluding remarks}

In conventional superconductivity of isotropic 3D many-electron problems the objects that pair are the 
quasiparticles of the Fermi-liquid theory \cite{Schrieffer}. The low-energy eigenstates are described by 
occupancy configurations of such quasiparticles so that their momenta are good quantum numbers
and their interactions are residual. That property simplifies enormously the description of the many-electron 
physics \cite{Ander07}. In turn, in the VEP quantum liquid the charge $c$ fermions and spin-neutral 
two-spinon $s1$ fermions are for $x\in (x_c,x_*)$ the constituents 
of the phase-coherent virtual-electron pair configurations. The momentum occupancies of such quantum objects generate the 
energy eigenstates of the Hubbard model on the square lattice in the one- and two-electron subspace 
\cite{companion2}. For the VEP quantum liquid studied in this paper such $c$ and $s1$ band 
momenta are close to good quantum numbers. The $c-s1$ fermion interactions play a central role
in the pairing mechanism of that quantum liquid, consistently with the evidence that unconventional 
superconductivity is in different classes of systems mediated by magnetic fluctuations \cite{Yu-09,spin-super}.
 
Our studies consider the fluctuations of the phases associated with the long-range antiferromagnetic,  
short-range spin, and long-range superconducting orders occurring at zero temperature for
$x\in (0,x_0)$, $x\in (x_0,x_*)$, and $x\in (x_c,x_*)$, respectively. For $x\in (x_0,x_c)$
the short-range spin order coexists with Anderson insulating behavior
brought about by intrinsic disorder. For hole concentrations in
the range $x\in (x_c,x_*)$ the short-range spin order and long-range superconducting order
coexist. Fortunately, for that $x$ range the effects of intrinsic disorder or superfluid-density
anisotropy are very weak. This justifies the success of our oversimplified description 
of such effects in terms of a single suppression coefficient $\gamma_d^{min}=(1-\alpha_d)$. The
results of this paper focus on the physics associated with the corresponding hole concentration
range $x\in (x_c,x_*)$. (The effects of intrinsic disorder are much stronger for $x\in (0,x_c)$, as
shortly discussed in Appendix B.)

The exotic $d$-wave superconducting phase of the VEP quantum liquid
involves virtual-electron pairs whose phases are coherent. Those are constituted by one zero-momentum 
$c$ fermion pair and one spin-singlet two-spinon $s1$ fermion of momentum 
$\vec{q}\in Q^{s1}_{cp}$. Hence a virtual electron pair involves 
the two charges $-e$ of its $c$ fermions and the spin-singlet configuration
of the two spin-$1/2$ spinons of its composite $s1$ fermion. 
The phases of the virtual-electron pairs read $\theta=\theta_0 +\theta_1$.
Here $\theta_0$ are overall centre-of-mass phases and
the phases $\theta_{1}$ are related to the internal pairing degrees of freedom.
The corresponding macroscopic condensate of zero-momentum 
$c$ fermion pairs is associated with the phase coherence occurring
for $x\in (x_c,x_*)$. The fluctuations of the phases $\theta_0$ and $\theta_1$ become
large for $0<(x-x_c)\ll 1$ and $0<(x_*-x)\ll 1$, respectively.
For the intermediate interaction range $U/4t\in (u_0,u_1)$
the critical hole concentrations are given by $x_c\approx 0.05$ and $x_*\in (0.23,0.28)$.
For it such fluctuations are controlled by the weak 3D uniaxial anisotropy 
effects and electronic correlations through the mass ratios
$1/\varepsilon^2=M/m_{c}^*$ and $r_c =m_c^{\infty}/m_{c}^*\approx 2r_s\approx 2e^{-4t\,u_0/U}$, respectively.
The direct relation to the problem investigated in Ref. \cite{duality} has simplified
the study of the fluctuations of the phases $\theta_0$ and $\theta_1$.

The virtual-electron pairing energy is supplied to the zero-momentum $c$ fermion
pair by the short-range spin correlations associated with the spin-singlet spinon pairing.
This occurs through the $c$ - $s1$ fermion interactions within each virtual-electron pair. 
The interrelated spinon and $c$ fermion pairings involved
in the exotic coherent virtual-electron pairing are associated with the coexisting two-gap scenario
consistent with the unusual properties of the representative hole doped cuprates
\cite{two-gaps,Delta-no-super,two-gap-Tanaka,two-gap-Tacon}: A superconducting energy 
scale $2\vert\Omega\vert$ and pseudogap $2\vert\Delta\vert$, over the whole dome
$x\in (x_c,x_*)$, as illustrated in Fig. 1 of Ref. \cite{cuprates0}. The energy parameter
$2\vert\Delta\vert$ is the maximum magnitude reached at $\phi=0,\pi/2$ of the spinon
pairing energy $2\vert\Delta_{s1} ({\vec{q}}^{\,d}_{Bs1})\vert \approx 2\vert\Delta\vert\vert\cos 2\phi\vert$ 
appearing in Eq. (\ref{EF}). The energy scale $2\vert\Omega\vert$ 
is the maximum magnitude reached at $\phi=\pi/4$ of the superconducting virtual-electron pairing energy
 
$2\vert\Omega_{s1} ({\vec{q}}^{\,d}_{Bs1})\vert \approx 2\vert\Omega\vert
\vert\sin 2\phi\vert$ of Eq. (\ref{lim-omega-gene}). The latter pairing energy is associated with the effective coupling
of the corresponding zero-momentum coherent $c$ fermion pairs. Such zero-momentum $c$ 
fermion pairs couple to charge probes independently of 
the $s1$ fermions, which couple to spin probes. It follows that concerning charge excitations the zero-momentum
coherent $c$ fermion pairs behave as independent objects relative to the virtual-electron
pairs. In turn, the virtual-electron pairs exist in virtual intermediate states generated
by one-electron and spin excitations, which break such pairs. 

The pseudogap state occurs both for hole concentrations 
$x\in (x_0,x_c)$ at temperatures below the pseudogap temperature $T^*$
and for a temperature-dependent hole concentration range centered
at $x=x_{op}$ for finite temperatures in the range $T\in (T_c,T^*)$. For 
such a state the $s1$ fermion spinon pairing energy remains
finite and there are $c$ fermion pairing correlations, yet there is no phase coherence.   
Normal ground states may be reached by applying a magnetic field $H$ aligned 
perpendicular to the planes. Our results are inconclusive on whether for $H=0$ the ground state
of the $\varepsilon^2=m_{c}^*/M=0$ Hubbard model on the square lattice is superconducting.
They seem to indicate that some small 3D uniaxial anisotropy  
is needed for the emergence of superconductivity. Indeed, at constant $U/4t$ values
the hole-concentration width $(x_*-x_c)$ of the superconducting dome 
decreases upon decreasing the small 3D uniaxial anisotropy parameter 
$\varepsilon^2=m_{c}^*/M$. 

Concerning previous related studies on the large-$U$ Hubbard model 
and $t-J$ model and Heisenberg model on a square 
lattice involving for instance Jordan-Wigner 
transformations \cite{Wang} or the slave particle formalism 
\cite{2D-MIT,Ander87,Baska,Baska-01,Xiao-Gang}, here the single occupancy 
constraint is naturally implemented for all $U/4t$ finite values. Indeed,
the spin-$1/2$ spinons refer to the rotated electrons of the singly 
occupied sites of the energy eigenstates configurations 
\cite{companion1b,companion2}.
In the one- and two-electron subspace where the VEP quantum liquid  
is defined only the charge $c$ fermions and spin-neutral two-spinon $s1$ 
fermions play an active role. The main difference relative
to the above related schemes is that for them the spinless fermions 
arise from individual spin-$1/2$ spins
or spinons. In contrast, according to the studies of Ref.
\cite{companion2} the $s1$ fermions of the representation 
used in the studies of this paper emerge 
from an extended Jordan-Wigner transformation performed 
on spin-neutral two-spinon composite $s1$ bond particles. 
A property of physical importance for the unusual scattering
properties of the VEP quantum liquid investigated in Ref. \cite{cuprates} is that the $s1$ band
is full for $m=0$ ground states and contains a single hole
for one-electron excitations.

When expressed in terms of the rotated-electron operators, the VEP quantum liquid
microscopic Hamiltonian has basic similarities to that considered in Ref. \cite{duality} in terms of electron operators.
The advantage of our scheme is that it applies to intermediate $U/4t$ values. Indeed,
the results of Refs. \cite{cuprates0,cuprates} strongly suggest that the
physics of several classes of hole-doped cuprates with critical hole
concentrations $x_c\approx 0.05$ and $x_*\approx 0.27$ 
\cite{two-gaps,k-r-spaces,duality,2D-MIT,Basov,ARPES-review,Tsuei,pseudogap-review}
corresponds to $U/4t\approx u_* =1.525\in (u_0,u_1)$. 
The effective superconductivity approach for such cuprates introduced in 
this paper is consistent with their physics being closely related
to the doping of a Mott insulator \cite{2D-MIT}.
There are previous studies where spin-charge
interactions also lead to superconducting effective pair 
coupling \cite{Yu-09,Pines-91,Maki,Scala,Scala-09}. 
Our VEP quantum liquid acts on the one- and
two-electron subspace \cite{companion2}. For it rotated-electron double occupancy is exactly zero. 
This implies that the mechanisms behind its long-range superconducting order are different from those of the $SO(5)$ theory 
\cite{SO5}.

Finally, in Refs. \cite{cuprates0,cuprates} evidence is provided that the use of the theoretical VEP
quantum liquid scheme introduced in this paper leads to quantitative agreement with 
experiments concerning the universal properties of the cuprate superconductors   

%%%%%%%%%%%%%%%%%%%%%%%%%%%%%%%%%%%%%%%%%%%%%%%%%%%%%%%%%%%%%%%%%%%%%%%%%%
\begin{acknowledgments}
I thank M. A. N. Ara\'ujo, J. Chang, A. Damascelli, R. G. Dias, F. Guinea, P. Horsch, 
P. A. Lee, J. Mesot, C. Panagopoulos, T. C. Ribeiro, P. D. Sacramento, M. J. Sampaio, M. Shi, 
Z. Te${\rm\check{s}}$anovi\'c, and K. Yamada for discussions. The support of the ESF Science Program 
INSTANS and grant PTDC/FIS/64926/2006 is acknowledged.  
\end{acknowledgments}
%%%%%%%%%%%%%%%%%%%%%%%%%%%%%%%%%%%%%%%%%%%%%%%%%%%%%%%%%%%%%%%%%%%%%%%%%%
%%%%%%%%%%%%%%%%%%%%%%%%%%%%%%%%%%%%%%%%%%%%%%%%%%%%%%%%%%%%%%%%%%%%%%%%%%
\appendix

\section{Basic information on the $c$ and $s1$ fermion description}

In this Appendix some basic information on the $c$ and $s1$ fermion description
used in the studies of Refs. \cite{companion2,companion1b} is provided.
The critical hole concentration $x_*$ expression given in Eq. (\ref{x-c}) refers
to the interaction range $U/4t>u_0\approx 1.302$ and involves the spin ratio $r_s$. That ratio can be
defined for the whole range of $U/4t$ values and increases smoothly upon 
increasing $U/4t$. The charge and spin ratios are for the approximate range $U/4t\in (u_0, u_1)$
given by $r_c\approx 2r_s\approx 2e^{-4t\,u_0/U}$. Here $u_1\approx 1.600$. For $U/4t>0$ such ratios
have the following limiting behaviors,
\begin{eqnarray}
r_c & \equiv & {m^{\infty}_c \over m^{*}_c} 
\, ; \hspace{0.35cm} 
r_s \equiv {\Delta_0\over 4W_{s1}^0} = e^{-\lambda_s} \, ,  
\nonumber \\
r_c & = & r_s = 0
\, , \hspace{0.10cm} U/4t\rightarrow 0
\, ; \hspace{0.35cm} r_s \approx e^{-\pi\sqrt{4t/U}}
\, , \hspace{0.10cm} U/4t\ll 1 \, ,
\nonumber \\
r_c & \approx & 2r_s \approx 2\,e^{-1} \approx 0.736 
\, , \hspace{0.35cm} U/4t \approx u_0\approx 1.302 \, ,
\nonumber \\
r_c & \approx  & 2r_s \approx 2e^{-4t\,u_0/U} \, , 
\hspace{0.35cm} U/4t\in (u_0, u_1)  \, ,
\nonumber \\ 
r_c & = & r_s = 1
\, , \hspace{0.35cm} U/4t\rightarrow
\infty \, .
\label{m*c/mc-UL}
\end{eqnarray}
For the range $U/4t\in (u_0, u_1)$ the critical hole concentration $x_*$ is given by $x_*\approx 2r_s/\pi$.
For $U/4t\geq u_0$ its limiting magnitudes are, 
\begin{eqnarray}
x_*  & \approx  & {2\over\pi}\,e^{-4t\,u_0/U} \, , 
\hspace{0.50cm} U/4t\in (u_0, u_1)  \, ,
\nonumber \\ 
& \approx & 0.23 \, , \hspace{0.50cm} U/4t = u_0 \approx 1.302 \, ,
\nonumber \\
& \approx &  0.27 \, , \hspace{0.50cm} U/4t \approx 1.525 \, ,
\nonumber \\
& \approx & 0.28 \, , \hspace{0.50cm} U/4t = u_1 \approx 1.600 \, ,
\nonumber \\
& = & {1\over \pi} \approx
0.32 \, , \hspace{0.50cm} U/4t= u_{\pi}>u_1  \, ,
\nonumber \\
& \leq & {2\over \pi} \approx
0.64 \, , \hspace{0.50cm} U/4t\rightarrow\infty  \, .
\label{xc-range}
\end{eqnarray}

An important energy scale is the
the $x=0$ and $m=0$ Mott-Hubbard gap $2\mu^0$ 
given by \cite{companion2},
\begin{eqnarray}
2\mu^0 & \approx & 64\,t\,e^{-\pi\sqrt{4t\over U}} \, ,
\hspace{0.35cm} U/4t\ll 1 \, ,
\nonumber \\
& \approx & {4e^1\,t\over \pi}\sqrt{1+(U/4t -u_0)}  \, ,
\hspace{0.35cm} U/4t\in (u_0, u_1) \, ,
\nonumber \\
& \approx  & [U - 8t] \, ;
\hspace{0.35cm}  U/4t\gg 1 \, .
\label{DMH}
\end{eqnarray}

Another energy scale that plays a key role in our studies
is that of Eq. (\ref{Delta-0-gen}), $\lim_{x\rightarrow 0}\vert\Delta\vert =\Delta_0$. It
has the following approximate limiting behaviors \cite{companion2},
\begin{eqnarray}
\Delta_0 & \approx &16t\,e^{-\pi\sqrt{4t/U}} \, ,
\hspace{0.35cm}  U/4t\ll 1 \, ,
\nonumber \\
& = & {\rm max}\,\{\Delta_0\} \approx t/\pi \, ,
\hspace{0.35cm}  U/4t = u_0 \, ,
\nonumber \\
& \approx & e^{(1-4t\,u_0/U)}[t/\pi]\,[1 - (U/4t-u_0)\,e^{-{u_*-U/4t\over u_*-u_0}\ln (u_0)}] \, ,
\hspace{0.35cm} U/4t \in (u_0,u_1) \, ,
\nonumber \\
& \approx & \pi\,[4t]^2/U \, ,
\hspace{0.35cm} U/4t\gg 19 \, ,
\label{Delta-0}
\end{eqnarray}
where $u_0\approx 1.3$, $u_*=1.525$, and $u_1\approx 1.6$. The $U/4t$ dependence reported here for approximately $U/4t \in (u_0,u_1)$ corresponds
to an interpolation function used to connect the following limiting
behaviors valid for $0\leq (U/4t-u_0)/(u_1-u_0)\ll 1$ and $U/4t\approx u_*=1.525$ \cite{companion2},
\begin{eqnarray}
\Delta_0 & \approx & e^{(1-4t\,u_0/U)}{t\over \pi}\left(2-{U\over 4t\,u_0}\right) \, ,
\hspace{0.35cm}  0\leq (U/4t-u_0)/(u_1-u_0)\ll 1 \, ,
\nonumber \\
& \approx & e^{(1-4t\,u_0/U)}{t\over \pi}\,\left[1 - \left({U\over 4t}-u_0\right)\right] \, ,
\hspace{0.35cm} U/4t \approx u_* =1.525 \, ,
\label{Delta-0-limits}
\end{eqnarray}
respectively. 

Let us consider a Brillouin zone centered at
the momentum $-\vec{\pi}$. The hole Fermi momentum $\vec{k}_F^h$ is related 
to the Fermi momentum $\vec{k}_F$ as follows, 
\begin{eqnarray}
\vec{k}^{\,h}_F & = & \vec{k}_F +\vec{\pi} = k_F^h (\phi)\,{\vec{e}}_{\phi} 
\, ; \hspace{0.35cm}
\phi = \arctan \left({k^h_{Fx_2}\over k^h_{Fx_1}}\right) \, ,
\nonumber \\
\phi & \in & \left(\phi_{AN}, {\pi\over 2} -\phi_{AN}\right) 
\, ; \hspace{0.25cm} \phi_{AN}=0 \, , \hspace{0.10cm}
x\leq x_h \, .
\label{phi-F}
\end{eqnarray}
Here the expression of the Fermi angle $\phi$, which defines the direction of the 
hole Fermi momentum $\vec{k}_F^h$, also holds for $\phi\in (0,2\pi)$. The angle
$\phi_{AN}$ appearing in Eq. (\ref{phi-F}) is small for the hole concentration range $x\in (x_h,x_*)$ for which the
Fermi line is particle like. The precise magnitude of the hole concentration $x_h>x_{c2}$ remains unknown.
The studies of Ref. \cite{companion2} consider that it belongs to the range $x_h \in (x_{c2},x_*)$. 
The hole Fermi momentum $\vec{k}_F^h$ can be expressed in terms of
the $c$ Fermi line hole momenta and $s1$ boundary momenta as
given in Eq. (\ref{kF-qFc-qBs1}). For $x\in (x_{c1},x_{c2})$ the
corresponding angles $\phi^{d}_c$ and $\phi^{d}_{s1}$  are provided  
in Eq. (\ref{phiF-c-s1}).

For one-electron excited states, the $s1$ band momentum,
\begin{equation}
{\vec{q}}_{Bs1} \equiv [A^{d}_{s1}]^{-1}\,{\vec{q}}^{\,d}_{Bs1}
= q_{Bs1}(\phi)\,{\vec{e}}_{\phi+\pi} \, .
\label{qFc-qBs1}
\end{equation}
is called auxiliary momentum of the $s1$ boundary-line momentum
${\vec{q}}^{\,d}_{Bs1}$. Both for $x<x_{c1}$ and $x>x_{c2}$ the form
of the matrix $[A^{d}_{s1}]^{-1}$ is not known. For approximately $x\in (x_{c1},x_{c2})$ it
reads $[A^{d}_{s1}]^{-1}\approx A^{-d}_{F}$ where the $F$ rotation matrix $A^{d}_{F}$ 
is given in Eq. (\ref{A-c-s1}).

Among the $s1$ boundary-line momenta ${\vec{q}}^{\,d}_{Bs1}$, it is useful to consider 
those whose corresponding auxiliary $s1$ boundary-line momenta ${\vec{q}}_{Bs1}$ point in the nodal and anti-nodal directions.
The nodal and anti-nodal $s1$ boundary-line momenta
${\vec{q}}^{\,d\,N}_{Bs1}$ and ${\vec{q}}^{\,d\,AN}_{Bs1}$, respectively, are defined as those
whose corresponding auxiliary momenta  ${\vec{q}}^{N}_{Bs1}$ and ${\vec{q}}^{AN}_{Bs1}$
have for instance for the quadrant such that $q_{0x_1}\leq 0$ and $q_{0x_2}\leq 0$ 
the following Cartesian components,
\begin{equation} 
{\vec{q}}^{N}_{Bs1} =  
-\left[\begin{array}{c}
q^N_{Bs1}/\sqrt{2} \\ 
q^N_{Bs1}/\sqrt{2}
\end{array} \right]
\, ; \hspace{0.5cm}
{\vec{q}}^{AN}_{Bs1} =
- \left[\begin{array}{c}
q^{AN}_{Bs1} \\ 
0
\end{array} \right] \, ; 
- \left[\begin{array}{c}
0 \\
q^{AN}_{Bs1} 
\end{array} \right] \, .
\label{q-A-q-AN-c-s1}
\end{equation}
Here $q^N_{Bs1}$ and $q^{AN}_{Bs1}$ are the
absolute values of both the auxiliary momenta ${\vec{q}}^{N}_{Bs1}$ and ${\vec{q}}^{AN}_{Bs1}$
and corresponding momenta ${\vec{q}}^{\,d\,N}_{Bs1}$ and ${\vec{q}}^{\,d\,AN}_{Bs1}$,
respectively. 

For $m=0$ ground states and their two-electron excited states the $s1$ boundary-line momentum 
reads ${\vec{q}}^{\,d}_{Bs1}={\vec{q}}_{Bs1}=q_{Bs1}(\phi)\,{\vec{e}}_{\phi+\pi}$ for
$x \in (x_c,x_*)$. Thus it equals the auxiliary momentum. For the sake of generality, we often use the 
notation ${\vec{q}}^{\,d}_{Bs1}$ for the $s1$ boundary-line momenta of the $s1$ fermion 
occupancy configurations of these states. For such states the hole $c$ Fermi momentum 
is independent of the doublicity $d$ introduced in Ref. \cite{companion2}. It is given by,
\begin{equation} 
{\vec{q}}_{Fc}^{\,h} = {\vec{q}}_{Fc} + \vec{\pi} 
= q^h_{Fc} (\phi)\,{\vec{e}}_{\phi_c} 
\, ; \hspace{0.35cm} \phi_{c} = \phi \, .
\label{q-Fc-h-m0}
\end{equation}

The $c$ and $s1$ energy dispersions 
appearing in the general energy functional of Eq. (\ref{DeltaE-plus})
depend on the Cartesian components of the $c$ band hole momentum and
$s1$ band momentum through the elementary functions $e_c (q)$ and $e_{s1} (q)$,
respectively. Those are known in some limits \cite{companion2}. 
The $c$ fermion energy dispersion $\epsilon_{c} ({\vec{q}}^{\,h})$ 
is for $U/4t>0$, $m=0$, and $x\in (x_c,x_*)$ given by,
\begin{equation}
\epsilon_c ({\vec{q}}^{\,h}) = \epsilon_c^0 ({\vec{q}}^{\,h}) =
\sum_{i=1,2}[e_{c} (q_{x_i}^{h}) - e_{c} (q_{Fc x_i}^{h\,d})] \, .
\label{c-band}
\end{equation}
For $x\in (x_c,x_*)$ and $U/4t\geq u_0$ the $c$ Fermi line is approximately circular
and the $c$ fermion energy dispersion (\ref{c-band}) and the chemical potential $\mu$
are given by \cite{companion2},
\begin{equation}
\epsilon_{c} (\vec{q}^{\,h}) \approx -{\vert{\vec{q}}^{\,h}\vert^2-\vert{\vec{q}}^{\,h\,d}_{Fc}\vert^2\over 
2m^{*}_{c}} \, ; \hspace{0.35cm} \vert\epsilon_{c} (\vec{q}^{\,h}) \vert<W_c^h\vert_{x=x_*}
 \, ; \hspace{0.35cm} W_c^h \approx {2x\pi\over m_c^*} 
\, ; \hspace{0.35cm} 
\mu \approx \mu_0 + W_c^h \, .
\label{bands}
\end{equation}
Provided that $\mu_0$ is replaced by the energy scale ${\breve{\mu}}^0$
given in Eq. (\ref{mu-x-AL}) of Appendix B, the $\mu$ expression given here is a 
good approximation for the VEP quantum liquid at $x\in (x_c,x_*)$ and $U/4t\in (u_0,u_1)$.

The $s1$ fermion energy dispersion
$\epsilon^0_{s1} ({\vec{q}})$ and pairing energy per 
spinon $\vert\Delta_{s1} ({\vec{q}})\vert$ appearing
in Eq. (\ref{band-s1}) are given by,
\begin{equation}
\epsilon^0_{s1} ({\vec{q}}) = 
\epsilon^{0,\parallel}_{s1} (\vec{q}_0) \, ;
\hspace{0.35cm}
\vert\Delta_{s1} ({\vec{q}})\vert = \vert\Delta^{\parallel}_{s1} ({\vec{q}}_0)\vert \, .
\label{bands-bipt}
\end{equation}
Here ${\vec{q}}_0$ obeys Eq. (\ref{qs1-qhc}),
$\epsilon^{0,\parallel}_{s1} (\vec{q}_0) = \sum_{i=1,2}
[e_{s1} (q_{0x_i}) - e_{s1} (q_{Bs1 x_i})]$,
$ \vert\Delta^{\parallel}_{s1} ({\vec{q}}_0)\vert = {\breve{g}}_1\,\Delta_0\,
F^{\parallel}_{s1} ({\vec{q}}_0)$, 
$F^{\parallel}_{s1} ({\vec{q}}_0) = [\vert e_{s1} (q_{0x_1}) 
- e_{s1} (q_{0x_2})\vert]/W_{s1}$, and
the energy bandwidth $W_{s1}$ 
decreases for increasing values of $U/4t$.
 
The elementary function $e_{s1} (q)$ 
is such that $\Delta_{s1} ({\vec{q}}^{\,d\,N}_{Bs1})=0$,
$\vert e_{s1} (q^{AN}_{Bs1}) - e_{s1} (0)\vert =W_{s1}$,
and $F^{\parallel}_{s1} ({\vec{q}}_{Bs1}) \approx 
\vert\cos 2\phi\vert$. Hence at $T=0$ the maximum magnitude $2\vert\Delta\vert$ of the spinon pairing energy
$2\vert\Delta_{s1} ({\vec{q}})\vert$ is reached at ${\vec{q}}={\vec{q}}^{\,d\,AN}_{Bs1}$,
\begin{equation}
2\vert\Delta\vert\vert_{T=0} = {\breve{g}}_1\,2\Delta_0 = 2\vert\Delta_{s1} ({\vec{q}}^{\,d\,AN}_{Bs1})\vert
\, ; \hspace{0.15cm} x \in (x_c, x_*) \, , \hspace{0.10cm} m=0 \, .
\label{D-x}
\end{equation}

The shapes of the $c$ Fermi line and
$s1$ boundary line are fully determined by the 
form of the auxiliary energy dispersions $\epsilon^0_{c} (\vec{q}^{\,h})$
and $\epsilon^0_{s1} ({\vec{q}})$ given in Eqs.
(\ref{c-band}) and (\ref{bands-bipt}), respectively, as follows,
\begin{eqnarray}
& & {\vec{q}}_{Fc}^{\,h\,d} \in {\rm hole} 
\hspace{0.10cm} c \hspace{0.10cm} {\rm Fermi} 
\hspace{0.10cm} {\rm line}
 \hspace{0.15cm} \Longleftrightarrow  \hspace{0.10cm}
\epsilon^0_{c} ({\vec{q}}_{Fc}^{\,h\,d})=0 \, ,
\nonumber \\
{\vec{q}}^{\,d}_{Bs1} & \in & s1 \hspace{0.10cm} {\rm boundary} 
\hspace{0.10cm} {\rm line}
\hspace{0.15cm} \Longleftrightarrow  \hspace{0.10cm}
\epsilon^0_{s1} ({\vec{q}}^{\,d}_{Bs1})=0 \, .
\label{g-FS}
\end{eqnarray}

The $c$ and $s1$ fermion group velocities derived from the 
energy dispersions of Eqs. (\ref{band-s1}), (\ref{c-band}), (\ref{bands}),
and (\ref{bands-bipt}) read, 
\begin{equation}
{\vec{V}}_{c} (\vec{q}^{\,h}) = {\vec{\nabla}}_{\vec{q}^{\,h}}\,\epsilon_{c} (\vec{q}^{\,h})
\, ; \hspace{0.35cm}
{\vec{V}}^{\,\Delta}_{s1} (\vec{q}) =
-{\vec{\nabla}}_{\vec{q}}\vert\Delta_{s1} ({\vec{q}})\vert \, ; \hspace{0.35cm}
{\vec{V}}_{s1} (\vec{q}) = {\vec{\nabla}}_{\vec{q}}\,\epsilon_{s1} (\vec{q}) 
\, ; \hspace{0.35cm}
{\vec{V}}^{\,0}_{s1} (\vec{q}) =
{\vec{\nabla}}_{\vec{q}}\,\epsilon^0_{s1} (\vec{q}) \, .
\label{g-velocities-def}
\end{equation}
We call their unit vectors ${\vec{e}}_{\phi_{s1}(\vec{q})}$, ${\vec{e}}_{\phi_{s1}^{0}(\vec{q})}$,
${\vec{e}}_{\phi_{s1}^{\Delta}(\vec{q})}$, and ${\vec{e}}_{\phi_{c}(\vec{q}^{\,h})}$,
respectively.

Finally, the $c$ fermion elementary charge current is given by \cite{companion2},
\begin{equation}
{\vec{j}}_c ({\vec{q}}^{\,h}) = -e\,\alpha_U\,{\vec{V}}_{c} ({\vec{q}}^{\,h}) \, ; \hspace{0.35cm}
{\vec{j}}_c ({\vec{q}}^{\,h}_{Fc}) = -e\,{q^{h}_{Fc}\over m_c^{\rho}}\,{\vec{e}}_{\phi_{c}+\pi} \, ; \hspace{0.35cm}
\alpha_U \equiv {m_c^*\over m_c^{\rho}} \approx
{1\over r_c^2} \, , \hspace{0.35cm} U/4t>u_0\approx 1.302 \, .
\label{jc}
\end{equation}
Here ${\vec{V}}_{c} (\vec{q}^{\,h})$ is the $c$ fermion velocity of Eq.
(\ref{g-velocities-def}), $m_c^{\rho}$ a renormalized transport
mass, and $\phi_{c} = \phi$.

\section{The hole-trapping effects}

In this Appendix the strong effects of intrinsic disorder on the square-lattice quantum-liquid physics of
Ref. \cite{companion2} in the hole concentration
range $x \in (0,x_c)$ are briefly discussed. In contrast to the range $x\in (x_c,x_*)$ referring to
the problem studied in this paper, the effects reported here change such a physics qualitatively.

As discussed in the following, intrinsic disorder is strongest in the range $x\in (0,x_0)$.
It has qualitatively different effects in that range and in the range
$x\in (x_0,x_c)$, respectively. Here $x_0 = x_*/[1 + (2\pi x_*)^2\,(t/\Delta_0)]$ 
is the hole concentration of Eq. (\ref{x-h}) at which the Fermi-energy anisotropy coefficient 
$\eta_0= \vert\Delta\vert/W_c^h$ given in that equation reads $\eta_0=1$ for the square-lattice quantum liquid. 
It is given by $x_0 \approx 0.013$ and $x_0 \approx 0.024$ for $x_*=0.27$, $t=295$\,meV, and
the magnitudes $\Delta_0=42$\,meV and $\Delta_0=84$\,meV found in Ref. \cite{cuprates0} for LSCO and 
the remaining representative systems, respectively.
For the square-lattice quantum liquid of Ref. \cite{companion2} 
there occurs a quantum phase transition from a Mott-Hubbard insulator with long-range 
antiferromagnetic order at the hole concentration $x=0$ to a short-range incommensurate-spiral spin 
ordered state for $0<x\ll 1$ \cite{companion2,AleMura}. An energy scale that could be used to characterize such
a transition is the $c$ fermion unfilled sea bandwidth $W_c^h$ of Eq. (\ref{bands}) of Appendix A. It is associated with the 
hole kinetic energy. For the square-lattice quantum liquid it vanishes at $x=0$ and is finite for finite hole concentrations. 

Within the effective and simplified description of the problem considered here, for hole concentrations 
in the range $x\in (0,x_c)$ the holes are trapped in the vicinity of randomly distributed impurities, 
so that their kinetic energy vanishes and the $c$ fermion hole energy bandwidth $W_c^h$ is not well defined.  
Nevertheless, the critical hole concentration $x_0$ of the N\'eel-state - spin-glass-state sharp quantum-phase transition
is that at which the ratio $\vert\Delta\vert/W_c^h$ of the corresponding square-lattice quantum liquid is given by one. Such
a ratio is for the Fermi-energy anisotropy coefficient $\eta_0= \vert\Delta\vert/W_c^h$ of Eq. (\ref{x-h})
introduced in Ref. \cite{companion2}. In turn, the absolute value $\vert\epsilon_{s1} (\vec{q})\vert$ of the $s1$ 
fermion energy dispersion of Eq. (\ref{bands}) of Appendix A changes from 
$\vert\epsilon_{s1} ({\vec{q}}^{\,d\,N}_{Bs1})\vert=0$ for $s1$ boundary-line auxiliary momenta pointing in the
nodal directions to $\vert\epsilon_{s1} ({\vec{q}}^{\,d\,AN}_{Bs1})=\vert\Delta\vert$ for such
momenta pointing in the anti-nodal directions. Therefore, the energy 
parameter $\vert\Delta\vert$ gives a measure of the $s1$ boundary-line anisotropy. 

Consistently with $\vert\Delta\vert$ providing a measure of the $s1$ boundary-line anisotropy, for the 
square-lattice quantum liquid at hole concentrations below $x_0$ for which $\vert\Delta\vert>W_c^h$ 
the corresponding anisotropy effects dominate over those of the isotropic $c$ Fermi line. 
It follows from Eq. (\ref{kF-qFc-qBs1}) that the Fermi line involves contributions from both that line and the $s1$ 
boundary line. Hence for the square-lattice quantum liquid the Fermi line is strongly anisotropic for hole concentrations below $x_0$. 
This weakens the hole kinetic-energy effects associated 
with the energy scale $W_c^h$. The hole-trapping effects profit from such weakening of the hole kinetic-energy 
effects: They give rise to the extension of the long-range antiferromagnetic order 
from $x=0$ to $x\in (0,x_0)$ for the VEP quantum liquid. As a result, the critical concentrations $x_0$ and $x_c$ are those at 
which the energy scales $\vert\Delta\vert$ and $W_c^h$ emerge for the latter quantum problem, being ill 
defined for hole concentrations below $x_0$ and below $x_c$, respectively. In contrast, for the
square-lattice quantum liquid they are well defined for the whole range $x\in (0,x_*)$.

The hole-trapping effects lead to two related phenomena: (i) the extension of the long-range antiferromagnetic 
order of the Mott-Hubbard insulator from $x=0$ to the range $x\in (0,x_0)$; (ii) The 
coexistence for $x\in (x_0,x_c)$ of the short-range spin order emerging at $x=x_0$ with Anderson insulating behavior. 
The latter behavior is brought about by intrinsic disorder. For hole concentrations in the range $x\in (x_0,x_c)$
such a disorder and associate hole-trapping effects are not strong enough to remove the 
short-range spin order. Finally, the only effect of the hole-trapping effects on the long-range superconducting order 
of the quantum liquid with weak 3D uniaxial anisotropy is 
shifting its emergence by $\delta x_c=x_0$. This changes the magnitude of the critical hole
concentration $x_c$ from $x_c ={\rm Gi}$ for that quantum liquid to $x_c ={\rm Gi} + x_0$
for the VEP quantum liquid. In turn, the magnitude of the hole concentration $x_*$ remains unaltered. 

For the square-lattice quantum liquid the hole concentrations $x_0$ and $x_{c1}=1/8$ of Eq. (\ref{x-h}) do not mark sharp quantum
phase transitions. For intermediate values of $U/4t$ they refer instead to crossovers between hole-concentration ranges
where the interplay of Fermi-line anisotropy and electronic correlations leads to a different physics.
Here that remains true for $x_{c1}=1/8$. In turn, the hole-tapping effects
render $x_0$ a critical hole concentration. It marks the sharp quantum phase transition between the
Mott-Hubbard insulator with long-range antiferromagnetic order and the Anderson insulator
with short-range spiral incommensurate spin order. 
\begin{table}
\begin{tabular}{|c|c|c|c|c|c|c|c|c|c|c|} 
\hline
x & $0.05$ & $0.11$ & $1/8$ & $0.15$ & $0.16$ & $0.21$ & $0.23$ & $0.24$ & $0.27$ \\
\hline
$\eta_{\Delta}$ & $0.12$ & $0.06$ & $0.05$ & $0.04$ & $0.03$ & $0.02$ & $0.01$ & $0.01$ & $0$ \\
\hline
$\eta_0$ & $0.44$ & $0.14$ & $0.11$ & $0.08$ & $0.07$ & $0.03$ & $0.02$ & $0.01$ & $0$ \\
\hline
\end{tabular}
\caption{Magnitudes of the Fermi-velocity anisotropy coefficient $\eta_{\Delta}$ and
Fermi-energy anisotropy coefficient $\eta_0$ of Eq. (\ref{x-h}) for several hole concentrations $x$,
$U/4t\approx 1.525$, and thus $\Delta_0\approx 0.285\,t$. It is considered that the 
Fermi-velocity anisotropy is small when $\eta_{\Delta}<2x_0\approx 0.05$. Consistently, the crossover hole
concentration $x_{c1}= 1/8$ approximately is that at which $\eta_{\Delta}\approx 2x_0$ for 
$U/4t\in (u_0,u_1)$. The magnitudes given here refer both to the square-lattice quantum liquid and the four representative
systems other than LSCO.}
\label{table5}
\end{table} 

The dependence (and independence) on the Fermi angle $\phi$ of the $s1$ fermion velocity $V^{\Delta}_{Bs1}$ of Eq. (\ref{pairing-en-v-Delta}) 
(and $c$ fermion velocity $V_{Fc}$ also given in that equation) implies an anisotropic (and isotropic) character for the $s1$ boundary line
(and $c$ Fermi line.) The Fermi angle $\phi$ given in Eq. (\ref{phi-F}) of Appendix A defines the direction of the hole Fermi momentum 
$\vec{k}_F^h$ also provided in that equation. According to the analysis of Ref. \cite{companion2},
the Fermi line anisotropy is measured both by the Fermi-velocity anisotropy 
coefficient $\eta_{\Delta}= {\rm max}\,\{r_{\Delta}\}= {\rm max}\,\{V^{\Delta}_{Bs1}\}/V_{Fc}$ and Fermi-energy 
anisotropy coefficient $\eta_0= {\rm max}\,\{\delta E_F\}/W_c^h$. For the square-lattice and VEP quantum liquids, 
those are well defined for the ranges $x\in (0,x_*)$ and $x\in (x_c,x_*)$, respectively. 
The velocity $V^{\Delta}_{Bs1}$ and anisotropic part of the Fermi energy $\delta E_F$
of Eq. (\ref{DE-3proc}) achieve their maximum magnitudes
for $s1$ band auxiliary momenta pointing in the nodal and anti-nodal directions, respectively. Hence the coefficients 
$\eta_{\Delta}$ and $\eta_0$ contain complementary yet different information. 
For approximately $U/4t \in (u_0,u_1)$ they are given in Eq. (\ref{x-h}).
The hole concentration $x_0$ is defined in that equation as that at which $\eta_0=1$ for the square-lattice quantum liquid.
For such an intermediate $U/4t$ range it is considered that the Fermi-velocity anisotropy is small 
for hole concentrations larger than $x_{c1}=1/8$. 

For $U/4t\approx 1.525$ the expressions provided in 
Eq. (\ref{x-h}) lead to $x_0\approx 0.024$ for the square-lattice quantum liquid for which $\Delta_0/t\approx 0.285$ 
and four representative systems other than LSCO and to $x_0\approx 0.013$  
for LSCO. Indeed, according to the analysis of Ref. \cite{cuprates0}, $\Delta_0/t\approx 0.142$ for the latter
system. The magnitudes of the coefficients $\eta_{\Delta}$ and $\eta_0$ are  
for $U/4t\approx 1.525$ and several values of $x$ given in 
Table \ref{table5} for these four representative systems and in 
Table \ref{table6} for LSCO. As given in the latter table, the anisotropy coefficients are smaller for LSCO. However, intrinsic 
disorder and randomness are larger for that compound. The Fermi-velocity anisotropy is then considered 
small at a smaller $\eta_{\Delta}=2x_0\approx 0.03$ value than that $\eta_{\Delta}=2x_0\approx 0.05$
of the other four representative systems. Indeed, due to
the LSCO larger intrinsic disorder and randomness, an anisotropy associated with $\eta_{\Delta}\approx 0.05$ 
has more impact on the physics than for the remaining four representative cuprates.
However, in both cases the hole concentration at which $\eta_{\Delta}=2x_0\approx 0.03$
and $\eta_{\Delta}=2x_0\approx 0.05$, respectively, is approximately $x_{c1}= 1/8$. 

As confirmed by the data of the Tables \ref{table5}
and \ref{table6}, the coefficient $\eta_0$ decreases upon increasing $x$ slower than $\eta_{\Delta}$.
For the VEP (and square-lattice) quantum liquid the Fermi-line anisotropy is strongest at $x=x_c$ 
(and for $x\rightarrow 0$), when $\eta_{\Delta}$ and $\eta_{0}$ are largest (and $\eta_{\Delta},\eta_{0}\rightarrow\infty$).
Such an anisotropy vanishes for $x\rightarrow x_*$, when $\eta_{\Delta},\eta_{0}\rightarrow 0$ and
the Fermi line becomes fully isotropic. Following the $x$ dependence of $\eta_{\Delta}$ and $\eta_0$, it 
is considered that the Fermi line is anisotropic for 
hole concentrations in the range $x\in (x_c,x_{c1})$. For the $x$ range $x\in (x_{c1},x_{c2})$ it has some 
Fermi-energy anisotropy, yet the Fermi velocity is nearly isotropic. The Fermi line is nearly isotropic for 
the hole concentration range $x\in (x_{c2},x_*)$. That for $x\rightarrow x_*$ the Fermi line becomes as isotropic as 
that of an isotropic Fermi liquid is consistent with the physics tending to that of a Fermi liquid 
as that critical hole concentration is approached.

Both for the $x$ range $x\in (0,x_c)$ and hole concentrations obeying the inequality $0<(x-x_c)\ll 1$ 
the hole-trapping effects change the chemical potential $\mu \approx \mu_0 + W_c^h$ of
Eq. (\ref{bands}) of Appendix A. For the range $x\in (0,x_c)$ it is pinned 
and given approximately by $\mu \approx \mu^0$. In turn, for $0<(x-x_c)\ll 1$ it is shifted by
$-\delta \mu \approx -x_c 2\pi/m^*_c = -4\pi r_c x_c t$. The emergence of its dependence on $x$ is then shifted from 
the hole concentration $x=0$ to $x=x_c$. The suppression of the
chemical potential shift is for the hole concentration range $x\in (0,x_c)$ due to its pinning by the intrinsic-disorder
impurity potential. For $U/4t\geq u_0\approx 1.302$ it is approximately given by,
\begin{eqnarray}
\mu & \approx & \mu^0 + [W_c^h -\delta\mu]\theta (x-x_c) \approx
\mu^0 + {(x-x_c)2\pi\over m^*_c}\,\theta (x-x_c) \, , \hspace{0.35cm} 0\leq x\leq x_c
\hspace{0.10cm}{\rm and}\hspace{0.10cm} 0<(x-x_c)\ll 1 \, ,
\nonumber \\
\mu & \approx & {\breve{\mu}}^0 + W_c^h \, ; \hspace{0.35cm} 
{\breve{\mu}}^0 = \mu^0 - \delta \mu 
\, ; \hspace{0.35cm}
\delta \mu = 4\pi r_c x_c t \, , \hspace{0.35cm} 
x\in (x_c,x_*) \, .
\label{mu-x-AL}
\end{eqnarray}
Here the second chemical potential 
expression refers to the hole-concentration range $x\in (x_c,x_*)$ mostly considered in the studies of
this paper. For it the only modification of the chemical potential 
$\mu \approx \mu_0 + W_c^h$ of Eq. (\ref{bands}) of Appendix A is the replacement of the $x$
independent energy term $\mu_0$ by ${\breve{\mu}}^0 = [\mu^0 - 4\pi r_c x_c t]$,
the $x$ dependent term $W_c^h$ remaining unaltered. In turn, the first $\mu$ expression
of Eq. (\ref{mu-x-AL}) is consistent with for hole concentrations below $x_c$ the
energy scale $W_c^h$ being ill defined for the VEP quantum liquid. 
For the particular case of LSCO, in addition to suppressing
the magnitude of the energy scale $\Delta_0$ by a factor two, the cation-random
effects considered in Ref. \cite{cuprates0} cause a further pinning of the chemical potential up to approximately $x\approx 0.14$
and lessen the magnitude of $W_c^h$. Hence the formula (\ref{mu-x-AL}) does not apply to that 
random alloy.
\begin{table}
\begin{tabular}{|c|c|c|c|c|c|c|c|c|c|c|} 
\hline
x & $0.05$ & $0.09$ & $1/8$ & $0.15$ & $0.16$ & $0.20$ & $0.21$ & $0.23$ & $0.27$ \\
\hline
$\eta_{\Delta}$ & $0.06$ & $0.04$ & $0.03$ & $0.02$ & $0.02$ & $0.01$ & $0.01$ & $0.01$ & $0$ \\
\hline
$\eta_0$ & $0.22$ & $0.10$ & $0.06$ & $0.04$ & $0.03$ & $0.02$ & $0.01$ & $0.01$ & $0$ \\
\hline
\end{tabular}
\caption{Magnitudes of the same coefficients as in Table \ref{table5} for several hole concentrations $x$,
$U/4t\approx 1.525$, and $\Delta_0\approx 0.142\,t$ for LSCO. Again, it is considered that the Fermi-velocity anisotropy is small when 
$\eta_{\Delta}<2x_0$. In the present case $2x_0\approx 0.03$ also implies that $x\approx x_{c1}= 1/8$ for $U/4t\in (u_0,u_1)$.}
\label{table6}
\end{table} 

\section{Hamiltonian terms that control the fluctuations
of the $c$ fermion-pair phases}

Here it is shown that the Hamiltonian terms (\ref{H-bonds-c-s1-b}) are 
approximately equivalent to those given in Eq. (\ref{H-bonds-c-s1-f}).
In order to reach that goal we start by expressing (\ref{H-bonds-c-s1-b}) in terms of the phases
$\theta_{j}$,
\begin{equation}
{\hat H}^{bonds}_{eff} = 
\sum_{j=1}^{N_{s1}} e^{i\theta_{j}}\sum_{j',j''[j-const]} \Delta_0\,
f_{\vec{r}_{j'},c}^{\dag}\,f_{\vec{r}_{j''},c}^{\dag}
\,b_{\vec{r}_{j'j''},s1,d,l,g}^{\dag} + {(\rm h. c.)} \, .
\label{H-bonds-c-s1-b-mean-00}
\end{equation}
Long-range superconducting order implies the occurrence of
coherent $c$ fermion pairing, so that  
$\langle f_{\vec{r}_{j'},c}^{\dag}\,f_{\vec{r}_{j''},c}^{\dag}\rangle$
is finite. We thus use a mean-field approximation
for which the Hamiltonian (\ref{H-bonds-c-s1-b}) is replaced by,
\begin{equation}
{\hat H}^{bonds}_{eff} = 
\sum_{j=1}^{N_{s1}} e^{i\theta_{j}}\sum_{j',j''[j-const]} \,\Delta_0\,
\langle f_{\vec{r}_{j'},c}^{\dag}\,f_{\vec{r}_{j''},c}^{\dag}\rangle
\,b_{\vec{r}_{j'j''},s1,d,l,g}^{\dag} + {(\rm h. c.)} \, .
\label{H-bonds-c-s1-b-mean}
\end{equation}

According to Eq. (\ref{singlet-conf-simpl-0}), local $c$ fermion pairs with
real-space coordinates $\vec{r}_{j'}$ and $\vec{r}_{j''}$ equal to those
of a given two-site bond refer to the same local rotated-electron
pair. Hence within our approach for pairs with the same centre of mass
$\vec{r}_{j} =[\vec{r}_{j'}+\vec{r}_{j''}]/2\approx [\vec{r}_{j'}+\vec{r}_{j''}]/2 -{\vec{r}_{d,l}}^{\,0}$
and belonging to the same family and thus 
having the same values for the indices $d$ and $l$ (where $d=1,2$ is not
the doublicity) the following relation holds,
\begin{equation}
{\langle f_{\vec{r}_{j'''},c}^{\dag}\,f_{\vec{r}_{j''''},c}^{\dag}\rangle_g
\over \vert\langle f_{\vec{r}_{j'},c}^{\dag}\,f_{\vec{r}_{j''},c}^{\dag}\rangle_0\vert} 
\approx  {2h^*_g\over \vert 2h_0\vert} \, ; \hspace{0.35cm}
\sum_{g=0}^{[N_{s1}/4-1]} \vert 2h_{g}\vert^2 = 1 \, .
\label{ff}
\end{equation}
Here the index $g$ was added to specify the two-site link type.
Hence the expectation values amplitudes are controlled
by the coefficient $h^*_{g}$ of the corresponding
spin two-site bond in the $s1$ fermion
operator defined in Eq. (\ref{g-s1+general-gb}). Such
a relation follows from the two $c$ fermions and
the corresponding two-site bond stemming from the same
spin-singlet rotated-electron pair. The expectation values amplitudes
decrease upon increasing the distance between the
two $c$ fermions of a pair. As discussed in Section III-C,
the physics behind the relation
(\ref{ff}) is that the generator of the spin degrees of freedom of
the overall occupancy configuration generated
by each of the $N/2$ spin-singlet rotated-electron pair
operators of the $j$ summation of Eq. (\ref{H-r-el})
is a $s1$ bond-particle operator.

Indeed, the use of the relation (\ref{ff}) and $s1$ bond-particle 
operator expression (\ref{g-s1+general-gb}) in 
the effective Hamiltonian of Eq. (\ref{H-bonds-c-s1-b-mean})
allows after some straightforward algebra to
express it in terms of a $s1$ bond-particle operator
as follows,
\begin{equation}
{\hat H}^{bonds}_{eff} = 
\sum_{j=1}^{N_{s1}} \sum_{\langle j',j''\rangle} e^{i\theta_{j}}
{\Delta_0 \over 4\vert h_0\vert}
\langle f_{\vec{r}_{j'},c}^{\dag}\,f_{\vec{r}_{j''},c}^{\dag}\rangle
\,b_{\vec{r}_{j},s1}^{\dag} + {(\rm h. c.)} \, .
\label{H-bonds-c-s1-bond}
\end{equation}
The use of the transformation (\ref{JW-f+}) and replacement of
$\phi_{j,s1}$ by the phase $\phi^0_{j,s1}$ of Eq. (\ref{theta-cp})
allows expressing (\ref{H-bonds-c-s1-bond}) 
in terms of the $s1$ fermion operator $f_{\vec{r}_{j},s1}^{\dag}$ as follows,
\begin{equation}
{\hat H}^{bonds}_{eff} = 
\sum_{j=1}^{N_{s1}} \sum_{\langle j',j''\rangle} 
e^{i\theta_{cp}}{\Delta_0\over 4\vert h_0\vert}
\langle f_{\vec{r}_{j'},c}^{\dag}\,f_{\vec{r}_{j''},c}^{\dag}\rangle
\,f_{\vec{r}_{j},s1}^{\dag} + {(\rm h. c.)} \, .
\label{H-bonds-c-s1-f-mean}
\end{equation}

Finally, note that (\ref{H-bonds-c-s1-f-mean}) is the effective Hamiltonian 
obtained by replacing in Eq. (\ref{H-bonds-c-s1-f}) the operator
$f_{\vec{r}_{j'},c}^{\dag}\,f_{\vec{r}_{j''},c}^{\dag}$
by $\langle f_{\vec{r}_{j'},c}^{\dag}\,f_{\vec{r}_{j''},c}^{\dag}\rangle$. 
This is alike for the effective Hamiltonians (\ref{H-bonds-c-s1-b-mean-00}) and
(\ref{H-bonds-c-s1-b-mean}). Since (\ref{H-bonds-c-s1-b}) and
(\ref{H-bonds-c-s1-b-mean-00}) are the same Hamiltonian terms,
one concludes that those are 
approximately equivalent to (\ref{H-bonds-c-s1-f}).

\section{The energy range of the $c$ - $s1$ fermion interactions behind $c$ fermion strong
effective coupling}

Here we derive the $c$ fermion energy range of the $c$ - $s1$ fermion interactions within a virtual-electron pair
configuration. Our results refer to the interactions that lead to $c$ fermion strong effective coupling. That range 
is expressed as a function of the $c$ fermion hole momenta $\pm {\vec{q}}^{\,h}$ and $s1$ fermion momentum ${\vec{q}}$.  
We start by considering the square-lattice quantum liquid perturbed by weak 3D uniaxial anisotropy
without hole-trapping effects. For it there is short-range spin order for $0<x\ll 1$ and the critical hole
concentration $x_c$ above which there is long-range superconducting order reads $x_c\approx {\rm Gi}$ 
rather than $x_c \approx {\rm Gi} + x_0$. Thereafter, we extrapolate our analysis to the range $x\in (x_c,x_*)$ 
of the VEP quantum liquid. 

The absolute maximum magnitude of the ratio 
$\vert\Omega_{s1} ({\vec{q}}^{\,d}_{arc})\vert/\Delta_0$ where $\vert\Omega_{s1} ({\vec{q}}^{\,d}_{arc})\vert$ 
is the pairing energy per electron of Eq. (\ref{omega-gene}) gives for $\gamma_d=1$ the maximum  
rate concerning the energy that the short-range spin correlations can supply to $c$ fermion strong effective coupling through the 
$c$ - $s1$ fermion interactions within a virtual-electron pair configuration. Indeed, in expression (\ref{omega-gene})
the suppression coefficient $\gamma_d$ accounts for the part of that energy used in phase-coherent pairing. 
For $\gamma_d=1$ such a ratio reaches its absolute maximum magnitude $\vert\Omega_{s1} ({\vec{q}}^{\,d\,N}_{Bs1})\vert/\Delta_0 = \gamma_c/4$
at zero temperature, ${\vec{q}}^{\,d}_{arc}={\vec{q}}^{\,d\,N}_{Bs1}$, and $x=x_{op}$. Importantly,
as confirmed by Eq. (\ref{W-ec}) the ratio $\Delta_0/W_{ec}=\gamma_c/4$ exactly equals that maximum magnitude.

In the absence of hole trapping effects there is $c$ fermion strong effective coupling for $0<x<x_*$,
yet it leads to phase-coherent pairing only for $x>x_c$. The above rate ratio 
$\vert\Omega_{s1} ({\vec{q}}^{\,d\,N}_{Bs1})\vert/\Delta_0 =\Delta_0/W_{ec}=\gamma_c/4$
plays an important role in the $c$ - $s1$ fermion interactions. Indeed the $c$ fermion energy range of 
these interactions contributing to strong effective coupling is for $0<x\ll 1$ controlled by a function 
given by,
\begin{equation}
\vert\Delta_{ec} ({\vec{q}}^{\,h})\vert ={\Delta_0\over W_{ec}}\,\vert\epsilon_c ({\vec{q}}^{\,h})\vert
\, ; \hspace{0.35cm}
\epsilon_c ({\vec{q}}^{\,h})=
-W_{ec}\,{\vert\Delta_{ec} ({\vec{q}}^{\,h})\vert\over \Delta_0}  \, .
\label{epsilon-c-W-c}
\end{equation}
It is such that,
\begin{equation}
\vert\Delta_{ec} ({\vec{q}}^{\,h})\vert = \vert\Delta_{ec} (-{\vec{q}}^{\,h})\vert\in (0,\Delta_0) \, ; \hspace{0.35cm}
\vert\Delta_{ec} ({\vec{q}}_{Fc}^{\,h})\vert = 0
\, ; \hspace{0.35cm}
\vert\Delta_{ec} ({\vec{q}}^{\,h}_{ec})\vert = \Delta_0 \, .
\label{Deltacp-limits}
\end{equation}
The momentum ${\vec{q}}^{\,h}_{ec}$ appearing here belongs to the
zero-temperature $ec$-pairing line defined in Eq. (\ref{g-Ul}).
We recall that the energy scale $W_{ec}$ of Eq. (\ref{W-ec}) is both the maximum energy bandwidth corresponding
to the hole momentum domain $Q^{c}_{ec}$ of $c$ fermions with
strong effective coupling and the maximum magnitude of the energy bandwidth of the superconducting-ground-state 
sea of $c$ fermions contributing to phase-coherent pairing. 
Alike the maximum pairing energy $2\vert\Omega_{s1} ({\vec{q}}^{\,d\,N}_{Bs1})\vert$ is smaller than
the energy $2\Delta_0$ associated with the short-range spin correlations, also
the energy $\vert\Delta_{ec} ({\vec{q}}^{\,h})\vert +\vert\Delta_{ec} (-{\vec{q}}^{\,h})$
is smaller than the $c$ fermion energy $\vert\epsilon_c ({\vec{q}}^{\,h})\vert+\vert\epsilon_c (-{\vec{q}}^{\,h})\vert$.
Indeed, for $\gamma_d=1$ the maximum magnitude of the 
ratio $2\vert\Omega_{s1} ({\vec{q}}^{\,d\,N}_{Bs1})\vert/2\Delta_0$ exactly
equals $[\vert\Delta_{ec} ({\vec{q}}^{\,h})\vert +\vert\Delta_{ec} (-{\vec{q}}^{\,h})]/
[\vert\epsilon_c ({\vec{q}}^{\,h})\vert+\vert\epsilon_c (-{\vec{q}}^{\,h})\vert]$.

In the present $0<x\ll 1$ limit
the energy scale $2\Delta_0$ of Eq. (\ref{Delta-0}) of Appendix A is both the maximum 
magnitude of the $s1$ fermion spinon-pairing energy and the maximum magnitude 
of the $c$ fermion energy scale $\vert\Delta_{ec} ({\vec{q}}^{\,h})\vert +
\vert\Delta_{ec} (-{\vec{q}}^{\,h})\vert$ of Eq. (\ref{Deltacp-limits}). For $0<x\ll 1$ it controls
the energy bandwidth of the following ranges,
\begin{equation}
2\vert\Delta_{s1} ({\vec{q}})\vert \in (0,2\Delta_0) \, ;
\hspace{0.35cm} \vert\Delta_{ec} ({\vec{q}}^{\,h})\vert
+ \vert\Delta_{ec} (-{\vec{q}}^{\,h})\vert \in (0,2\Delta_0)  \, .
\label{2Deltac-limits}
\end{equation}

Consistently with the ranges (\ref{2Deltac-limits}) and the above analysis,
the energy scale $2\Delta_0$ is for $0<x\ll 1$ the maximum energy
of the short-range spin correlations. It is the source of the energy supplied 
through the $c$ - $s1$ fermion interactions within each virtual-electron pair
configuration, to ensure $c$ fermion strong effective coupling. The occurrence 
of such a strong effective coupling is a necessary condition for 
virtual-electron pairing phase coherence. This role of the energy scale $2\Delta_0$ 
is consistent with: i) $2\Delta_0=\lim_{x\rightarrow 0}2\vert\Delta\vert$
being the maximum magnitude of the $s1$ fermion spinon pairing energy
associated with the short-range spin correlations; ii) The presence of $\Delta_0$ in the Hamiltonian terms of 
Eq. (\ref{H-bonds-c-s1-b}), which control the elementary processes associated 
the $c$ - $s1$ fermion interactions.

Only at zero temperature and for $x\rightarrow 0$ all phases $\theta_{j,1}$ line up.
Only then $g_1 = \vert\langle e^{i\theta_{j,1}}\rangle\vert\approx 1$ and
the pseudogap energy $2\vert\Delta\vert = g_1\,2\Delta_0$ of Eq. (\ref{2Delta})
reaches its maximum magnitude $2\Delta_0$. Hence for finite 
hole concentrations and/or temperatures the pseudogap energy  
$2\vert\Delta\vert = g_1\,2\Delta_0$ gives the suppressed magnitude
of the energy scale $2\Delta_0$ due to the fluctuations of the phases $\theta_{j,1}$.
That pseudogap is the order parameter of the short-range spin
correlations. As a result of $2\Delta_0$ being the maximum energy of the short-range spin correlations,
at small hole concentrations $0<x \ll 1$ the overall quantity $\vert\Delta_{ec} ({\vec{q}}^{\,h})\vert +
\vert\Delta_{ec} (-{\vec{q}}^{\,h})\vert + 2\vert\Delta_{s1} ({\vec{q}})\vert$ must be equal or
smaller than $2\Delta_0$. The ranges provided in Eq. (\ref{2Deltac-limits}) correspond to two
limiting cases of such an overall inequality: i) The range $2\vert\Delta_{s1} ({\vec{q}})\vert \in (0,2\Delta_0)$
holds for $c$ band hole momenta $\pm{\vec{q}}^{\,h}$ at the $c$ Fermi line,
so that $\vert\Delta_{ec} ({\vec{q}}^{\,h})\vert =\vert\Delta_{ec} (-{\vec{q}}^{\,h})\vert=0$;
ii) The range $\vert\Delta_{ec} ({\vec{q}}^{\,h})\vert
+ \vert\Delta_{ec} (-{\vec{q}}^{\,h})\vert \in (0,2\Delta_0)$
holds for $s1$ band momenta ${\vec{q}}$ pointing in the nodal directions,
so that $2\vert\Delta_{s1} ({\vec{q}})\vert=0$.

Hence for $0<x \ll 1$ the residual interactions of $c$ fermions with general 
hole momenta ${\vec{q}}^{\,h}$ and $-{\vec{q}}^{\,h}$ and energy
$\epsilon_c ({\vec{q}}^{\,h})=\epsilon_c (-{\vec{q}}^{\,h})=
-W_{ec}\,[\vert\Delta_{ec} ({\vec{q}}^{\,h})\vert/\Delta_0]$
with a $s1$ fermion of general momentum ${\vec{q}}$ and
spinon-pairing energy $2\vert\Delta_{s1} ({\vec{q}})\vert$
contribute to the strong effective pairing coupling of the
former two objects provided that,
\begin{equation}
0\leq\vert\Delta_{ec} ({\vec{q}}^{\,h})\vert +
\vert\Delta_{ec} (-{\vec{q}}^{\,h})\vert +
2\vert\Delta_{s1} ({\vec{q}})\vert \leq 2\Delta_0 \, .
\label{range-0-ec}
\end{equation}
This inequality is equivalent to,
\begin{equation}
\vert\epsilon_c (\vec{q}^{\,h})\vert \leq 
W_{ec}\left(1-{\vert\Delta_{s1} ({\vec{q}})\vert\over\Delta_0}\right) \, .
\label{range-1-ec}
\end{equation}

Generalization of the $0<x \ll 1$ inequality (\ref{range-1-ec})
to finite hole concentrations below $x_*$ and $U/4t\in (u_0,u_{\pi})$
involves the replacement of the energy parameter $\Delta_0=\lim_{x\rightarrow 0}\vert\Delta\vert$ 
by the general spin energy scale $\vert\Delta\vert$ in the ratio 
$[\vert\Delta_{s1} ({\vec{q}})\vert/\Delta_0]$. For
finite hole concentrations below $x_*$ it then becomes the inequality provided in
Eq. (\ref{range-2-ec}). Due to the hole trapping effects discussed in Appendix B, which 
are strongest for $x\in (0,x_0)$, for the VEP quantum liquid the latter inequality 
is valid for $x\in (x_0,x_*)$ rather than for $x\in (0,x_*)$.

Strong effective coupling is required for the occurrence of phase-coherent virtual-electron 
pairing. However, strong effective coupling also occurs in the pseudogap state. That
$2\vert\Omega\vert =0$ does not affect though the validity of the inequalities provided in
Eqs. (\ref{range-2-ec}) and (\ref{range-1-ec}). It defines the $c$ fermion energy range of
the $c$ - $s1$ fermion interactions that lead to strong effective coupling
independently on whether it is associated with phase-coherent virtual-electron pairing or not.

\section{The $s1$ boundary line nodal and anti-nodal momenta}

The controlled approximations used to derive the following expressions of the
$s1$ boundary-line absolute-value momenta $q^N_{Bs1}$ and $q^{AN}_{Bs1}$ rely
on the change of the $s1$ boundary line shape and size of the square-lattice quantum liquid of Ref. \cite{companion2}
from a square of edge magnitude $\sqrt{2}\pi$ and length $\sqrt{2}\,4\pi$ for $x\rightarrow 0$ to a circle of
radius $\sqrt{(1-x)2\pi}$ and length $(1-x)2\pi^2$ for $x\rightarrow 1$. For $U/4t\in (u_0,u_1)$ the latter shape 
is a good approximation for $x>x_{c3}$ where $x_{c3}\approx (x_{c1} + x_*)\approx 0.40$ is defined below and 
$x_{c1}= 1/8$. The obtained approximate expressions read,
\begin{eqnarray}
q^N_{Bs1} & \approx & {\pi\over\sqrt{2}}(1-x) \, , \hspace{0.35cm} x\in (x_c,x_{c1}) \, ,
\nonumber \\
& \approx & {7\pi\over\sqrt{2}\,8} = \sqrt{(1-x_{c3})2\pi} \, , \hspace{0.35cm} x\in (x_1,x_{c3}) \, ,
\nonumber \\
& \approx & \sqrt{(1-x)2\pi} \, , \hspace{0.35cm} x\in (x_{c3},1) \, ;
\nonumber \\
q^{AN}_{Bs1} & \approx & \pi \left[1 -{1\over 2\pi}\tanh \left(\sqrt{\pi\,x\over x_{c1}}\right)\right] \, , \hspace{0.35cm} x\in (x_c,x_{c1}) \, ,
\nonumber \\
& \approx & C_2\,\sqrt{(1-x)2\pi}
\, , \hspace{0.35cm} x\in (x_{c1},x_{c3}) \, ,
\nonumber \\
& \approx & \sqrt{(1-x)2\pi} \, , \hspace{0.35cm} x\in (x_{c3},1) \, ,
\label{qN-qAN}
\end{eqnarray}
respectively. Here,
\begin{eqnarray}
C_1 & \equiv & C_A\vert_{x=x_{c1}} = {8\over 7}\left[1 -{1\over 2\pi}\tanh (\sqrt{\pi})\right] 
\, ; \hspace{0.35cm}  
C_2 = {C_1\,\sqrt{7\pi}\over 4}\,\left({x_3 -x\over x_3 -x_{c1}}\right) \, ,
\nonumber \\
x_{c3} & = & 1 - {49\pi\over 256} \approx (x_{c1}+x_*) \approx 0.40 \, ; \hspace{0.35cm}
x_3 = x_{c3} + {(x_{c3}-x_{c1})\over ({C_1\sqrt{7\pi}\over 4}-1)} \, .
\label{x3-C1}
\end{eqnarray}
The coefficient $C_A \equiv q^{AN}_{Bs1}/\sqrt{2}q^N_{Bs1}$ has then the following approximate limiting behaviors,
\begin{eqnarray}
C_A \equiv {q^{AN}_{Bs1}\over\sqrt{2}q^N_{Bs1}} & \approx & {1 -{1\over 2\pi}\tanh (\sqrt{\pi\,x/x_{c1}})\over 1-x} \, , \hspace{0.35cm} x\in (x_c,x_{c1})  \, ,
\nonumber \\
& \approx & C_1\,\sqrt{8\over 7}\left({x_3 -x\over x_3 -x_{c1}}\right)\sqrt{1-x} \, , \hspace{0.35cm} x\in (x_{c1},x_{c3}) \, ,
\nonumber \\
& \approx & {1\over\sqrt{2}} \, , \hspace{0.35cm} x\in (x_{c3},1) \, .
\label{C-A}
\end{eqnarray}

As discussed in Ref. \cite{companion2}, for approximately $U/4t  >u_0$ and hole concentrations in the
range $x\in (0,x_{c1})$ the shape of the $s1$ boundary line of the square-lattice quantum liquid is independent of $U/4t$. This 
justifies the independence of $U/4t$ for that $x$ range of the $q^N_{Bs1}$ and $q^{AN}_{Bs1}$ 
expressions given in Eq. (\ref{qN-qAN}) for $x\in (x_c,x_{c1})$. Indeed, the results of that reference
apply to the latter $x$ range. (The behavior $q^{AN}_{Bs1} \approx \pi - \sqrt{2\pi\,x}$ 
reported in Ref. \cite{companion2} is valid for $x\ll 1$ rather than for $x<x_{c1}$ \cite{missprint}; Note
also that for $U/4t>u_0$ the expression $q^{AN}_{Bs1} \approx \pi (1 -[1/2\pi]\tanh (\sqrt{8\pi\,x}))$ of 
Eq. (\ref{qN-qAN}) applies; In turn, the crossover hole concentration magnitude $x_{c1}=1/8$ 
refers only to the range $U/4t\in (u_0,u_1)$, for which that expression
reads $q^{AN}_{Bs1} \approx \pi (1 -[1/2\pi]\tanh (\sqrt{\pi\,x/x_{c1}}))$, as given in Eq. (\ref{qN-qAN}).) 

For the square-lattice quantum liquid of Ref. \cite{companion2}
the coefficient $C_A = q^{AN}_{Bs1}/\sqrt{2}q^N_{Bs1}$ of Eq. (\ref{C-A})
exactly reads $C_A=1$ in the limit $x\rightarrow 0$. In the range 
$x\in (0,x_{c1})$ for that liquid and $x\in (x_c,x_{c1})$ for the VEP quantum liquid
the overall decreasing of $q^{AN}_{Bs1}$ and $q^N_{Bs1}$ is very similar, so that $C_A =C_1\approx 0.97$
at $x=x_{c1}= 1/8$. At that $x$ value the magnitude of $q^N_{Bs1}$ equals the radius $\sqrt{(1-x_{c3})2\pi}$
of the nearly circular $s1$ boundary line at a larger hole concentration 
$x=x_{c3}\approx (x_{c1}+x_*)$. In the range $x\in (x_{c1},x_{c3})$ the $q^N_{Bs1}$ 
magnitude decreases very little. This justifies why within our approximation $q^N_{Bs1}\approx [\pi/\sqrt{2}](1-x_{c1})=
\sqrt{(1-x_{c3})2\pi}$ for $x\in (x_{c1},x_{c3})$, as given in Eq. (\ref{qN-qAN}). Since for $x>x_{c3}$ it
is a reasonable good approximation to consider that the $s1$ boundary line is a circle of
radius $\sqrt{(1-x)2\pi}$, it is assumed that $q^{AN}_{Bs1}\approx C_2\,\sqrt{(1-x)2\pi}$ for
the range $x\in (x_{c1},x_{c3})$, $C_2$ reaching the value $C_2=1$ at $x=x_{c3}$. For the latter $x$ range the 
magnitudes of both $q^{AN}_{Bs1}$ and $C_2$ are decreasing functions of $x$. While the exact 
form of the decreasing function $C_2 =C_2 (x)$ remains an open problem, that 
$C_2\approx \sqrt{\pi(1-x_{c1})/2}$ and $C_2=1$ for $x=x_{c1}$ and $x=x_{c3}$, respectively, is expected to be 
a good approximation. As given in Eq. (\ref{x3-C1}), here we assume that the coefficient
$C_2$ decreases linearly as $C_2\propto (x_3 -x)$ where $x_3$ is provided in that equation. 
That is the simplest curve connecting the above $C_2$ magnitudes at $x=x_{c1}$ and $x=x_{c3}$,
respectively.

%%%%%%%%%%%%%%%%%%%%%%%%%%%%%%%%%%%%%%%%%%%%%%%%%%%%%%%%%%%%%%%%%%%%%%%%%%


\begin{references}
\bibitem{missprint}
	The $q^N_{Bs1}$ and $q^N_{s1,arc}$ expressions given in Eq. (99) of Ref. \cite{companion2}
	are valid for $x<x_{c1}$. However, the $q^{AN}_{Bs1}$ expression provided in that
	equation holds for $x\ll 1$. That as given in that equation it is valid for $x<x_{c1}$ is
	a misprint.
\end{references}
\end{document}